\RequirePackage{lineno}
\documentclass[prc,floatfix,twocolumn,superscriptaddress,letter]{revtex4}

\usepackage{color}
\usepackage[normalem]{ulem}
\input epsf

\usepackage{amsmath}
\usepackage{graphicx}
\usepackage{url}
\usepackage{color}
\usepackage{subfigure}
\usepackage{longtable}
\usepackage{dcolumn}

\newcommand{\GPDtE}{\langle \tilde{E} \rangle}
\newcommand{\GPDtH}{\langle \tilde{H} \rangle}
\newcommand{\GPDHT}{\langle H_T \rangle}
\newcommand{\GPDETbar}{\langle \bar{E}_T \rangle}

\begin{document}
\setpagewiselinenumbers


\title{Exclusive $\pi^0$ electroproduction at $W>2$ GeV with CLAS}

\newcommand*{\ANL}{Argonne National Laboratory, Argonne, Illinois 60439}
\newcommand*{\ANLindex}{1}
\affiliation{\ANL}
\newcommand*{\ASU}{Arizona State University, Tempe, Arizona 85287-1504}
\newcommand*{\ASUindex}{2}
\affiliation{\ASU}
\newcommand*{\CSUDH}{California State University, Dominguez Hills, Carson, CA 90747}
\newcommand*{\CSUDHindex}{3}
\affiliation{\CSUDH}
\newcommand*{\CMU}{Carnegie Mellon University, Pittsburgh, Pennsylvania 15213}
\newcommand*{\CMUindex}{4}
\affiliation{\CMU}
\newcommand*{\CUA}{Catholic University of America, Washington, D.C. 20064}
\newcommand*{\CUAindex}{5}
\affiliation{\CUA}
\newcommand*{\SACLAY}{CEA, Centre de Saclay, Irfu/Service de Physique Nucl\'eaire, 91191 Gif-sur-Yvette, France}
\newcommand*{\SACLAYindex}{6}
\affiliation{\SACLAY}
\newcommand*{\CNU}{Christopher Newport University, Newport News, Virginia 23606}
\newcommand*{\CNUindex}{7}
\affiliation{\CNU}
\newcommand*{\UCONN}{University of Connecticut, Storrs, Connecticut 06269}
\newcommand*{\UCONNindex}{8}
\affiliation{\UCONN}
\newcommand*{\FU}{Fairfield University, Fairfield CT 06824}
\newcommand*{\FUindex}{9}
\affiliation{\FU}
\newcommand*{\FIU}{Florida International University, Miami, Florida 33199}
\newcommand*{\FIUindex}{10}
\affiliation{\FIU}
\newcommand*{\FSU}{Florida State University, Tallahassee, Florida 32306}
\newcommand*{\FSUindex}{11}
\affiliation{\FSU}
\newcommand*{\GWUI}{The George Washington University, Washington, DC 20052}
\newcommand*{\GWUIindex}{12}
\affiliation{\GWUI}
\newcommand*{\ISU}{Idaho State University, Pocatello, Idaho 83209}
\newcommand*{\ISUindex}{13}
\affiliation{\ISU}
\newcommand*{\INFNFE}{INFN, Sezione di Ferrara, 44100 Ferrara, Italy}
\newcommand*{\INFNFEindex}{14}
\affiliation{\INFNFE}
\newcommand*{\INFNFR}{INFN, Laboratori Nazionali di Frascati, 00044 Frascati, Italy}
\newcommand*{\INFNFRindex}{15}
\affiliation{\INFNFR}
\newcommand*{\INFNGE}{INFN, Sezione di Genova, 16146 Genova, Italy}
\newcommand*{\INFNGEindex}{16}
\affiliation{\INFNGE}
\newcommand*{\INFNRO}{INFN, Sezione di Roma Tor Vergata, 00133 Rome, Italy}
\newcommand*{\INFNROindex}{17}
\affiliation{\INFNRO}
\newcommand*{\ORSAY}{Institut de Physique Nucl\'eaire ORSAY, Orsay, France}
\newcommand*{\ORSAYindex}{18}
\affiliation{\ORSAY}
\newcommand*{\ITEP}{Institute of Theoretical and Experimental Physics, Moscow, 117218, Russia}
\newcommand*{\ITEPindex}{19}
\affiliation{\ITEP}
\newcommand*{\JMU}{James Madison University, Harrisonburg, Virginia 22807}
\newcommand*{\JMUindex}{20}
\affiliation{\JMU}
\newcommand*{\KNU}{Kyungpook National University, Daegu 702-701, Republic of Korea}
\newcommand*{\KNUindex}{21}
\affiliation{\KNU}
\newcommand*{\LPSC}{LPSC, Universite Joseph Fourier, CNRS/IN2P3, INPG, Grenoble, France}
\newcommand*{\LPSCindex}{22}
\affiliation{\LPSC}
\newcommand*{\UNH}{University of New Hampshire, Durham, New Hampshire 03824-3568}
\newcommand*{\UNHindex}{23}
\affiliation{\UNH}
\newcommand*{\NSU}{Norfolk State University, Norfolk, Virginia 23504}
\newcommand*{\NSUindex}{24}
\affiliation{\NSU}
\newcommand*{\OHIOU}{Ohio University, Athens, Ohio  45701}
\newcommand*{\OHIOUindex}{25}
\affiliation{\OHIOU}
\newcommand*{\ODU}{Old Dominion University, Norfolk, Virginia 23529}
\newcommand*{\ODUindex}{26}
\affiliation{\ODU}
\newcommand*{\RPI}{Rensselaer Polytechnic Institute, Troy, New York 12180-3590}
\newcommand*{\RPIindex}{27}
\affiliation{\RPI}
\newcommand*{\URICH}{University of Richmond, Richmond, Virginia 23173}
\newcommand*{\URICHindex}{28}
\affiliation{\URICH}
\newcommand*{\ROMAII}{Universita' di Roma Tor Vergata, 00133 Rome Italy}
\newcommand*{\ROMAIIindex}{29}
\affiliation{\ROMAII}
\newcommand*{\MSU}{Skobeltsyn Institute of Nuclear Physics, Lomonosov Moscow State University, 119234 Moscow, Russia}
\newcommand*{\MSUindex}{30}
\affiliation{\MSU}
\newcommand*{\SCAROLINA}{University of South Carolina, Columbia, South Carolina 29208}
\newcommand*{\SCAROLINAindex}{31}
\affiliation{\SCAROLINA}
\newcommand*{\JLAB}{Thomas Jefferson National Accelerator Facility, Newport News, Virginia 23606}
\newcommand*{\JLABindex}{32}
\affiliation{\JLAB}
\newcommand*{\UTFSM}{Universidad T\'{e}cnica Federico Santa Mar\'{i}a, Casilla 110-V Valpara\'{i}so, Chile}
\newcommand*{\UTFSMindex}{33}
\affiliation{\UTFSM}
\newcommand*{\EDINBURGH}{Edinburgh University, Edinburgh EH9 3JZ, United Kingdom}
\newcommand*{\EDINBURGHindex}{34}
\affiliation{\EDINBURGH}
\newcommand*{\GLASGOW}{University of Glasgow, Glasgow G12 8QQ, United Kingdom}
\newcommand*{\GLASGOWindex}{35}
\affiliation{\GLASGOW}
\newcommand*{\VT}{Virginia Polytechnic Institute and State University, Blacksburg, Virginia   24061-0435}
\newcommand*{\VTindex}{36}
\affiliation{\VT}
\newcommand*{\VIRGINIA}{University of Virginia, Charlottesville, Virginia 22901}
\newcommand*{\VIRGINIAindex}{37}
\affiliation{\VIRGINIA}
\newcommand*{\WM}{College of William and Mary, Williamsburg, Virginia 23187-8795}
\newcommand*{\WMindex}{38}
\affiliation{\WM}
\newcommand*{\YEREVAN}{Yerevan Physics Institute, 375036 Yerevan, Armenia}
\newcommand*{\YEREVANindex}{39}
\affiliation{\YEREVAN}

\newcommand*{\NOWUTFSM}{Universidad T\'{e}cnica Federico Santa Mar\'{i}a, Casilla 110-V Valpara\'{i}so, Chile}
\newcommand*{\NOWUCONN}{University of Connecticut, Storrs, Connecticut 06269}
\newcommand*{\NOWGLASGOW}{University of Glasgow, Glasgow G12 8QQ, United Kingdom}
\newcommand*{\NOWROMAII}{Universita' di Roma Tor Vergata, 00133 Rome Italy}


\author {I.~Bedlinskiy}
\affiliation{\ITEP}

\author {V.~Kubarovsky} 
\affiliation{\JLAB}
\affiliation{\RPI}

\author {S.~Niccolai} 
\affiliation{\ORSAY}
\affiliation{\GWUI}
\author {P.~Stoler} 
\affiliation{\RPI}

\author {K.P.~Adhikari} 
\affiliation{\ODU}
\author {M.D.~Anderson} 
\affiliation{\GLASGOW}
\author {S.~Anefalos~Pereira} 
\affiliation{\INFNFR}
\author {H.~Avakian} 
\affiliation{\JLAB}
\author {J.~Ball} 
\affiliation{\SACLAY}
\author {N.A.~Baltzell} 
\affiliation{\ANL}
\affiliation{\SCAROLINA}
% ...
\collaboration{The CLAS Collaboration}
\noaffiliation

\begin{abstract}
Exclusive neutral-pion electroproduction ($ep\to e^\prime p^\prime \pi^0$) was measured at Jefferson Lab with a 5.75-GeV electron beam and the CLAS detector. Differential cross sections $d^4\sigma/dtdQ^2dx_Bd\phi_\pi$  and structure functions $\sigma_T+\epsilon\sigma_L, \sigma_{TT}$ and $\sigma_{LT}$ as functions of $t$ were obtained over a wide range of $Q^2$ and $x_B$. The data are compared with Regge and handbag theoretical calculations. Analyses in both  frameworks  find  that a large dominance of transverse processes is necessary to explain the experimental results.  For the Regge analysis it is found that the inclusion of vector meson rescattering processes is necessary to bring the magnitude of the calculated and measured structure functions into rough agreement. In the  handbag framework, there are two independent  calculations, both of which appear to roughly explain the magnitude of the structure functions in terms of transversity generalized parton distributions. 
\end{abstract}

\date{\today}

\maketitle

\section{Introduction}

Understanding nucleon structure in terms of the fundamental degrees of freedom of Quantum Chromodynamics (QCD) is one of the main goals in the theory of strong interactions.  The nucleon is a many-body system of quarks and gluons. How partons move and how they are distributed in space is still an open question on which new theoretical and experimental developments are starting to shed a new light. The study of deep inelastic scattering provides the distribution of longitudinal momentum and polarization carried by quarks and antiquarks within the fast moving hadron.  However, the spatial distribution of the partons in the plane perpendicular to the hadron motion is not accessible in these experiments. The role of the partons' orbital angular momenta in making up the total spin of the nucleon is one more unresolved question. In recent years it became clear that  exclusive reactions may provide such  information encoded in so-called Generalized Parton Distributions (GPDs) ~\cite{Ji,Radyushkin}. The GPDs describe the simultaneous distribution of partons with respect to both the partons'  transverse positions and longitudinal momenta. In addition to the information about  transverse  spatial density (form factors) and momentum density, these functions reveal the correlation of the spatial and momentum distributions, i.e. how the spatial shape of the nucleon changes when probing quarks of different longitudinal momenta. GPDs give access as well to the total angular momentum carried by partons, comprising the spin and orbital parts~\cite{Ji}.

\begin{figure*}
\includegraphics[width=\textwidth]{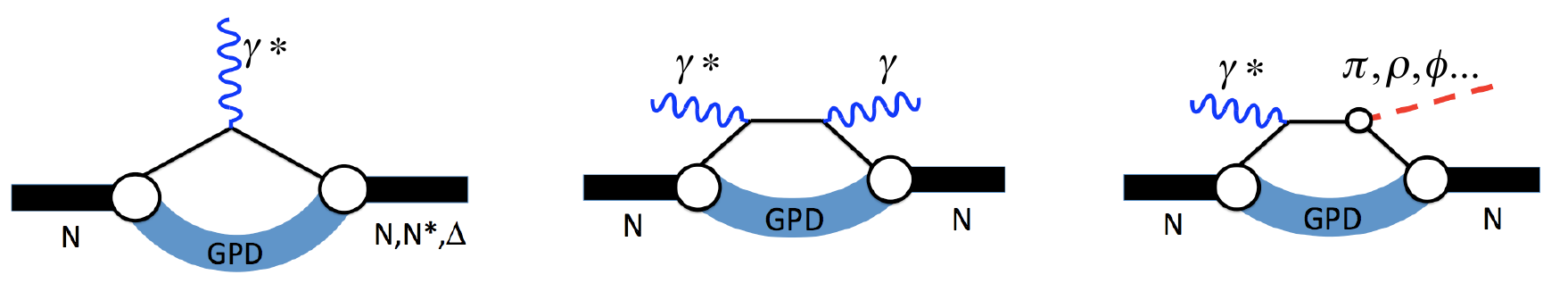}
\caption{\label{handbag} Schematic diagram of the lowest order handbag mechanism applied to: (left) elastic scattering, (middle) DVCS and (right) meson production.
}
\end{figure*}

The possibility to study GPDs in exclusive scattering processes rests on factorization theorems, which are  proven for  virtual Compton scattering
\cite{factorization-compton} and light meson electroproduction \cite{factorization-mesons} in the limit of $Q^2 \to \infty$,  at fixed $x_B$ and $t$.  
Here, $q^2 \equiv -Q^2$ is the square of the 4-momentum transferred  to the hadronic system by the scattered electron, $-t$ is the 4-momentum transferred to the recoiling proton
and $x_B$ is the Bjorken variable. 
These proofs are based on the properties of matrix elements represented by Feynman diagrams colloquially referred to as  handbags ~\cite{Ji,Radyushkin,Muller}.
The reaction  is factorized into two parts.
One part treats the elementary interaction with one of the partons in the nucleon perturbatively, while the non-perturbative remainder is embodied in GPDs.   
While the perturbative process between the virtual photon and the quark is reaction dependent, the information contained within the GPDs is universal.  Figure \ref{handbag} indicates the lowest order handbag mechanism applied to three reactions: elastic scattering, deeply virtual Compton scattering (DVCS) and deeply virtual meson electroproduction (DVMP), which is the subject of this article.

While the handbag mechanism should be mostly applicable at asymptotically  large photon virtuality $Q^2$, there is some experimental evidence ~\cite{dvcs-halla} that  the DVCS reaction at $Q^2$ as low as 1.5~GeV$^2$  appears to be applicable by the handbag mechanism.  This is not unexpected since both vertices of the Compton scattering reaction from a single quark  involve perturbative electromagnetic processes.  
On the other hand, for DVMP,  the second vertex ($\pi qq$ in the right plot of Fig.~\ref{handbag}) involves the exchange of at least one gluon, and the kinematic range of leading-order applicability of the handbag formalism is not as clearly determined. 

There are eight GPDs. Four correspond to  parton helicity-conserving (chiral-even) processes,  denoted 
by $H^q$,  $\tilde H^q$,  $E^q$ and  $\tilde E^q$, and 
four correspond to parton helicity-flip (chiral-odd) processes  \cite{diehl,ji},  $H^q_T$,  $\tilde H^q_T$,  $E^q_T$ and  $\tilde E^q_T$. 
At a given $Q^2$ the GPDs depend on three kinematic variables: $x$, $\xi$ and $t$. In  a symmetric frame,  $x$ is the   average longitudinal momentum fraction of the struck parton before and after the hard interaction and $\xi$ (skewness) is half of the longitudinal momentum fraction transferred to  the struck parton. The skewness can be expressed in terms of the   Bjorken variable $x_B$  as
$\xi\simeq x_B/(2-x_B)$. Here $x_B=Q^2/(2p\cdot q)$ and $t=(p-p^\prime)^2$, where $p$ and $p^\prime$ are the initial and final four-momenta of the nucleon. The GPDs encode both the longitudinal momentum distributions through their dependence on $x$ and the transverse position distributions through  their dependence on  $t$.

In the forward limit where $t\to 0$,  $H^q$ and  $\tilde H^q$   reduce to the parton density distributions $q(x)$ and parton helicity distributions $\Delta q(x)$, respectively. The first moments in  $x$ of the chiral-even GPDs are related to the elastic form factors of the nucleon:  the Dirac form factor   $F_1^q(t)$,  the Pauli form factor  $F_2^q(t)$,  the axial-vector form factor  $g_A^q(t)$ and the  pseudoscalar form factor $h_A^q(t)$ \cite{polyakov}. 

\begin{figure}[h]
\includegraphics[width=3.0in]{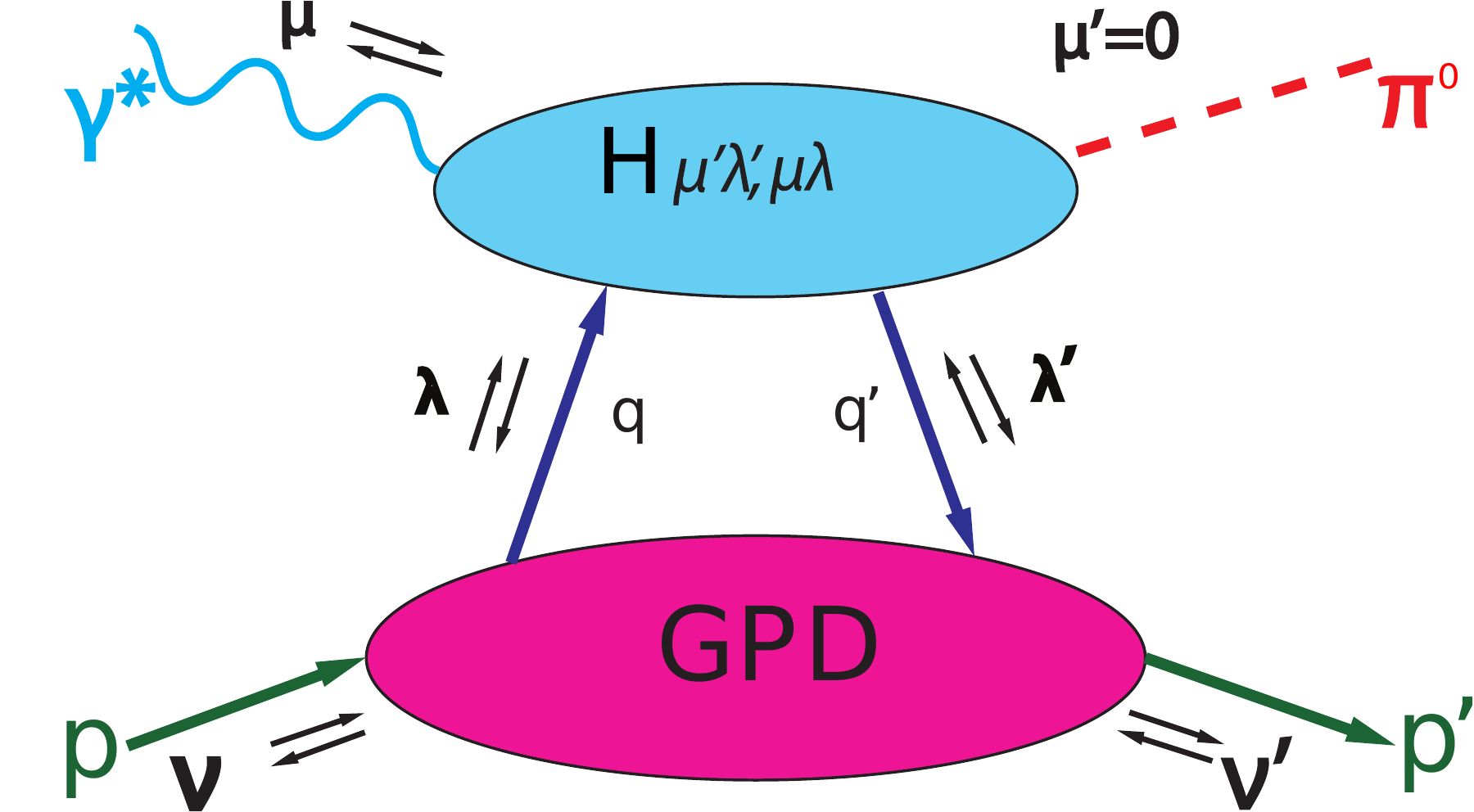}
\caption{\label{fig:handbag-pi0} Schematic diagram of the $\pi^0$ electroproduction amplitude in the framework of  the handbag mechanism.  The helicities of the initial and final nucleons are denoted by $\nu$ and $\nu^\prime$, of the incident photon and produced meson by $\mu$ and $\mu^\prime$ and of the active initial and final quark by $\lambda$ and $\lambda^\prime$. The arrows in the figure schematically represent the corresponding positive and negative helicities, respectively. For final-state photons or vector mesons $\mu^\prime = \pm 1$, while for pseudoscalar mesons $\mu^\prime =  0$.}
\end{figure}

The DVMP process specifically for  $\pi^0$ production is shown in more detail in Fig.~\ref{fig:handbag-pi0}. 

It was shown early-on~\cite{eides-98} that for  pion electroproduction the leading handbag approach  is valid at large $Q^2$ for longitudinal helicity-conserving virtual photons. Using Regge phenomenology as a guide for parametrization of the four longitudinal GPDs, Refs.~\cite{Kroll-Huang, G-K-09} calculated cross-section structure functions for longitudinal helicity-conserving virtual photons. Simultaneously, the CLAS Collaboration as well as other groups~\cite{bedlinskiy,Hall-A-pi0, hermes-transverse}, measured the differential cross sections for pion electroproduction and extracted structure functions, which are the subject of the present paper.
When the theoretical calculations for longitudinal virtual photons were compared with the JLab data, as well as with HERMES data,  they were  found  to  underestimate the measured cross sections by more than an order of magnitude in their accessible kinematic regions, even after including  finite--size  corrections through Sudakov form factors
\cite{G-K-09} . At JLab, sizeable beam-spin asymmetries for exclusive neutral pion electroproduction off the proton were  measured ~\cite{demasi} above the resonance region. These non-zero asymmetries imply that both transverse and longitudinal amplitudes participate in the process.

The failure to describe the experimental results with quark helicity-conserving operators ~\cite{polyakov, Kroll-Huang} stimulated a consideration of the role of the  chiral-odd quark helicity-flip processes. Pseudoscalar meson electroproduction, and in particular $\pi^0$ production in the reaction $ep\to e^\prime p^\prime \pi^0$, was 
identified~\cite{Ahmad:2008hp,G-K-09,G-K-11} 
as especially sensitive to the quark helicity-flip subprocesses. The produced meson has no intrinsic helicity so that the angular momentum of the incident photon is either transferred to the nucleon via a quark helicity-flip or involves  orbital angular momentum processes. 
Evidence of the contribution of helicity-flip subprocesses, especially  $H_T$,   to $\pi^+$ electroproduction in transverse target spin asymmetry data  \cite{hermes-transverse} was noted in Ref.~\cite{G-K-09}.  A disadvantage of   
$\pi^+$ production is that the interpretation is complicated by the dominance  of the longitudinal $\pi^+$-pole term, which is absent in $\pi^0$ production. 
 In addition, for $\pi^0$ production the structure of the amplitudes further suppresses the quark helicity-conserving amplitudes relative to the  helicity-flip  
 amplitudes~\cite{G-K-09}.  On the other hand, 
$\pi^0$ cross sections  over a large kinematic range are much more difficult to obtain than for  $\pi^+$ since the clean detection of $\pi^0$s requires the measurement of their two decay photons.  

During the past few years, two parallel theoretical approaches - \cite{Ahmad:2008hp, 
Goldstein:2010gu}~(GL) and \cite{G-K-09,G-K-11}~(GK) have been developed utilizing the  chiral-odd GPDs in the calculation of  pseudoscalar meson electroproduction. The GL and GK approaches, though employing different models of  GPDs, lead to {\it transverse} photon amplitudes that are much larger than the longitudinal amplitudes.

At the same time the most successful theoretical approaches for describing exclusive reactions in the past have been those based upon the  Regge  model, which was introduced in the 1960's. The  Regge model  \cite{Laget}  has continued to provide  insights into the nature of hadrons and their interactions. 

The comparison of the results of GL and GK with each other and with the results obtained by the analysis of some of the CLAS data was discussed in Ref.~\cite{bedlinskiy}. 

This paper presents  the complete results of  that experiment and a comprehensive description of the data analysis,  following the description of the experiment.  The experimental results  will  be compared with those of G-L  and G-K as well as  with the most advanced Regge model predictions~\cite{Laget} for the $\pi^0$ exclusive production over a wider range of kinematic intervals than previously available.

The main goal of the experiment was to measure the  differential cross section 
 $\frac {d^4\sigma}    {dQ^2 dx_B dt d\phi_\pi}$ of the reaction
$ep\to e^\prime p^\prime \pi^0$ in bins of $Q^2$, $x_B$, $t$ and $\phi_\pi$, where $\phi_\pi$ is the angle of the final-state hadronic plane relative to the electron scattering plane. 
Fits to the  $\phi_\pi$ dependence (see Appendix \ref{section:helicity_amp}
 Eq. \ref{eq:d4sigma}), in each bin of $Q^2$, $x_B$ and $t$, give access to the structure functions  $(\sigma_T+\epsilon \sigma_L)$, $\sigma_{TT}$ and $\sigma_{LT}$.


\section{Experimental setup}

\begin{figure}
\includegraphics[width=\columnwidth]{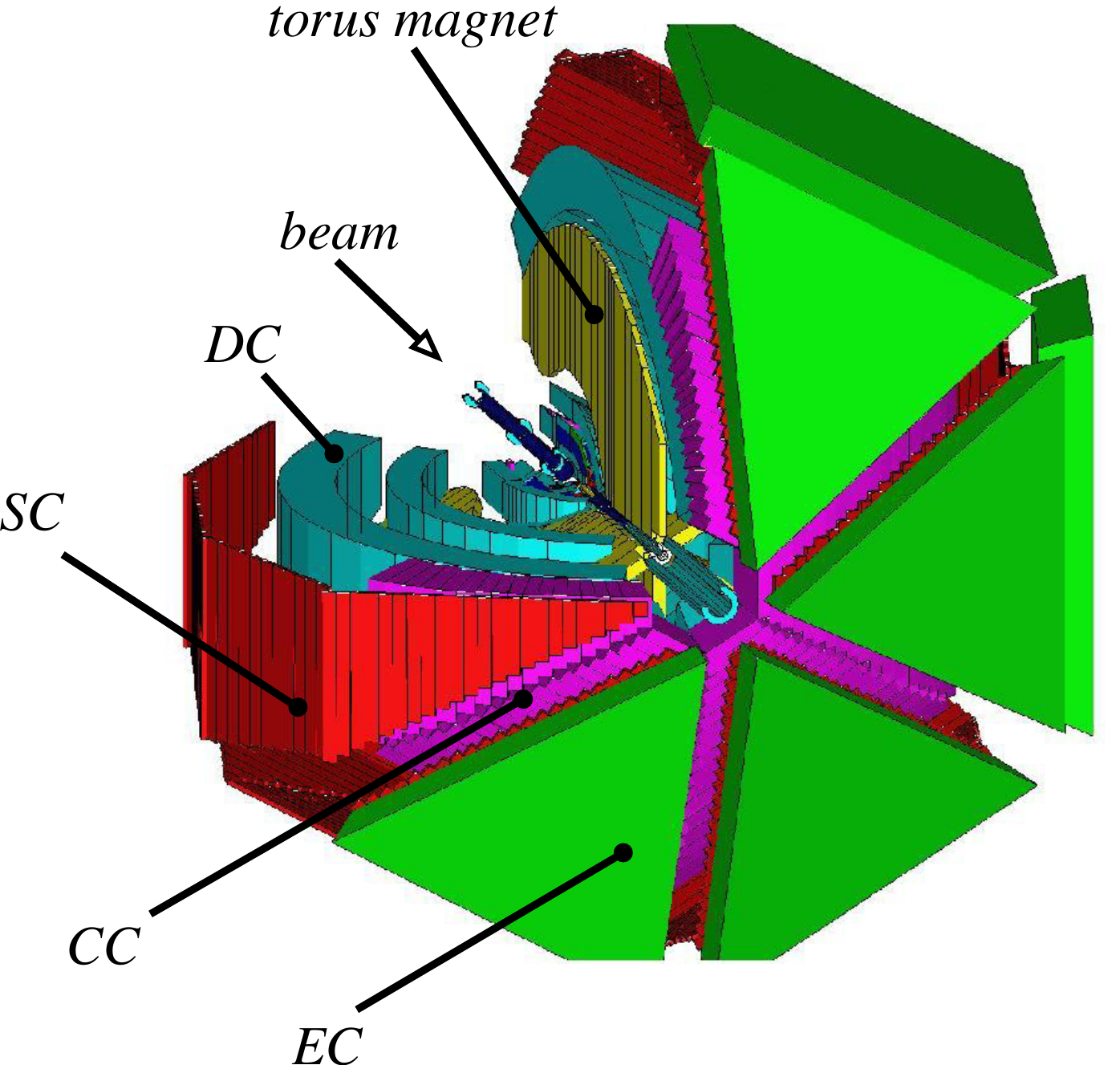}
\caption{(Color online) Three-dimensional schematic  view of the elements of the CLAS detector with the different subsystems labeled. A single sector of the detector has been cut away to enable a view of the inner subsystems. The diameter of the CLAS detector is $\sim$10 m. The notation is as follows: EC--Electromagnetic Calorimeter, CC--Cherenkov Counter, SC--Scintillation hodoscope, DC--Drift Chambers. }
\label{fig:clas_pic}
\end{figure}

\begin{figure*}
\includegraphics[width=\textwidth]{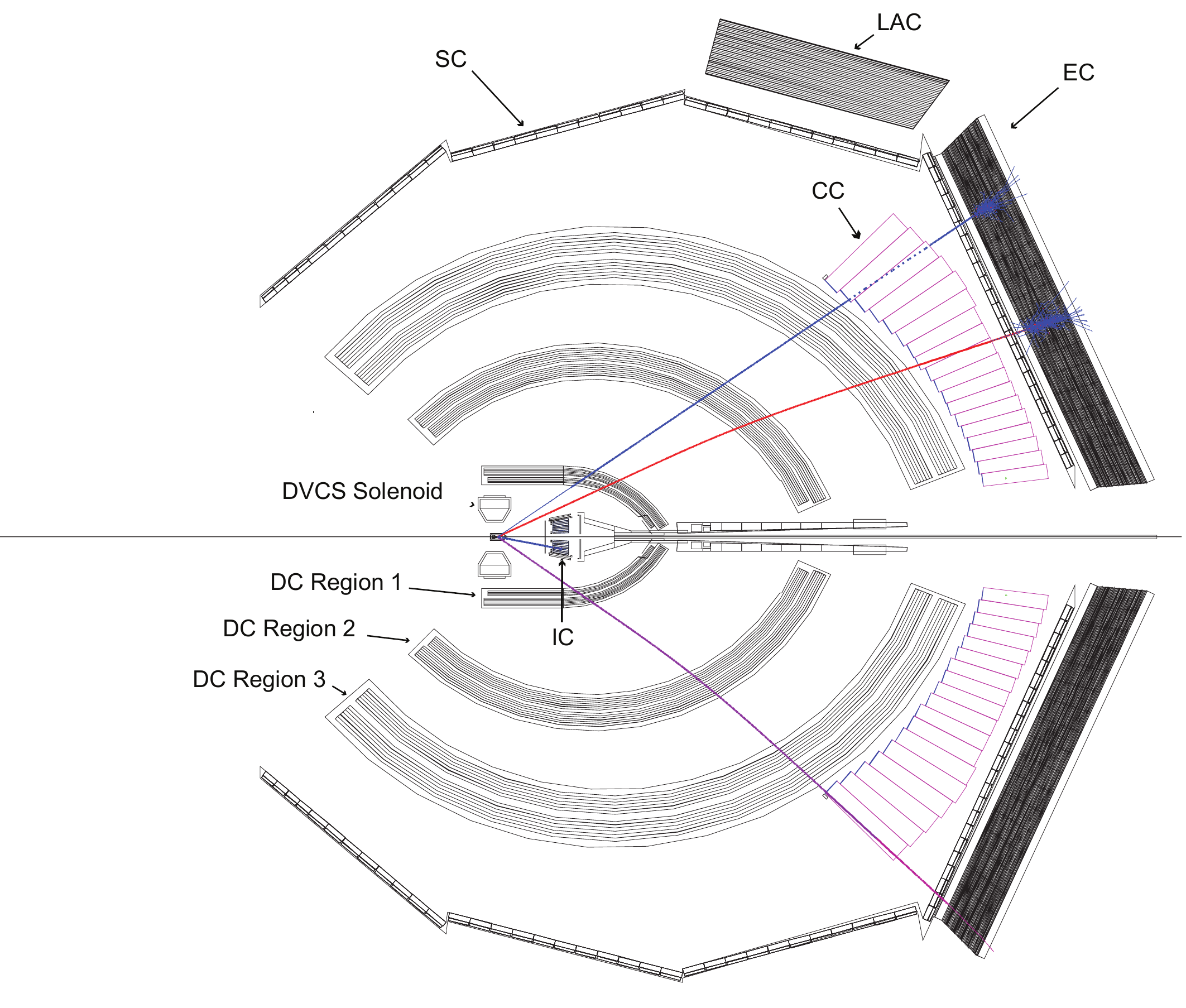}
\caption{(Color online) Schematic view of the CLAS detector constructed by the Monte-Carlo simulation program GSIM.  Note, IC--inner calorimeter, EC--electromagnetic calorimeter, LAC--large angle electromagnetic calorimeter, CC--Cherenkov counter, SC--scintillation hodoscope, DC--Drift Chambers.  The LAC was not used in this analysis. The tracks correspond, from top to bottom,  to a photon (blue online), an electron (red online) curving toward the beam line, and a proton (purple online) curving away from the beam line. }
\label{fig:clas}
\end{figure*}

\begin{figure}
\includegraphics[width=\columnwidth]{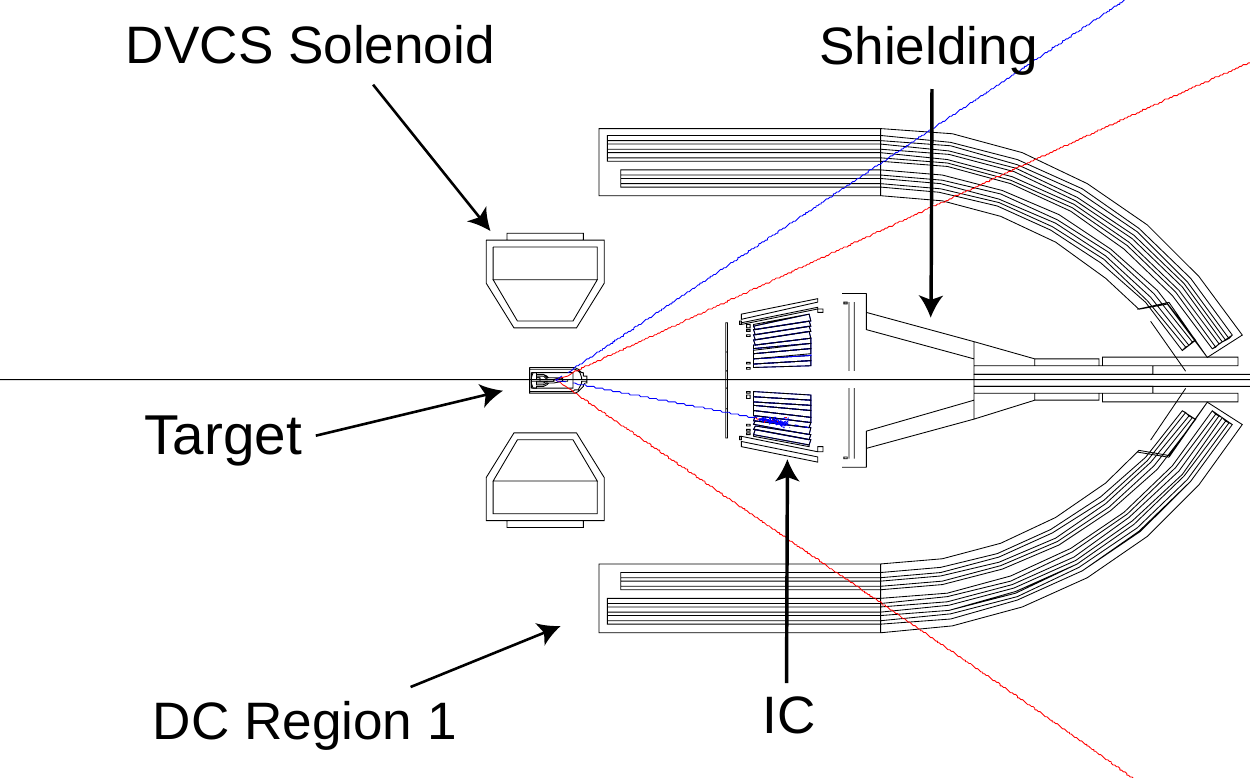}
\caption{(Color online) A blowup of Fig.~\ref{fig:clas} showing  the CLAS target region  in detail. IC is the inner calorimeter and  DC-region 1 represents the drift chambers closest to the target. 
}\label{fig:target}
\end{figure}

The measurements reported here were carried out with the CEBAF Large Acceptance Spectrometer (CLAS)~\cite{clas-detector} located in Hall B at Jefferson Lab. A three-dimensional view of  CLAS  with the different subsystems labeled is shown in Fig.~\ref{fig:clas_pic}.  
The data were taken with a 5.75-GeV electron beam and a 2.5-cm-long liquid-hydrogen target.
The target was placed 66 cm upstream of the nominal center of CLAS inside a solenoid magnet to shield the detectors from 
M{\o}ller electrons.
The spectromenter was operated at an instantaneous luminosity of $2\times10^{34}$ cm$^{-2}$s$^{-1}$.
The scheme of the CLAS geometry, as coded in the GEANT3-based CLAS simulation code GSIM,
is shown in Fig.~\ref{fig:clas}.
CLAS  consisted  of six identical sectors with an approximately toroidal magnetic field. Each sector was equipped with three regions of drift chambers (DC)~\cite{DC}
to determine the trajectory of charged particles, gas threshold Cherenkov counters (CC)~\cite{CC} 
for electron identification, a scintillation hodoscope~\cite{SC}  for time-of-flight (TOF) measurement of charged particles and an electromagnetic calorimeter (EC)~\cite{EC}  which was used for electron identification as well as detection of neutral particles. 
To detect photons at small polar angles (from 4.5$^\circ$ up to 17$^\circ$) an inner calorimeter (IC) was added to the standard CLAS configuration, 55 cm downstream from the target. Figure~\ref{fig:target} zooms in on the target area of Fig.~\ref{fig:clas} to better illustrate the deployment of the IC and solenoid relative to the target.
The IC  consisted of 424 PbWO$_4$ tapered crystals whose orientations were projected somewhat upstream of the target. Each crystal had a 13.3 $\times$13.3 mm$^2$ square front face, a $16 \times16$ mm$^2$  rear face and 160 mm of length.
The light from each crystal was collected with an avalanche photo-diode followed by a low-noise preamplifier. The temperature of the IC was stabilized with $<0.1^\circ$C precision. 
The toroidal magnet was operated at a current
corresponding to an integral magnetic field of about 1.36 T-m in the forward direction.
 The magnet polarity was set such that negatively charged particles were bent inward towards the electron beam line. The 
scattered electrons were detected in the CC and EC, which extended from 21$^\circ$ to 45$^\circ$. The lower limit was defined by the IC calorimeter located just after the target.
A totally-absorbing Faraday cup was used to determine the integrated beam charge passing through the target.

\begin{figure}
\begin{center}
\includegraphics[width=\columnwidth]{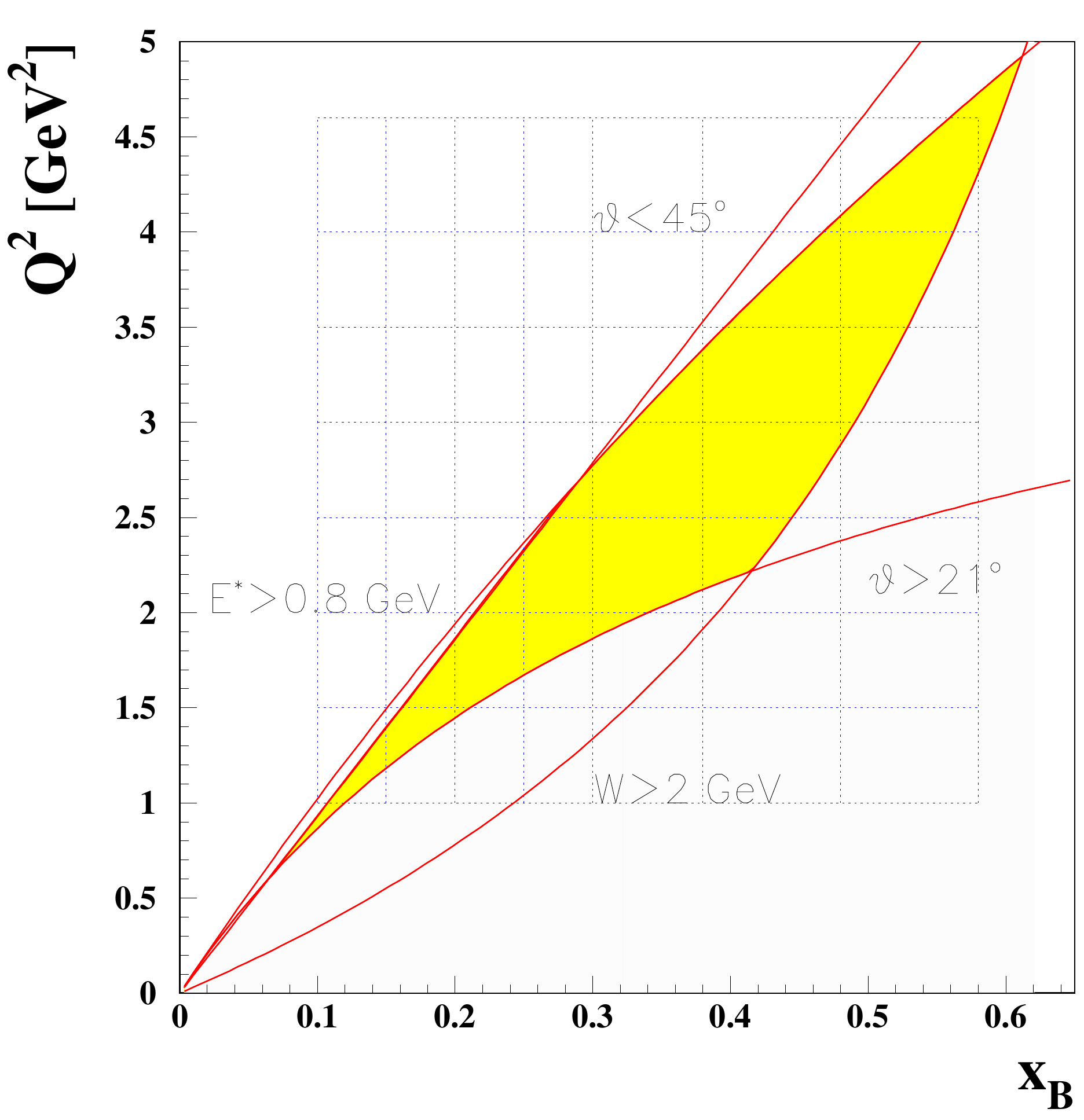}
\caption{(Color online) The kinematic coverage and binning as a function of $Q^2$ and $x_B$. 
The accepted region (yellow online) is determined by the following cuts: 
$W>2$ GeV, $E^\prime>$ 0.8 GeV, $21^\circ<\theta<45^\circ$.
$W$ is the $\gamma^*p$ center-of-mass energy, $E^\prime$ is the scattered electron energy and $\theta$ is  the  electron's polar angle in the lab frame. The dotted grid represents the kinematic regions for which the cross sections are calculated and presented. }
\label{fig:kin_cuts}
\end{center}
\end{figure}

\begin{figure*}
\centering
\includegraphics[width=0.43\textwidth]{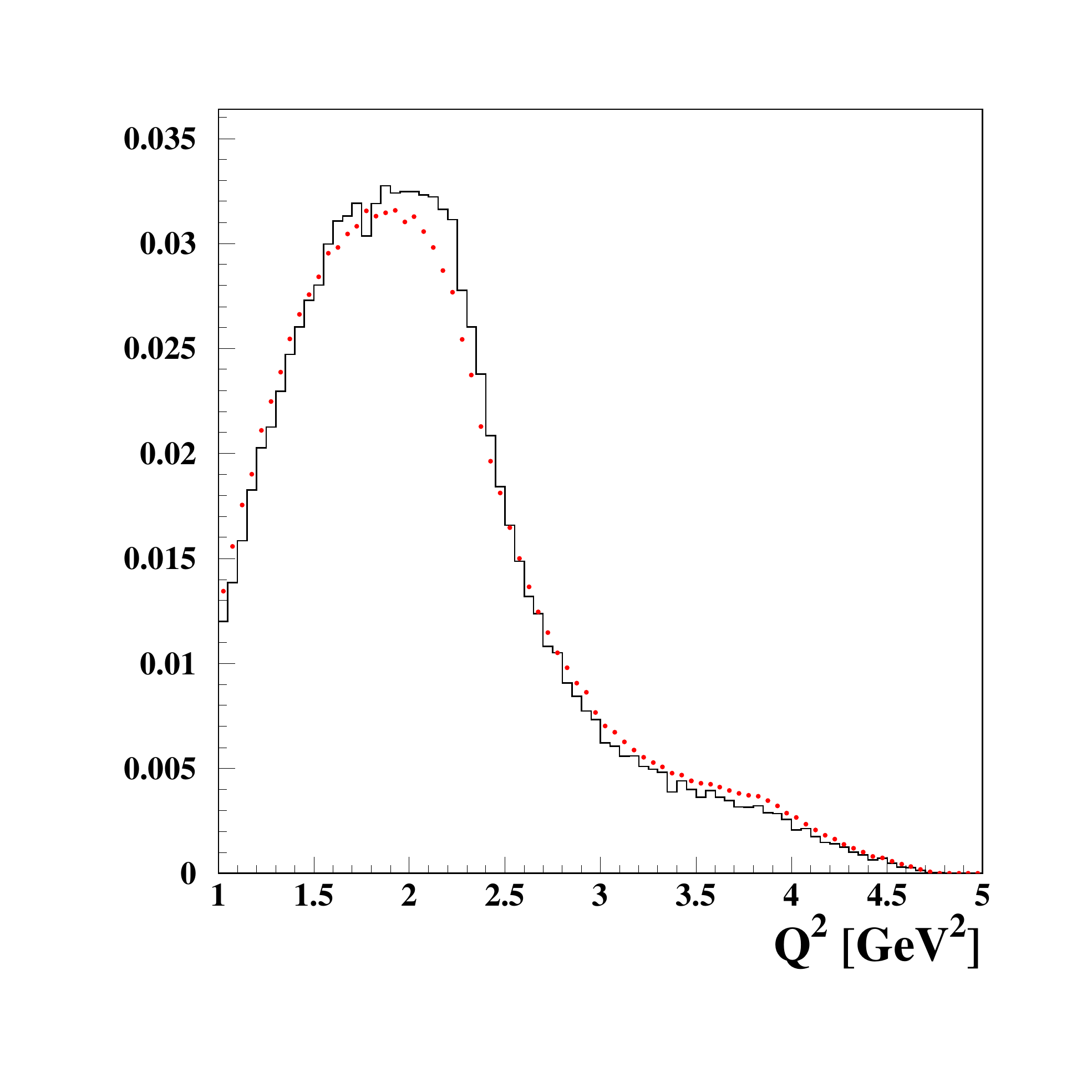}
\includegraphics[width=0.43\textwidth]{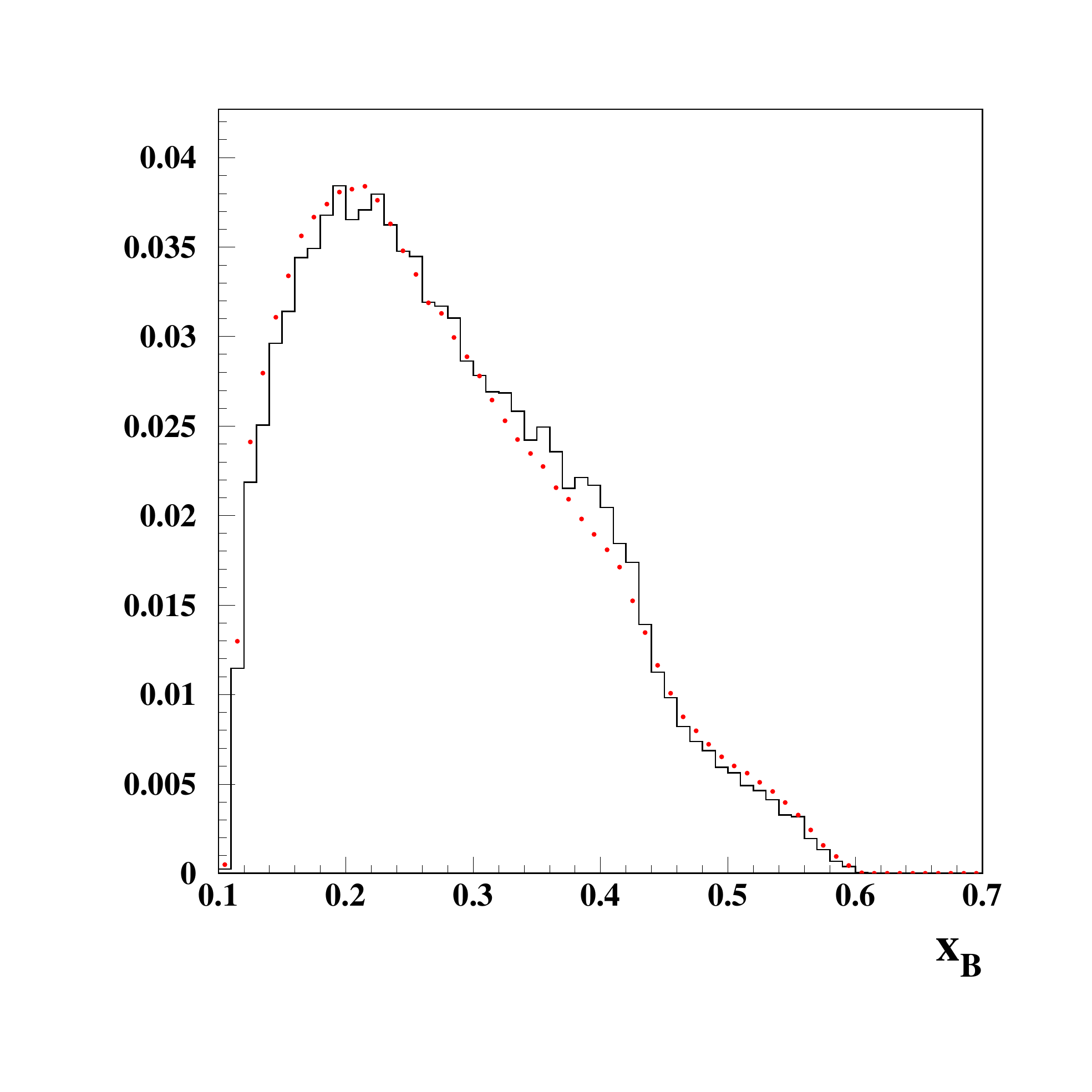}
\includegraphics[width=0.43\textwidth]{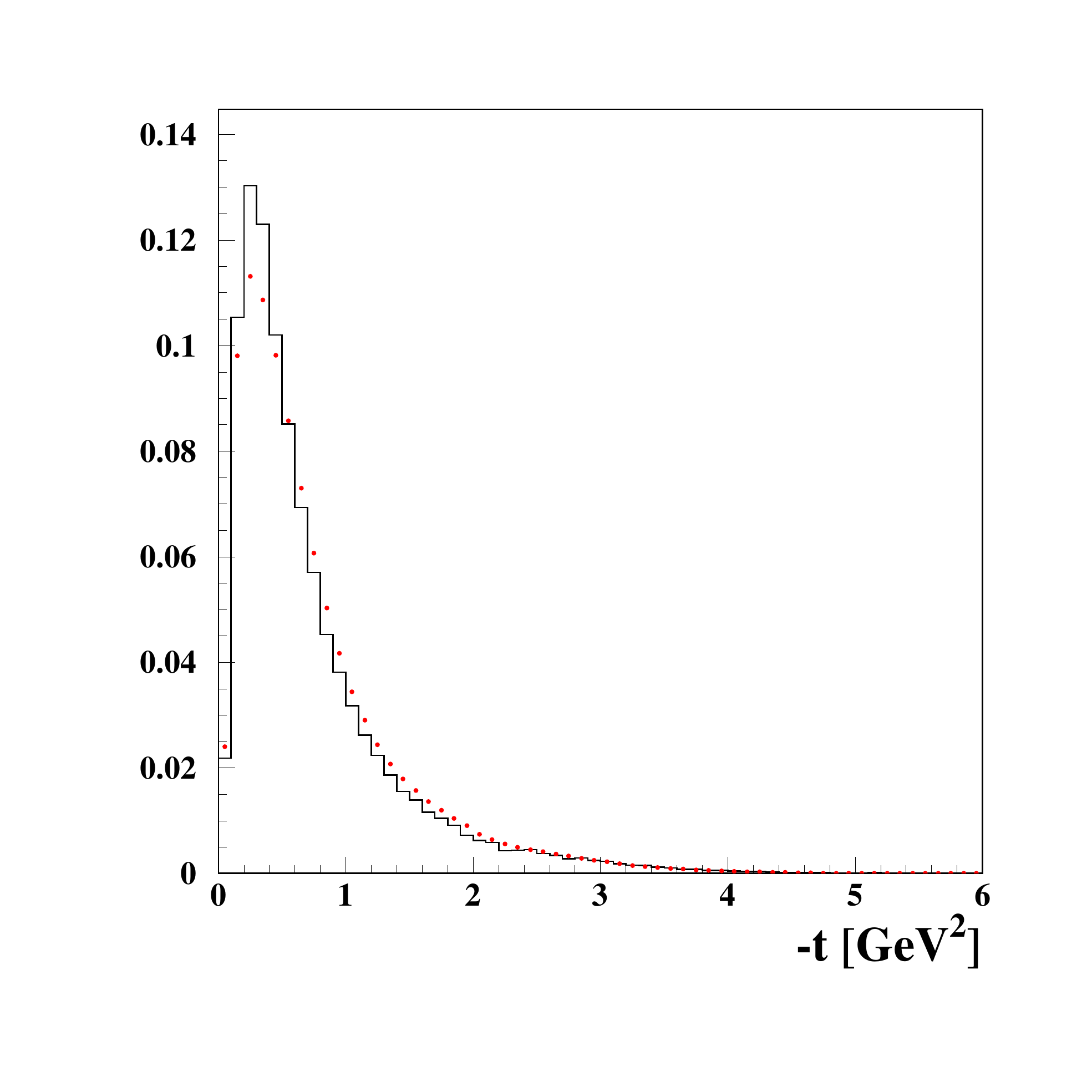}
\includegraphics[width=0.43\textwidth]{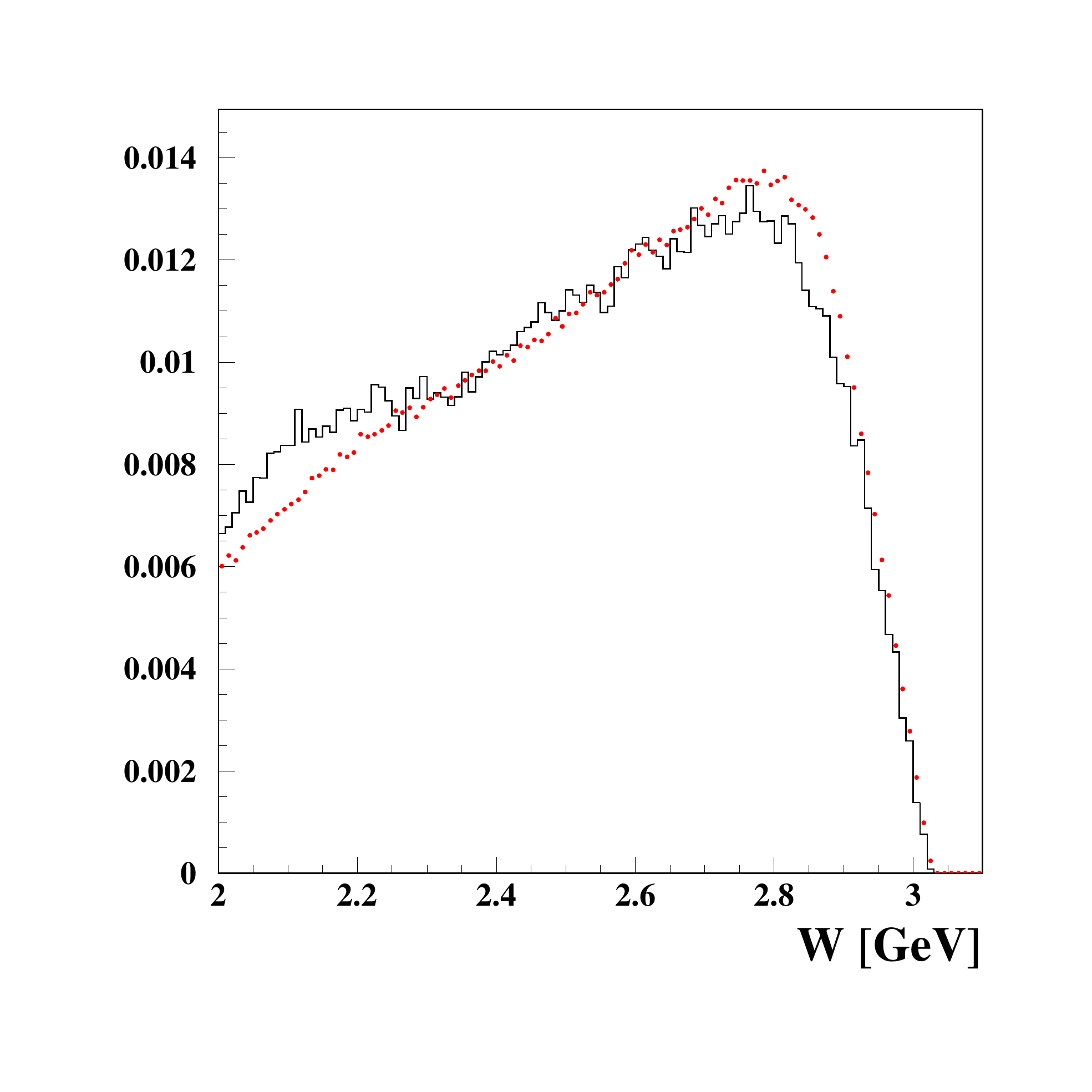}
\caption{(Color online) Distributions for kinematic variables $Q^2$, $x_B$, $-t$ and $W$ in arbitrary units. 
The data are in black (solid) and the results of Monte Carlo simulations are in red (dotted). The areas under the curves are normalized to each other. }
\label{fig:kinvar}
\end{figure*}

In the experiment, all four final state particles of the reaction $ep \to e' p' \pi^0,\  \pi^0 \to \gamma\gamma$ were detected. 
The kinematic coverage for this reaction is shown in Fig.~\ref{fig:kin_cuts},  and for the individual kinematic variables in Fig.~\ref{fig:kinvar}. For the purpose of physics analysis an additional cut on  $W>2$~GeV was applied as well, where W is the $\gamma^*p$ center-of-mass energy.

The basic configuration of the trigger included the coincidence between signals from two detectors in the same sector: the CC and the EC  with a threshold $\sim 500$ MeV. 
Out of a total of about $7 \times 10^9$  recorded  events, about $1 \times 10^5$ events for the reaction of interest were finally retained.
The specific experimental data set (``e1-dvcs")  used for this analysis was collected in 2005.
The integrated luminosity collected was 31.4 fb$^{-1}$.
However, not all data were used for the measurement of the cross section. After applying strict run-to-run 
stability criteria, the integrated luminosity corresponding to the data presented here was 
 was 19.9  fb$^{-1}$.

\section{ Particle Identification}

\subsection{Electron Identification} 
An electron was identified by requiring the track of a negatively charged particle in the DCs to 
be matched in time and space with hits in 
the  CC,  the  EC and the SC
in the same sector of CLAS.
This electron selection effectively suppresses $\pi^{-}$ contamination up to momenta $\sim$2.5 GeV. 
Additional requirements were used in the offline analysis to refine electron identification and to suppress the  remaining pions.
 Geometric ``fiducial'' cuts were applied in such a way that only regions in the  CC  and EC that had high electron efficiency  were used.

\begin{figure}
\includegraphics[width=\columnwidth]{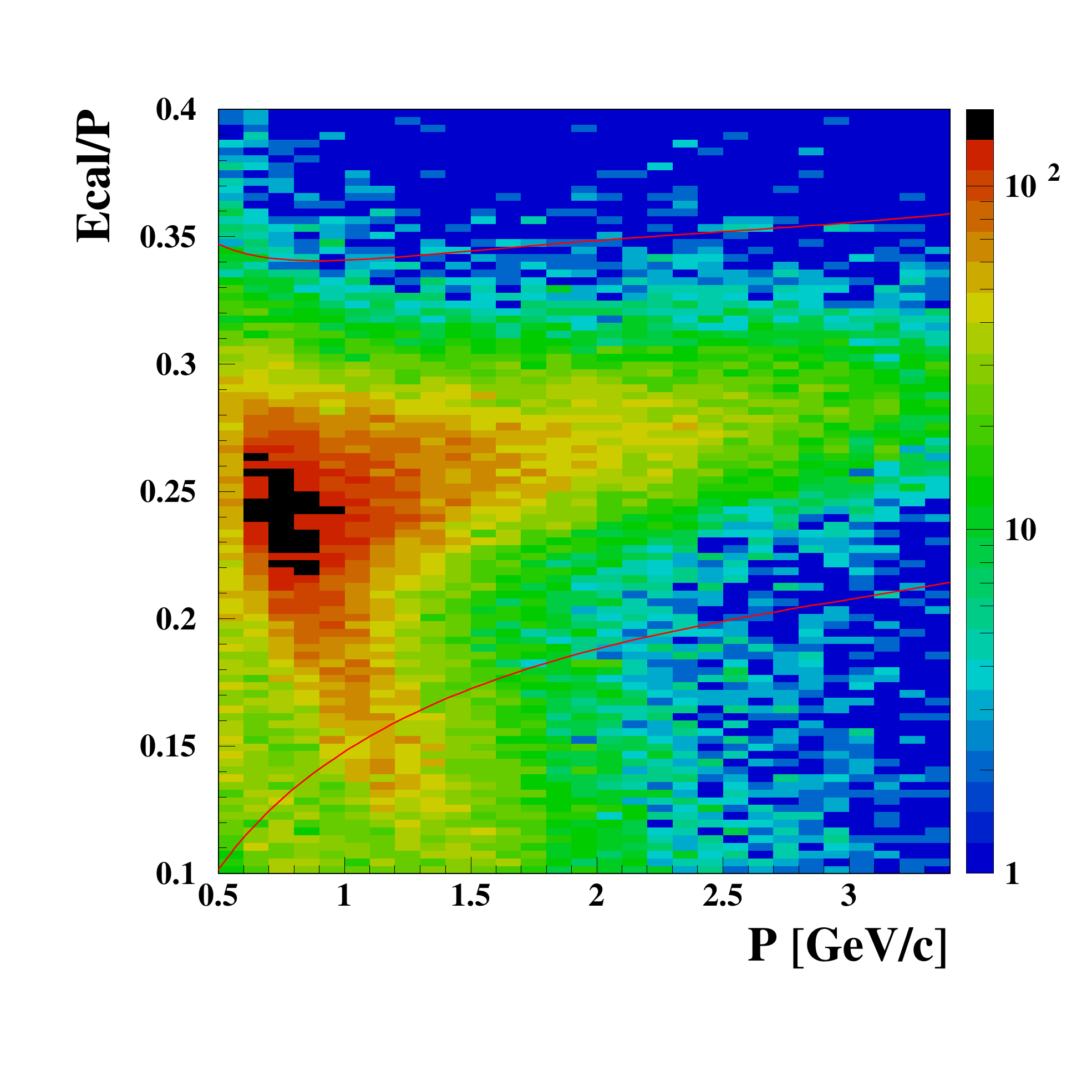} 
\caption{(Color online). 
Sampling fraction $E_{cal}/p$ of electrons in the EC as a function of electron momentum.
The solid lines show the $\pm$3.5$\sigma$ sampling-fraction cut used in this analysis. } 
\label{fig:ec_cut}
\end{figure}

Energy deposition cuts on the electron signal in the EC  also play an important role in suppressing background.
 An electron propagating through the  calorimeter produces an electromagnetic shower and deposits a large fraction of its energy in the calorimeter proportional to its momentum, while  pions typically lose a smaller fraction of their energy primarily   by ionization. 
For an electron, the observed energy to momentum ratio $E_{cal}/p$ is  known as the sampling fraction. The observed sampling fraction vs. momentum is shown in Fig.~\ref{fig:ec_cut}.
The  electron events are broadly clustered near $E_{cal}/p\sim$ 0.25. A cut was then applied to select events  within the cluster area. As shown in  Fig.~\ref{fig:ec_cut}, a $\pm$3.5$\sigma$ sampling fraction cut was  used in this analysis.

\begin{figure}
\centering 
\includegraphics[width=0.4\textwidth]{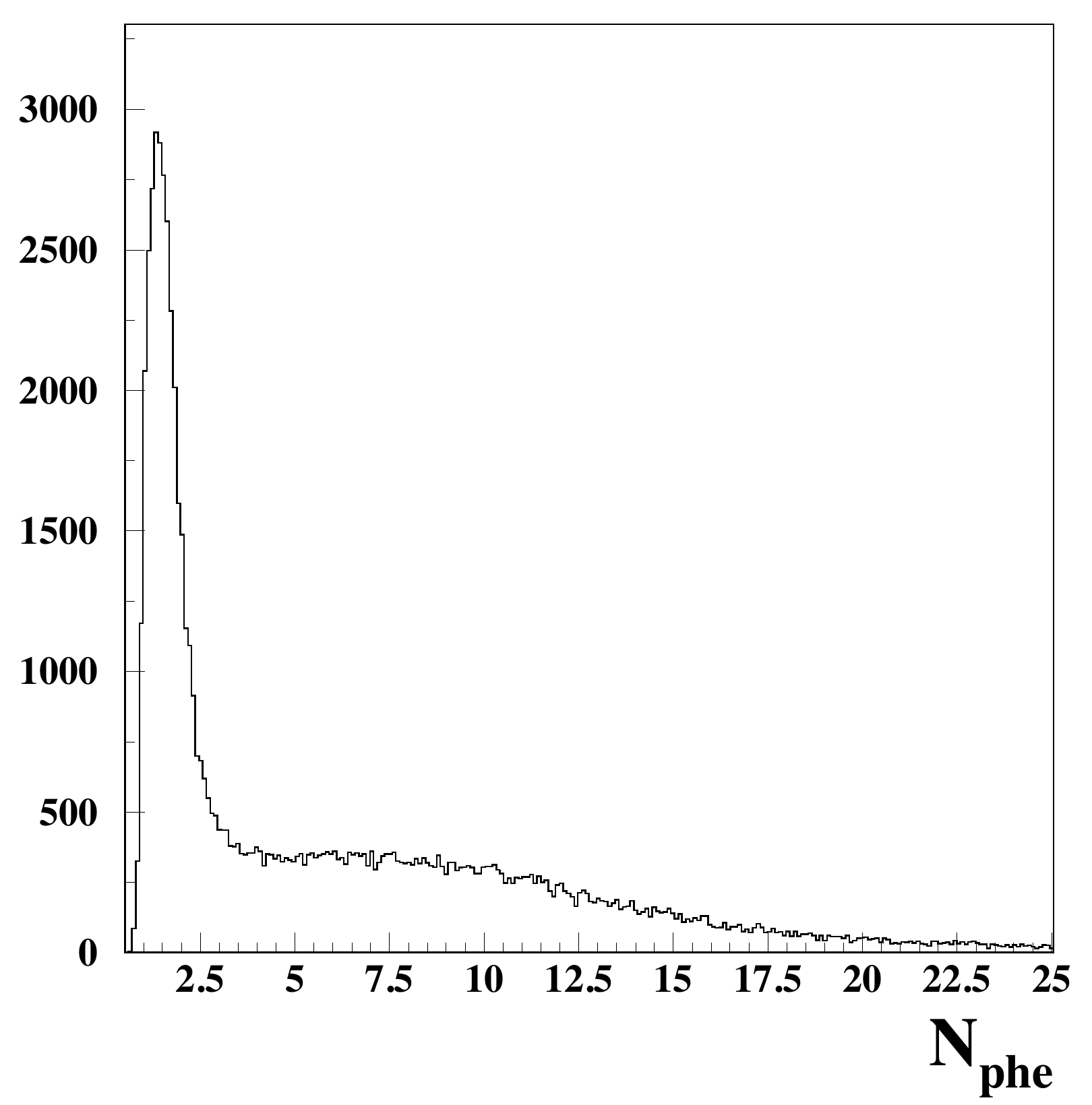}
\includegraphics[width=0.4\textwidth]{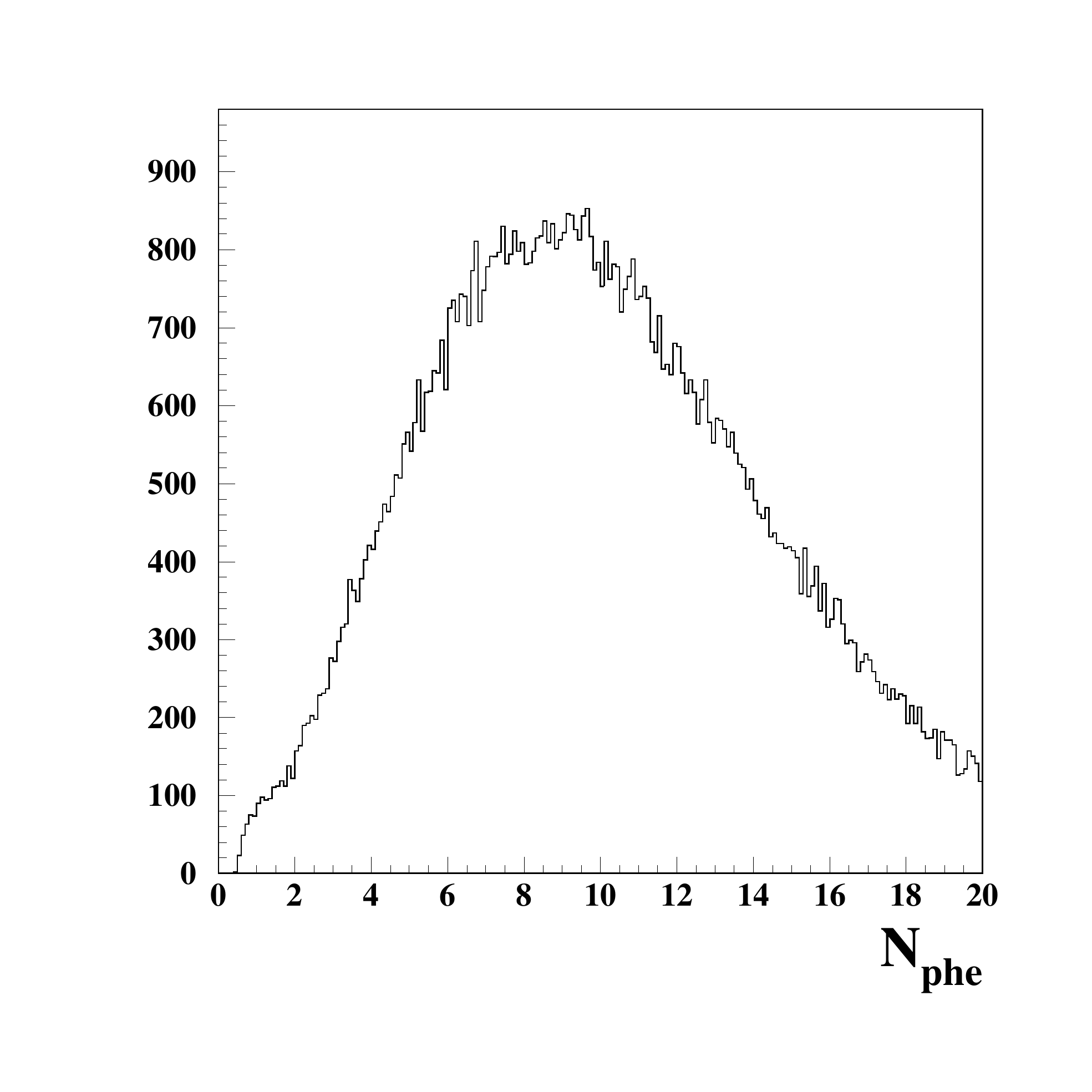}
\caption{
Upper panel: The number of CC photoelectrons for events before the various cuts such as CC angle matching, EC sampling fraction and  exclusivity cuts were applied. Lower panel: The number of CC photoelectrons for events that pass all cuts.}
\label{fig:cc_match}
\end{figure}

The distribution of the number of the photoelectrons in the CC is shown in 
Fig.~\ref{fig:cc_match}. The upper panel shows the distribution before the various cuts such as EC sampling fraction, and angle  and geometry matching  between the electron track and the hits in the CC. The peak around
$N_{phe}=1$ represents the pion contamination. The lower panel shows the same distribution after these cuts and 
the selection of the exclusive reaction (see Section \ref{sect:exclusivity_cuts}).
The single photoelectron peak becomes negligibly small.

\begin{figure*}
\subfigure[Before exclusivity cuts]
{\includegraphics[width=\columnwidth]{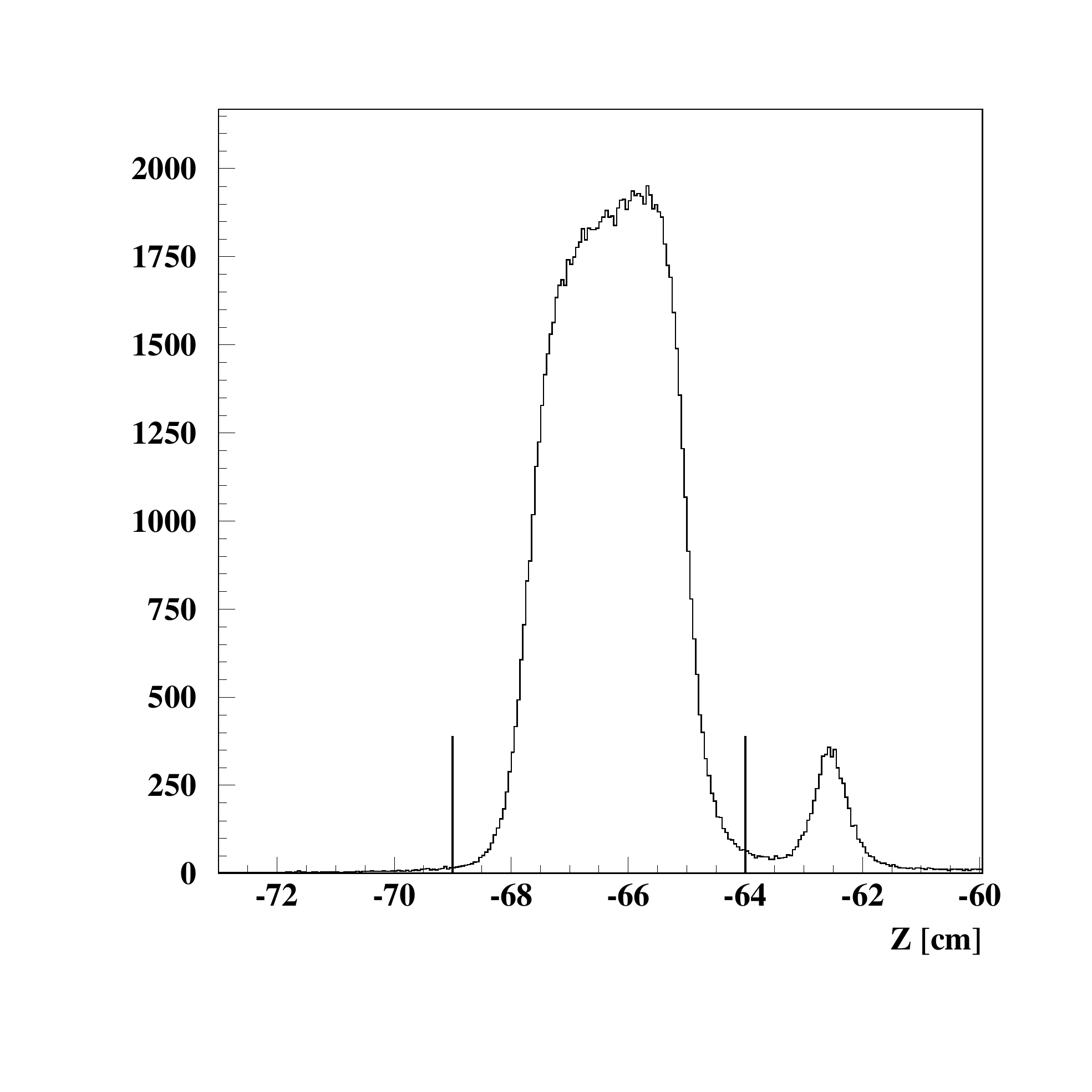}}
\subfigure[After exclusivity cuts]
{\includegraphics[width=\columnwidth]{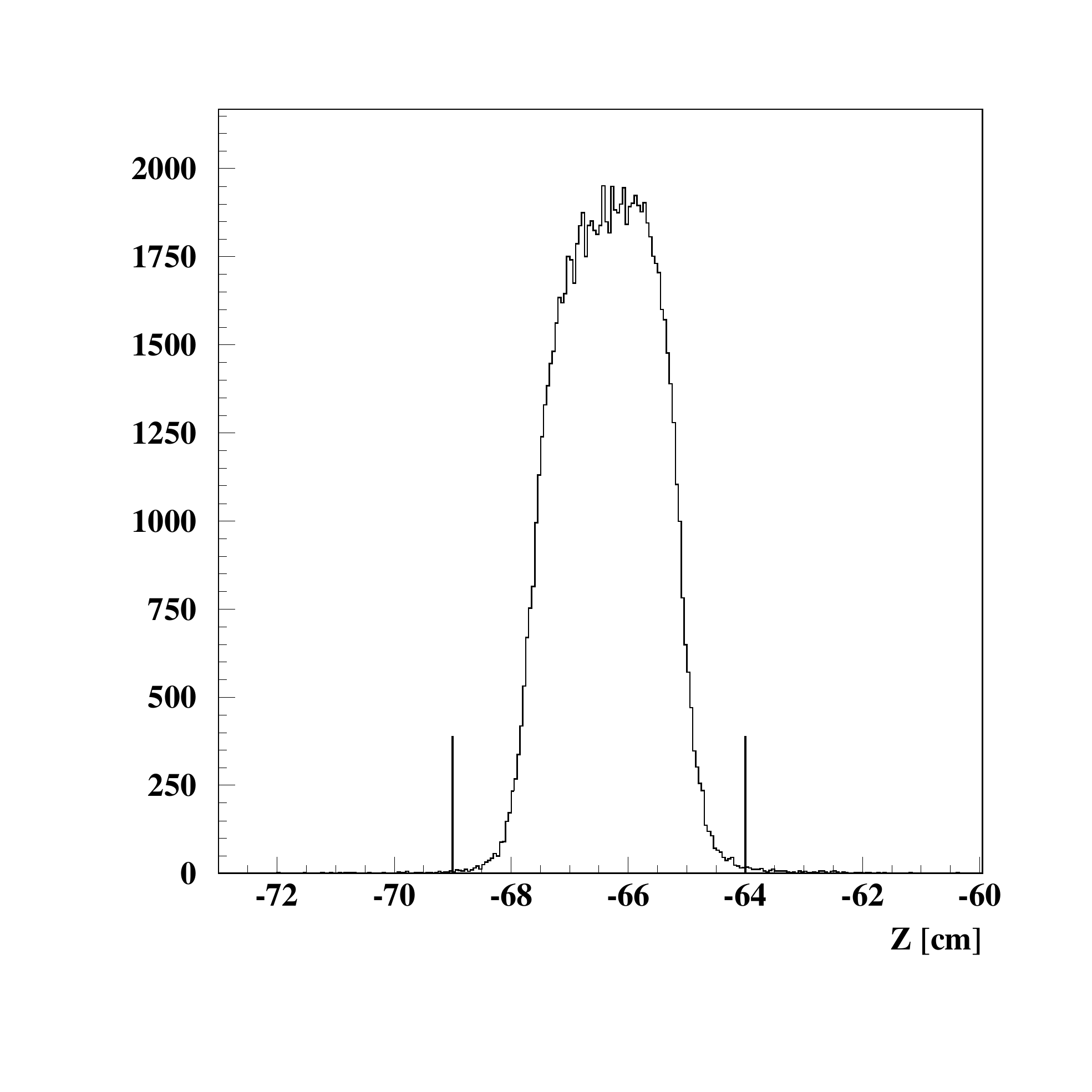}}
\caption{
The $z$-coordinate of the electron vertex. The vertical lines are the positions of the applied cuts. Note in (a) the small peak to the right of the target that is due to a foil placed at  $z=-62.5$ cm downstream of the target window.
In (b) the peak due to the foil is seen to disappear after the selection of the exclusive reaction.}
\label{fig:vertex2}
\end{figure*}

The charged particle tracks were reconstructed by the drift chambers. The vertex location was calculated by the intersection of the track with the beam line. 
 A cut was applied on the  $z$-component of the electron vertex  position to eliminate events  originating outside the target. The vertex distribution and cuts for one of the sectors is shown in Fig.~\ref{fig:vertex2}. The left plot shows the $z$-coordinate distribution before the exclusivity cuts, which are described below in Section~\ref{sect:exclusivity_cuts},  and the right plot  is the distribution after the exclusivity cuts. The peak at  $z=-62.5$ cm exhibits the interaction of the beam with an insulating foil.
 It is completely removed after the exclusivity cuts, demonstrating that these cuts very effectively exclude  the interactions involving nuclei of the surrounding non-target material.

\subsection{Proton identification} 
The proton was identified as a positively charged particle with the correct time-of-flight.
The  quantity of interest  ($\delta t=t_{SC}-t_{exp}$) is the difference in the time between the measured flight time from the event vertex to the SC system ($t_{SC}$) and that expected for the proton ($t_{exp}$). The quantity $t_{exp}$ was computed from the velocity of the particle and the track length. The velocity was determined from the momentum assuming the mass of the  particle equals that of a proton.  A cut at the level of $\pm 5 \sigma_t$ was applied around  $\delta t = 0$, where  $\sigma_t$  is the time-of-flight resolution.
Such a wide cut is possible because the exclusivity cuts very effectively suppressed the remaining pion contamination.

\subsection{Photon identification} 

Photons were detected in both calorimeters, the EC  and IC. 
In the EC, photons  were identified as {\it neutral} particles  with $\beta>0.8$ and $E>0.35$ GeV. 
Fiducial cuts were applied to avoid  the EC  edges. When a photon hits the boundary of the calorimeter, the energy cannot be fully reconstructed due to the leakage of the shower out of the detector.
Additional fiducial cuts on the EC  were applied to account for the shadow of the IC  (see Fig.~\ref{fig:clas}). 
The  calibration of the EC  was done using cosmic muons and  the photons from neutral pion decay ($\pi^0\to\gamma\gamma$).

In the IC   each detected cluster was considered  a photon. The assumption was made that this photon originated from the electron vertex. Additional geometric cuts were applied to remove low-energy clusters around the beam axis and photons near the edges of the IC, where the energies of the photons were incorrectly reconstructed due to the electromagnetic shower leakage.
The photons from $\pi^0\to \gamma\gamma$ decays were detected in the IC  in an angular range between $5^\circ$  and $17^\circ$   and in the EC   for angles greater than $21^\circ$. 
The reconstructed invariant mass of two-photon events was then subjected to various cuts to isolate exclusive $\pi^0$ events, with a small residual background, as discussed in the section on exclusivity cuts in Sec.~\ref{sect:exclusivity_cuts} below.  

\subsection{Kinematic corrections}
Ionization energy-loss corrections were applied to  protons and electrons in both data and Monte-Carlo events. These corrections were estimated using the GSIM Monte Carlo program.
Due to imperfect knowledge of the properties of the CLAS detector, such as the magnetic field distribution and the precise  placement of the components or detector materials, small empirical sector-dependent corrections had to be made on the momenta and angles of the detected electrons and protons. The corrections were determined by systematically studying the kinematics of the particles emitted from well understood kinematically-complete processes, e.g. elastic electron scattering. These corrections were on the order of 1\%.


\section{Event selection}

\subsection{Fiducial cuts}
Certain areas of the detector acceptance were not efficient due to  gaps in the DC,
problematic SC panels, and inefficient zones of the CC and the EC. These areas were removed from the analysis as well as the simulation by means of geometrical cuts,  which were momentum, polar angle and azimuthal angle dependent.

\subsection{Exclusivity cuts}
\label{sect:exclusivity_cuts}

To select the exclusive reaction $ep\rightarrow e'p'\pi^0$, each event was required to contain an electron, one proton and at least two photons in the final state. Then,  so called {\it exclusivity cuts} were applied to all combinations of an electron,  a proton and two  photons  to ensure energy and momentum conservation, thus eliminating events in which there were any additional undetected particles.

\begin{figure*}
\includegraphics[width=1.0\textwidth]{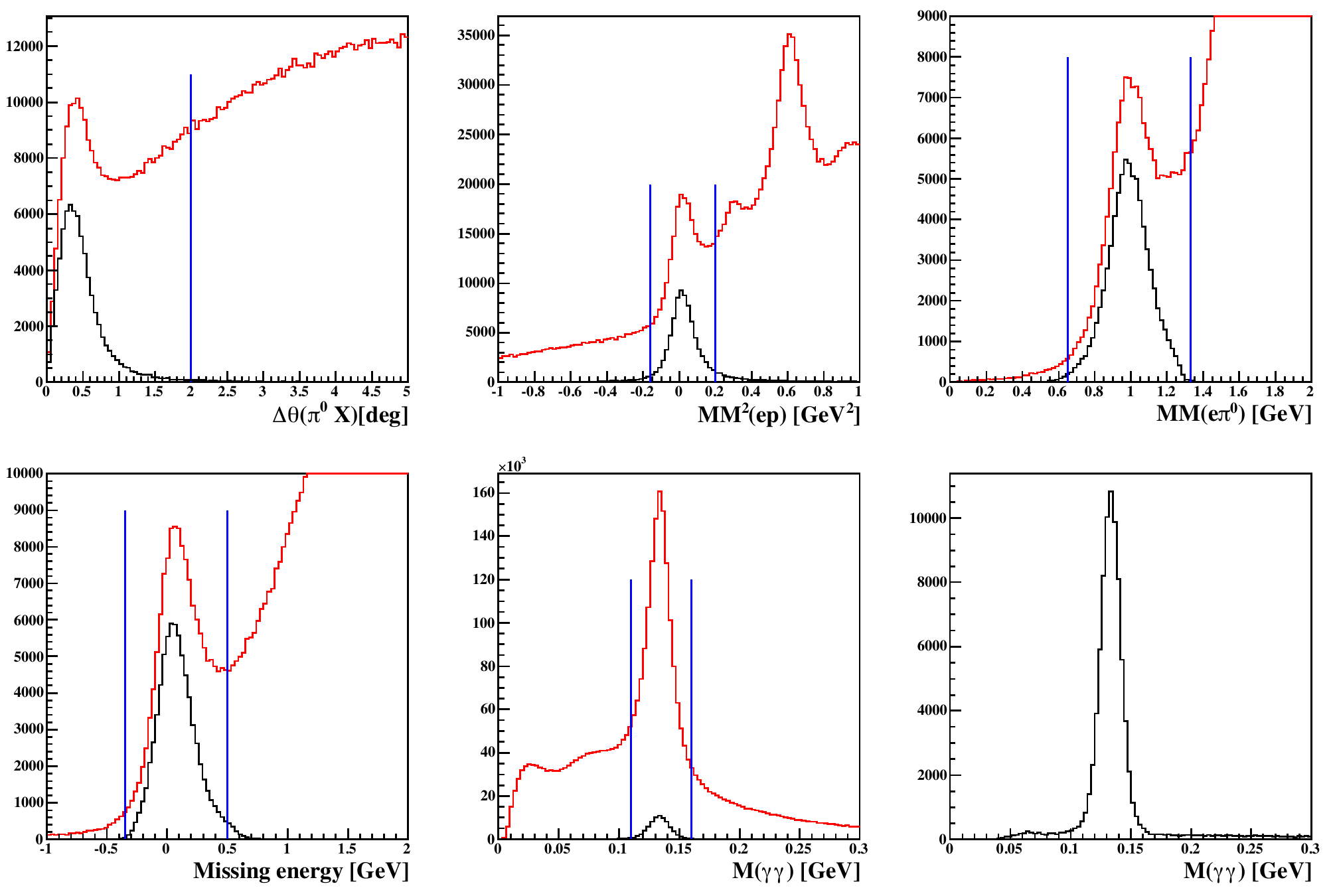}

\caption{
\label{fig11}
(Color online) The exclusivity cuts for $\pi^0$ production for the topology where both decay photons are detected in the IC calorimeter. The graph for each variable  shows  the number of events per channel plotted before (red) and after (black) the cuts on the other variables. Upper left: $\theta_X$ cut: angle between the reconstructed  
$\pi^0$ momentum vector and the missing momentum vector $ep\to e'p'X$. 
Upper middle: Missing mass $M_X^2(ep)$.  
Upper right: Missing mass $M_X(e\gamma\gamma)$. 
Lower left:  Missing energy $E_X(ep\gamma\gamma)$. 
Lower middle: Invariant  mass $M(\gamma\gamma)$.  
Lower right: Same as in lower middle  ($M(\gamma\gamma)$),  but magnified to illustrate the residual background. This background is subtracted from the pion distribution using the wings on either side of the peak, as explained in the text.
The vertical lines denote the positions of the applied cuts on each distribution.\\
}
\end{figure*}

Five cuts were used for the exclusive event selection (see Fig.~\ref{fig11}):
\begin{itemize}
\item  A cut, $\theta_X$,  on the angle between the reconstructed  
$\pi^0$ momentum vector and the missing momentum vector for the reaction $ep\to e'p'X$, in which $\theta_X<2^o$.
\item The missing mass squared of the $ep$--system  ($ep\to e'p'X$), with $| M_x^2(ep) - M^2_{\pi^0}| < 3\sigma$.
\item The missing mass of the $e\gamma\gamma$--system  ($ep\to e'\gamma\gamma X$), with $|M_x(e\gamma\gamma)-M_{p}| < 3\sigma$.
\item The missing energy ($ep\to e'p'\gamma\gamma X$), with $|E_x(ep\pi^0)- 0| < 3\sigma$.
\item$\gamma\gamma$ invariant mass - $|M(\gamma\gamma) - M_{\pi^0}| < 3\sigma$.
\end{itemize}
\noindent
Here $\sigma $ is the observed experimental resolution obtained as the variance from the mean value of the distributions of each quantity. Three sets of resolutions were determined independently for each of the three photon-detection topologies (IC-IC, IC-EC, EC-EC). The effects of these cuts on the various distributions and the positions of the applied cuts  are shown in Fig. 11 for the case where both photons were detected in the IC.
These distributions were generally broader than  in the Monte Carlo simulations so that the cuts for the data were typically broader than those used for the Monte Carlo simulations.
Similar results were obtained for the topology in which one photon was detected in the IC and one in the EC, as well as the case where both photons were detected in the EC.

\subsection{Background subtraction}

\begin{figure} 
\centering
\includegraphics[width=0.45\textwidth]{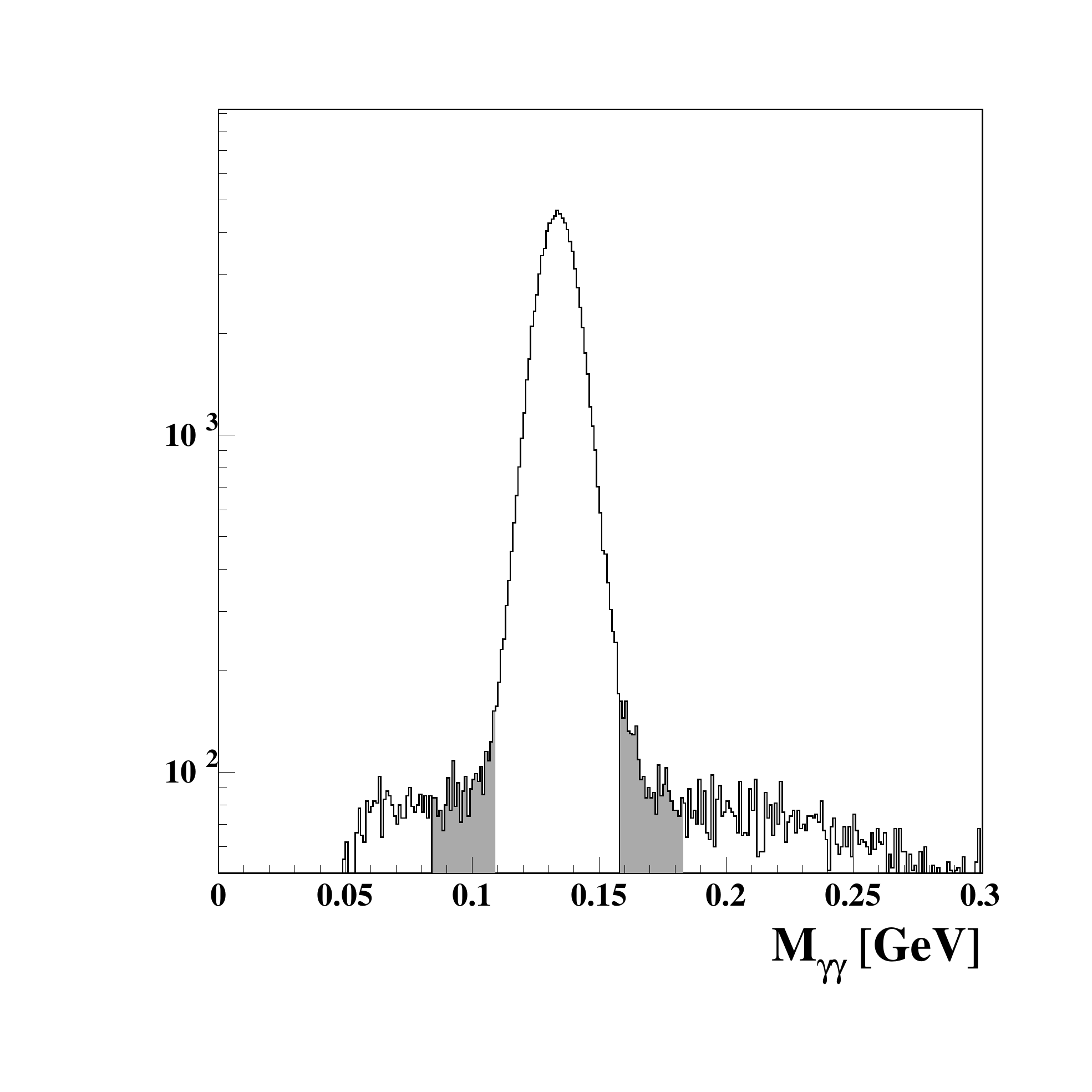}
\caption{The invariant mass distribution $M(\gamma\gamma)$ for all events in which all selection criteria were applied,
where both decay photons were detected in the IC (note the log scale). The shaded regions  were used to estimate the residual background on a kinematic bin-by-bin basis. 
}
\label{fig:backsub} 
\end{figure}

The $M(\gamma\gamma)$ distribution contains a small amount of  background under the $\pi^0$ peak even after the application of all exclusivity cuts shown in Fig.~\ref{fig11}.
The background under the $\pi^0$ invariant mass peak, typically 3--5\%, was subtracted for each kinematic bin using the data in the sidebands $(-6\sigma,-3\sigma) \cup  (3\sigma,6\sigma)$ in the $M(\gamma\gamma)$ distributions (lower right distribution in Fig.~\ref{fig11} and  in greater detail in Fig.~\ref{fig:backsub}). The same cuts were applied to all the kinematic bins. 

 \subsection{Kinematic binning}

The kinematics of the reaction  are defined by four variables: $Q^2$, $x_B$,  $t$ and $\phi_{\pi}$.
In order to obtain differential cross sections the data were divided into four-dimensional rectangular bins in these variables.
There are 8 bins in $x_B$, $Q^2$ and $t$ as shown 
in Tables \ref{q2-bins}--\ref{t-bins}. For each of these kinematic bins there are 20 bins in $\phi_{\pi}$ of equal angular width.
  The binning in $x_B$ and $Q^2$  is shown in Fig.~\ref{fig:kin_cuts}.

\begingroup
\squeezetable
\begin{table}
\caption{$Q^2$ bins}
\begin{ruledtabular}
\begin{tabular}{ccc}
Bin Number & Lower Limit & Upper limit \\ 
 & (GeV$^2$)   & (GeV$^2$) \\ \hline
1 & 1.0   &  1.5 \\ 
2 & 1.5   &  2.0 \\   
3 & 2.0   &  2.5 \\
4 & 2.5   &  3.0 \\ 
5 & 3.0   &  3.5 \\ 
6 & 3.5   &  4.0 \\  
7 & 4.0   &  4.6 \\
\end{tabular}
\label{q2-bins}
\end{ruledtabular}

\caption{$x_B$ bins}
\begin{ruledtabular}
\begin{tabular}{ccc}
Bin Number & Lower Limit & Upper limit \\ \hline
1 & 0.10   &  0.15 \\
2 & 0.15   &  0.20 \\  
3 & 0.20   &  0.25 \\
4 & 0.25   &  0.30 \\ 
5 & 0.30   &  0.38 \\
6 & 0.38   &  0.48 \\  
7 & 0.48   &  0.58 \\ 
\end{tabular}
\end{ruledtabular}
\label{x-bins}

\caption{$|t|$ bins}
\begin{ruledtabular}
\begin{tabular}{ccc}
Bin Number & Lower Limit & Upper limit \\ 
           & (GeV$^2$)   & (GeV$^2$) \\ \hline

1 & 0.09   &  0.15 \\ 
2 & 0.15   &  0.20 \\   
3 & 0.20   &  0.30 \\ 
4 & 0.30   &  0.40 \\
5 & 0.40   &  0.60 \\
6 & 0.60   &  1.00 \\ 
7 & 1.00   &  1.50 \\
8 & 1.50   &  2.00 \\
\end{tabular}
\end{ruledtabular}
\label{t-bins}
\end{table}
\endgroup

\section{Monte Carlo simulation}

The acceptance for each ($Q^2$, $x_B$, $t$, $\phi_\pi$) bin of the CLAS detector with the present setup for the reaction $ep\rightarrow e'p'\gamma\gamma$ was calculated using the Monte Carlo program GSIM. The event generator used an empirical parametrization of the cross section as a function  of $Q^2$, $x_B$ and $t$. The parameters were tuned using the MINUIT program to best match the  simulated $\pi^0$ spectra, including radiative effects, with the measured  electroproduction cross section. 
Two iterations were found to be sufficient to
describe the cross section with reasonable accuracy. 
The comparison of the experimental data and Monte Carlo simulated data is shown in Fig.~\ref{fig:kinvar}
for the variables $Q^2$, $x_B$, $-t$ and $W$.

Additional smearing factors for tracking and timing resolutions were included in the simulations to provide more realistic resolutions for charged particles. The Monte Carlo events were analyzed by the same code that was used to analyze the experimental data, and with the additional smearing and somewhat different exclusivity cuts, to account for the leftover discrepancies in calorimeter resolutions. Ultimately the number of reconstructed Monte Carlo events  was an order of magnitude higher  than the number of reconstructed experimental events.  Thus, the statistical uncertainty introduced by the acceptance calculation was typically much smaller than the statistical uncertainty of the data.

The efficiency of the event reconstruction depends on the level of noise in the detector, the greater the noise the lower the efficiency. It was found that the efficiency for reconstructing particles decreased linearly with increasing beam current.
To take this into account the background hits from random 3-Hz-trigger events  were mixed with the Monte Carlo events for all detectors - DC, EC, IC, SC and CC.
The acceptance for a given bin  $i$  was calculated as a ratio of the number of reconstructed events to the number of generated events, including the random background events as

\begin{equation}
\epsilon_i(Q^2,x_B,t,\phi_\pi)=\frac {N^{rec}_i(Q^2,x_B,t,\phi_\pi)}{N^{gen}_i(Q^2,x_B,t,\phi_\pi)}.
\end{equation}

Only areas of the 4-dimensional space with an acceptance equal to or greater than 0.5\% were used.
This cut was applied to avoid the regions where the calculation of the acceptance was not reliable. 

\section{Radiative Corrections}\label{sim:radcor}
\begin{figure*}
\includegraphics[width=2.cm]{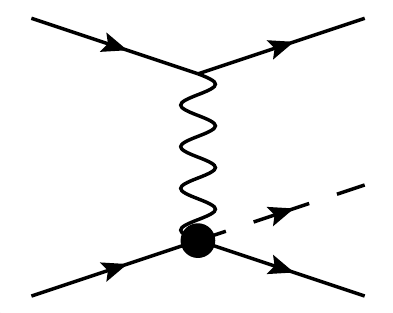}
\includegraphics[width=2.cm]{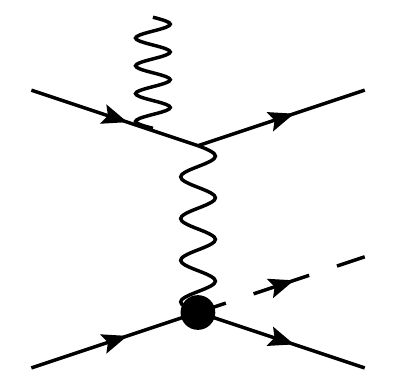}
\includegraphics[width=2.cm]{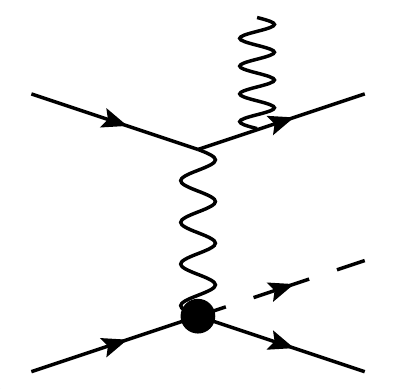}
\includegraphics[width=2.cm]{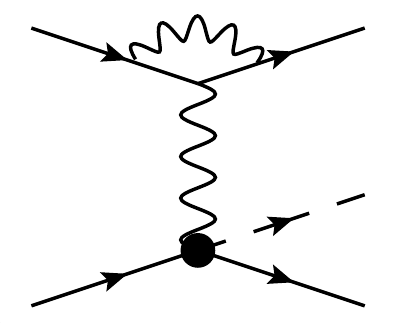}
\includegraphics[width=2.cm]{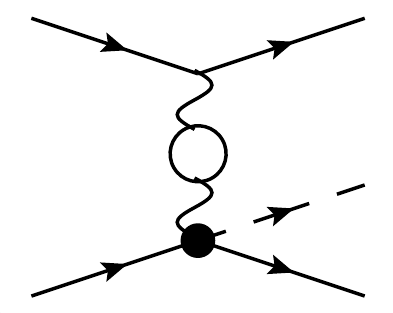}
\caption{\label{fig:rad_proc}	
Feynman diagrams contributing to the pion electroproduction
cross section. Left to right: Born process,
Brehmsstrahlung (by the initial and the final electron),  vertex correction, and
vacuum polarization.
}
\end{figure*}
Radiative processes which modify the observed cross section were taken into account. Some of these, illustrated in Fig.~\ref{fig:rad_proc},  include radiation of real photons, vacuum polarization and lepton-photon vertex corrections. Vacuum polarization refers
to the process where the virtual photon temporarily creates and annihilates
a lepton-anti-lepton pair. The lepton-photon vertex corrections are for
processes where a photon is emitted by the incoming electron and is absorbed
by the outgoing electron. These processes
give the largest contribution to the cross section at the next-to-leading-order level 
and can be calculated exactly from QED~\cite{exclurad}.  Thus, the measured cross section can be corrected to extract the Born term.
The  radiative correction, $\delta_{RC}$, connects the experimentally measured cross section to the basic non-radiative (Born) cross section as follows

\begin{equation}
	\sigma_{Born} = \frac{\sigma_{meas}}{\delta_{RC}}.
\end{equation}
Here, $\sigma_{meas}$ is the observed cross section from experiment and 
$\sigma_{Born}$ is the desired cross section after corrections.

\begin{figure}
\centering
\includegraphics[width=\columnwidth]{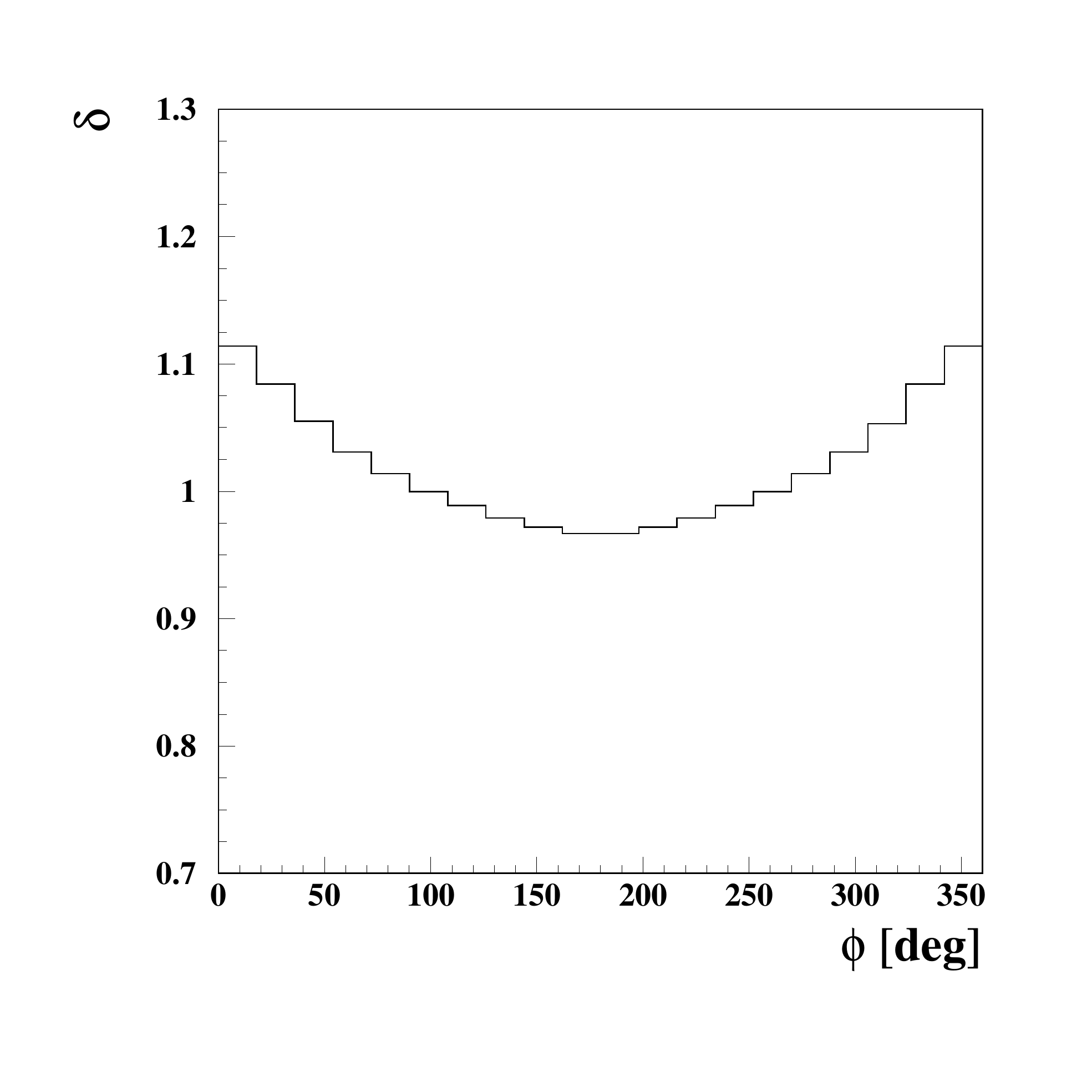}
\caption{Radiative corrections for  $\pi^0$ electroproduction as a function of $\phi_\pi$ for the bin
$(Q^2=1.25$~GeV$^2$, $x_B=0.125$, $t=-0.12$~GeV$^2)$.}
\label{fig:rc_2d}
\end{figure}

The corrections were obtained using the software package EXCLURAD \cite{exclurad}
which uses  theoretical models as input for the hadronic current. 
The same  analytical structure functions were implemented  in the EXCLURAD package as were used to generate  the $\pi^0$  electroproduction events in the Monte-Carlo simulation. 
The corrections were computed for each kinematic bin  ($Q^2$, $x_B$, $t$, $\phi_\pi$). 
They vary from 5\% to 10\%, depending on the kinematics.
For example, Figure~\ref{fig:rc_2d} shows the radiative corrections calculated 
for the first kinematic bin
as a function of the $\phi_\pi$ angle. 
Note that the correction increases near  
$\phi_\pi=0^\circ$ and $\phi_\pi=360^\circ$.

\section{Normalization Correction}
\label{normalization}
To check the overall absolute normalization   the cross section of  elastic electron-proton scattering was measured using the same data set.  The measured cross section was lower than the known elastic cross section by approximately 12\% over most of the elastic kinematic range. Studies made using additional other reactions where the cross sections are well known, such as $\pi^0$ production in the resonance region, and Monte Carlo simulations of the effects of random backgrounds, indicate that this was approximately true over a wide range of kinematics. Thus, a normalization factor $\delta_{Norm}\sim 0.89$ was  applied to the measured cross section. This value includes the efficiency of the SC counters which was estimated to be around  around 95\%, as well as other efficiency factors which are not accounted for in the analysis, such as trigger and CC efficiency effects. 
This correction comprises the largest single contribution to the systematic uncertainties in the extracted cross section.

\section{Systematic Uncertainties}
\label{systematics}

The determination of  the  differential cross section of the reaction $ep\to e^\prime p^\prime\pi^0$
requires the knowledge of  the yield and the acceptance, including various efficiency factors and radiative effects, for each kinematic bin ($Q^2$,$x_B$,$t$,$\phi_\pi$),  as well as the integrated luminosity of the experiment. These quantities are subject to systematic uncertainties which contribute to the uncertainty of the measured cross section in each kinematic bin.
 Each of these factors is subject to systematic uncertainty. The size of these systematic  uncertainties was estimated 
by  repeating the calculation of the cross section varying  each of the cut parameters within reasonable limits.
Table~\ref{table:syst-summary} contains a summary of the information on all the studied sources of systematic uncertainties. Some sources of uncertainty vary bin-by-bin, others are global.

The systematic uncertainty on the  proton  identification was studied by removing the cut  on the
difference between the measured and predicted flight times. The systematic uncertainty was estimated  in each  $(Q^2, x_B, t, \phi_\pi)$ bin to be on average $\sim 2.5$ \%.

 To estimate the systematic uncertainty introduced by the electron and proton fiducial cuts, we varied  the cuts  applied to the $\phi$ angles accepted in each sector. The $\phi$ acceptance of each  of the six sectors was  less than $60^\circ$, depending on $\theta$, due to the thickness of the toroid magnet coil cryostats.   In order to avoid tracks which are too close to the coils, a fiducial cut in $\Delta\phi$ was applied of nominally $40^\circ$ ($\pm 20^\circ$ from the sector mid-plane) at larger angles $\theta$, tapering down to smaller $\Delta\phi$  for smaller $\theta$ as the  $\phi$ acceptance decreases. For electrons an additional cut of $\pm 3^\circ$ from the mid-plane was applied to avoid known inefficiencies of the Cherenkov detector in the sector mid-plane. The average systematic uncertainty arising from the placement of these cuts was estimated to be around 4.7\%.

The lower limit on the photon's energy in the EC calorimeter was varied from 350 MeV to 300 MeV for the evaluation of the systematic uncertainties due to this selection criteria. The uncertainties  were calculated for each bin and on average were  estimated to be $\sim1.6$\%.

The systematic uncertainties due to the exclusivity cuts on
 $M_x(e\gamma\gamma)$,
$E_x(ep\pi^0)$, and
M($\gamma\gamma$)
were studied in detail for each cut independently. The cuts were changed from 3$\sigma$ to 2$\sigma$ and
systematic uncertainties  were calculated in each bin. The average uncertainties for each cut, shown in Table~\ref{table:syst-summary}, varied between 2.5--3.2\%.

The systematic uncertainty of the radiative corrections was estimated as follows. The missing mass of the $ep$ system $M_x(ep)$ exhibits a radiative tail. Thus, when making a cut on $M_x(ep)$ there is a loss of radiated events, which was corrected using the routine EXCLURAD ~\cite{exclurad},  which depends on the value of the cut. The correction procedure was applied with varied cuts on $M_x(ep)$  from 0.1~GeV to 0.25~GeV in the data analysis program, and the same
value of this cut was applied to the simulated data.
The obtained cross sections were compared to the original ones bin-by-bin. On average the uncertainty was estimated to be 2.9\%.

 The systematic uncertainty in the cross section due to the normalization correction factor was estimated by the comparison of the normalization factors extracted from the six independent measurements of the elastic cross section in the six different CLAS sectors. 
The absolute normalization correction reflects systematic uncertainties which were not accounted for and which may lead to normalization errors. This systematic uncertainty  was estimated to be 6\%.

The uncertainty in the incident electron beam energy  was determined to be about  0.017 GeV and
its contribution to the overall cross section is small.

Finally, the overall systematic uncertainty was estimated by adding all contributions in quadrature and is about 10\%.

\begingroup
\squeezetable
\begin{table*}
\caption{ Summary table of systematic uncertainties.  B denotes  bin-to-bin and O indicates overall uncertainties}
\centering
\begin{ruledtabular}
\begin{tabular}{lcc}
Source & Bin-to-bin or overall & Average Uncertainty \\
\hline 
Proton ID &B& $\sim 2.5 \%$ \\
Fiducial cut &B &$\sim  4.7\%$ \\
Cut on energy of photon detected in the EC & B & $\sim 1.6\%$ \\
Cut on missing mass of the $e\gamma\gamma$ & B& $\sim2.5\%$ \\
Cut on invariant mass of 2 photons & B& $\sim2.9\%$ \\
Cut on missing energy of the $ep\gamma\gamma$ & B & $\sim 3.2\%$ \\
Radiative corrections & B & $\sim 2.9\%$ \\
Total beam charge on target & O & $ < 1\%$  \\
Target length &  O & 0.2\% \\
Absolute normalization &  O  & $6.0\%$ \\
\end{tabular}
\end{ruledtabular}
\label{table:syst-summary}
\end{table*} 
\endgroup

\section{Cross sections for $\gamma^*p\to \pi^0p$}
\label{section:cross_section}
\begin{widetext}
\noindent The four-fold differential cross section as a function of the four variables $(Q^2,x_B,t ,\phi_\pi)$ was obtained from the expression
\begin{equation}
\frac{d^4 \sigma_{ep \rightarrow e^\prime p^\prime \pi^0}}{dQ^2 dx_B dt d\phi_\pi} =
\frac{N(Q^2,x_B,t,\phi)}{\mathcal{L}_{int} (\Delta Q^2  \Delta x_B \Delta t \Delta \phi)} \times
\frac{1}{\epsilon_{ACC}\delta_{RC} \delta_{Norm}Br(\pi^0\to\gamma\gamma)}.
\label{eq:sig_ep_eppippim}
\end{equation}
\end{widetext}

\noindent The definitions of the kinematic variables  are given in  Appendix~\ref{section:kinematics}. The definitions of the other quantities in Eq.~\ref{eq:sig_ep_eppippim}  are:

\begin{itemize}

\item $N(Q^2,x_B,t,\phi_\pi)$ is the  number of $ep \rightarrow e^\prime p^\prime \pi^0$ events in a given ($Q^2,x_B,t,\phi_\pi$) bin;

\item $\mathcal{L}_{int}$ is the integrated luminosity (which takes
into account the correction for the data-acquisition dead time);

\item $(\Delta Q^2 \Delta x_B \Delta t \Delta \phi_\pi)$ is the corresponding bin width  (see Tables~\ref{q2-bins}--\ref{t-bins}).
For bins not completely filled, because of cuts in $\theta_e$, $W$  and $E^\prime$, as seen in Fig.~\ref{fig:kin_cuts},  the phase space $(\Delta Q^2 \Delta x_B\Delta t\Delta\phi_\pi)$ includes a 4-dimensional correction to take this into account. The specified $Q^2$, $x_B$ and $t$  values are the mean values of the data for each variable for each 4-dimensional bin, as if the cross sections in each bin vary linearly in each variable in the filled portion of the accepted kinematic volume.

\item $\epsilon_{ACC}$ is the acceptance calculated for each bin $(Q^2,x_B,t,\phi_\pi)$;

\item $\delta_{RC}$ is the correction factor due to the radiative effects calculated for each $(Q^2,x_B,t,\phi_\pi)$ bin;

\item $\delta_{Norm}$ is the overall absolute normalization factor calculated from the elastic cross section measured in the same experiment (see  Sec.\ref{systematics} above);
 
\item $Br(\pi^0\to\gamma\gamma)=\frac  {\Gamma(\pi^o\to\gamma\gamma) }   {\Gamma_{total}}$ is the branching ratio for the $\pi^0 \to \gamma\gamma$ decay mode.

\end{itemize}

The reduced or ``virtual photon"  cross sections  were extracted from the data through:

\begin{widetext}
\begin{equation}
\frac{d^2\sigma_{\gamma^* p \rightarrow p^\prime \pi^0} (Q^2, x_B,t,\phi_\pi, E)}{dt d\phi} = 
\frac{1}{\Gamma_V(Q^2,x_B,E)} 
\frac{d^4\sigma_{ep\rightarrow e^\prime p^\prime \pi^0}}{dQ^2 dx_B dt d\phi_\pi}.
\end{equation}
\end{widetext} 

\noindent
The Hand convention~\cite{Hand} was adopted for the
definition of the virtual photon flux $\Gamma_V$ (see Eq.~\ref{eq:GammaV} in  Appendix~\ref{section:helicity_amp}). 
A table of  the 1867 reduced cross sections   can be obtained online in Ref. \cite{full_table}. 
As an example of the information available, Table \ref{numerical-cross-sections} presents the reduced cross section for one kinematical point ($Q^2$=1.15~GeV$^2$, ~$x_B$=0.132,~$t$=-0.12~GeV$^2$).

\begin{centering}
\begingroup
\squeezetable
\begin{table}[h]
\caption{$d^2\sigma/dtd\phi_\pi$ at $t=-0.18$ GeV$^2$, $x_B=0.22$ and $Q^2=$1.75 GeV$^2$. The complete numerical listing for all measured kinematic points is found in Ref.~\cite{full_table}.}
\begin{ruledtabular}
\begin{tabular}{cccc}

$\phi_\pi$  & $\frac{d^2\sigma}{dt d\phi_\phi}$   & Statistical Error & Systematic Error \\
 (deg)     &  (nb/GeV$^2$)                & (nb/GeV$^2$)       & (nb/GeV$^2$)     \\
\hline

 9   & 55.8       & 9.0 & 12.0     \\
27  & 45.5       & 6.1 & 0.7     \\
45  & 56.7     & 5.9 & 6.0     \\
63  & 62.0       & 6.3 & 6.6     \\
81  & 70.8      & 6.1 & 11.1     \\
99  & 85.2      & 6.5 & 7.0     \\
117 & 61.7      & 6.4 &  5.8  \\
135 & 41.2       & 5.9 & 4.6     \\
153 & 35.7      & 5.5 & 3.6     \\
171 & 44.8       & 7.8 & 0.5     \\
189 & 30.9        & 5.9 & 3.6     \\
207 & 41.0      & 5.9 & 5.6     \\
225 & 42.9       & 6.5 & 2.8     \\
243 & 51.8     & 5.8 & 8.8     \\
261 & 69.2       & 6.0 & 2.4     \\
279 & 82.3      & 7.3 & 3.6     \\
297 & 77.5     & 7.1 & 4.2     \\
315 & 57.8     & 5.5 & 9.8     \\
333 & 48.7    & 6.2 & 4.4     \\
351 & 37.3     & 7.8 & 8.2     \\

\end{tabular}
\end{ruledtabular}
\label{numerical-cross-sections}
\end{table}
\endgroup
\end{centering}

\subsection{Integrated virtual photon cross section  $\sigma_U=\sigma_{T}+\epsilon\sigma_L$}
The total virtual photon cross section  is defined as the  reduced differential cross section  integrated over $\phi_\pi$ and $t$:

\begin{equation}
\sigma_{U}=\sigma_T+\epsilon\sigma_L=\int \int  \frac{d^2 \sigma}{ dt d\phi_\pi} dt  d\phi_\pi,
\end{equation}
\noindent
where $\sigma_T$ and $\sigma_L$ are due to transverse and longitudinal photons respectively. $\sigma_U$ depends on two variables $Q^2$ and $x_B$. The variable $\epsilon$ is the ratio of fluxes of longitudinally and transversely polarized virtual photons (see Eq. \ref{epsilon} in the appendix).

Since the CLAS acceptance has limited coverage  in some areas of the 4-dimensional phase space $(Q^2,x_B,t,\phi_\pi)$, the integral could be carried out over a finite range of the total phase space.  For example, at high $Q^2$ and $x_B$, the acceptance around  $\phi_\pi=180^\circ$ is near zero, so the $\phi_\pi$ integral cannot be fully calculated using the present data.
 To account for regions with small acceptance, a model that was developed for the Monte Carlo generator to describe $d^2 \sigma^{MC}/ dt d\phi_\pi$  was used.
This generator was  tuned using our own $\pi^0$ experimental data. 
Thus the integrated cross sections have an additional factor  $1/\eta$, where

\begin{equation}
\eta=\frac{\int \int_{\Omega^\prime} \ \frac{d^2 \sigma}{ dt d\phi_\pi}^{MC} dt  d\phi_\pi}{\int\int _{\Omega}\frac{d^2 \sigma}{ dt d\phi_\pi}^{MC} dt  d\phi_\pi}, 
\end{equation}
\noindent
in which  $\Omega$ is the full phase space and $\Omega^\prime$ is the phase space where CLAS has non-zero acceptance.
Only data points were included for partially covered kinematic volumes in which 
 $\eta$ was  greater than 0.45  to avoid extrapolation to the regions where the acceptance is low.  
The  value of $\eta$ is model dependent, which introduces an additional systematic uncertainty of $\sim 15\%$. 
The integration over the variable $|t|$  extends from  $|t_{min}|$  to 2 GeV$^2$.  

\begin{figure*}
\begin{center}
\includegraphics[width=5 in]{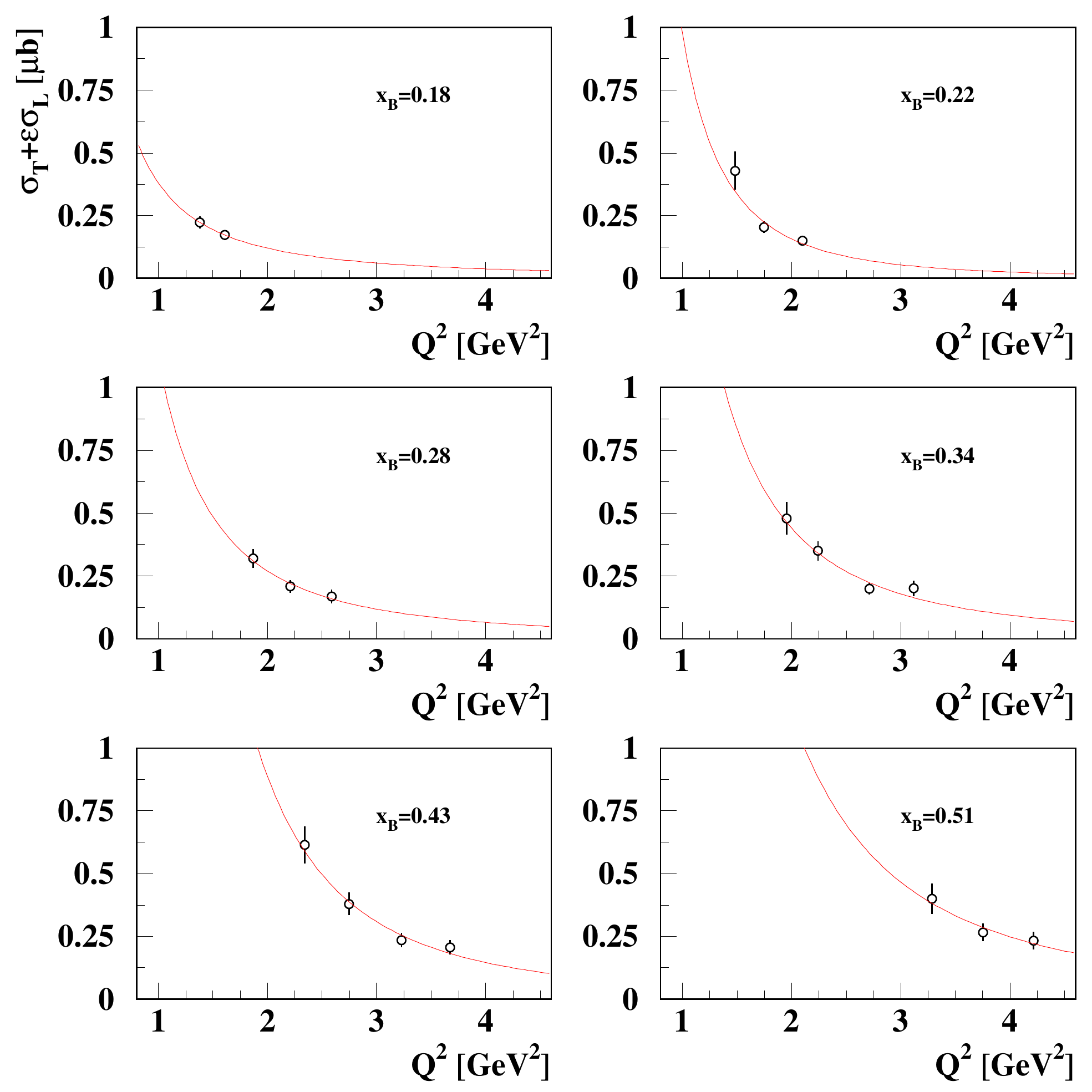}
\end{center}
\caption{(color online) 
The   $t$-integrated   ``virtual photon"  cross section $\sigma_T+\epsilon\sigma_L$ as a function of $Q^2$ for the reaction
$\gamma^*p\to p^\prime\pi^0$ for $x_B$=0.18, 0.22, 0.28, 0.34, 0.43 and 0.51. The curves are fits
to a power law $\sigma_U =A_{Q^2}/Q^n$ where $A_{Q^2}$ and $n$ are fit parameters. }
\label{fig:s_q2} 
\end{figure*}

Fig.~\ref{fig:s_q2} shows the integrated  cross section  $\sigma_U$  as a function of $Q^2$ for  different
values of $x_B$. The  cross sections were fit by the simple expression \mbox{$\sigma_U \sim 1/Q^n$} to estimate the
$Q^2$ dependence.  The weighted mean of the exponent parameters is $n=4.7\pm 0.7
$. Reference~\cite{Hall-A-pi0} finds $n= 4.78\pm0.16$ based upon two values of $Q^2$ (1.9 and 2.3~GeV$^2$).
The asymptotic prediction of the conventional GPD models is  $\sigma_L\sim1/Q^6$ and $\sigma_T\sim1/Q^8$.
The parameters of the fit are given in Table \ref{table-qdep}.

\begingroup
\squeezetable
\begin{table}
\caption{Parameters of $Q^2$-dependent fits to the $t$-integrated cross sections in Fig.~\ref{fig:s_q2} for different values of $x_B$.}
\begin{ruledtabular}
\begin{tabular}{ccc}
 $x_B$ & $A_{Q^2}$ & $n$ \\\hline

 0.18 &  0.38 $\pm$ 0.16 &  3.32 $\pm$ 2.04 \\
 0.22 &  0.97 $\pm$ 0.39 &  5.26 $\pm$ 1.34 \\
 0.28 &  1.11 $\pm$ 0.48 &  4.09 $\pm$ 1.12 \\
 0.34 &  2.06 $\pm$ 0.71 &  4.46 $\pm$ 0.77 \\
 0.43 &  5.41 $\pm$ 1.83 &  5.22 $\pm$ 0.63 \\
 0.51 &  5.19 $\pm$ 3.12 &  4.39 $\pm$ 0.91 \\
\end{tabular}
\end{ruledtabular}
\label{table-qdep}
\end{table}
\endgroup

\begingroup
\squeezetable
\begin{table}
\caption{Parameters of $W$-dependent fits  to the $t$-integrated cross sections in Fig.~\ref{fig:s_W_fit} for different values of $Q^2$.}
\begin{ruledtabular}
\begin{tabular}{ccc}
 $Q^2$ &  $A_{W}$  & $n$ \\\hline
    1.34 &  5.01  $\pm$   2.94 &   3.03 $\pm$   0.56\\
    1.79 &  7.82  $\pm$   2.77 &   3.64 $\pm$   0.37\\
    2.22 &  11.90 $\pm$   3.53 &   4.23 $\pm$   0.33\\
    2.68 &  5.76  $\pm$   2.64 &   3.61 $\pm$   0.52\\
    3.21 &  2.38  $\pm$   1.56 &   2.68 $\pm$   0.80\\
    3.71 &  1.30  $\pm$   1.24 &   2.12 $\pm$   1.20\\
\end{tabular}
\end{ruledtabular}
\label{table-wdep}
\end{table}
\endgroup

\begin{figure*}
\begin{center}
\includegraphics[width=5in]{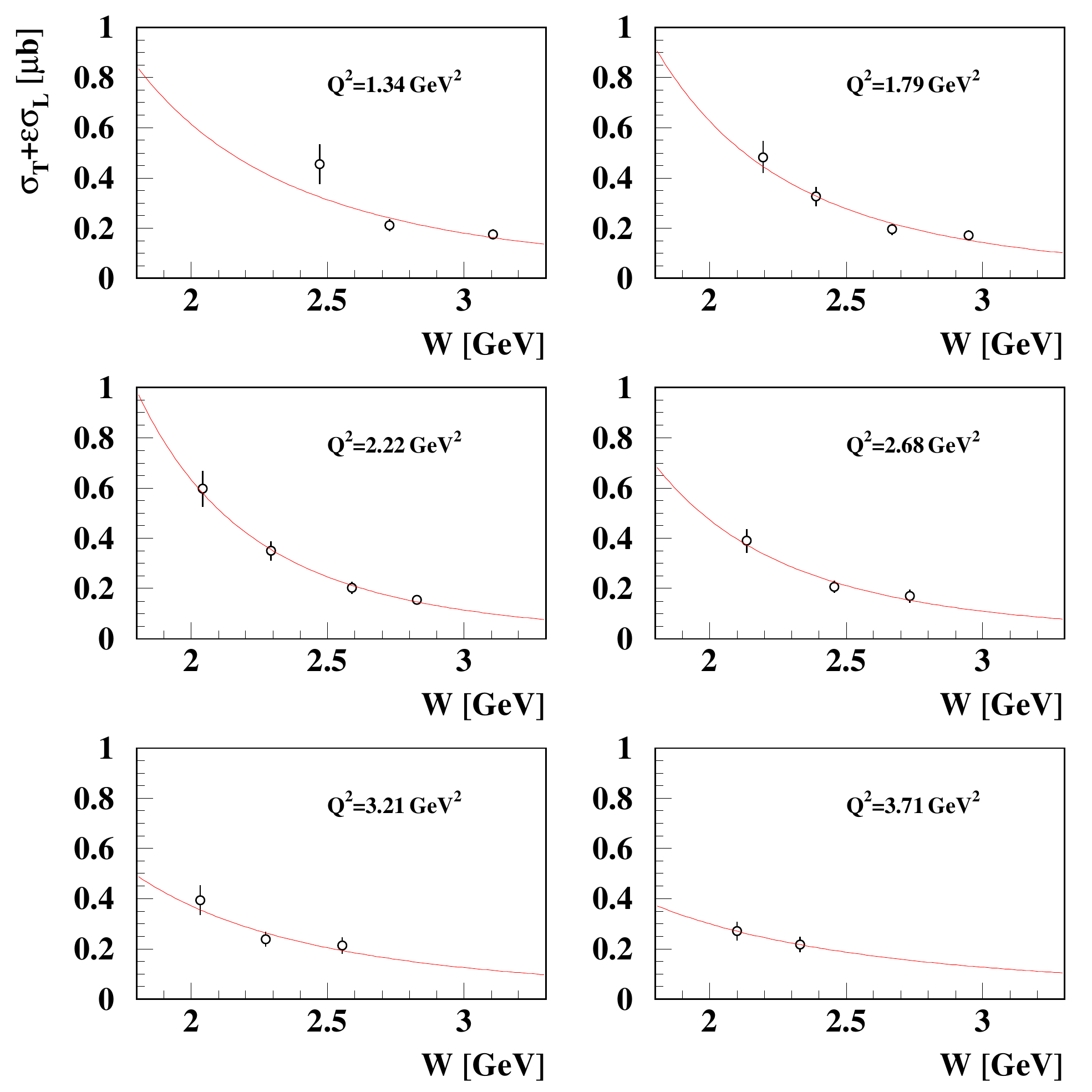}
\end{center}
\caption{(Color online) 
The $t$-integrated  ``virtual photon"  cross section $\sigma_T+\epsilon\sigma_L$ as a function of $W$ for the reaction
$\gamma^*p\to p^\prime\pi^0$ for $Q^2$=1.34, 1.79, 2.22, 2.68, 3.21 and 3.71 GeV$^2$. The curves are fits
to a power law $\sigma_U= A_W/W^n$ where $A_W$ and $n$ are fit parameters.
}
\label{fig:s_W_fit} 
\end{figure*}

The total cross section  $\sigma_U=\sigma_T+\epsilon\sigma_L$ as a function of $W$ for different values of $Q^2$ is shown in  Fig.~\ref{fig:s_W_fit}. The cross sections were fitted with the function $\sigma\sim1/W^n$. The weighted mean value of the exponent is $n=3.7\pm0.3$. Ref.~\cite{Hall-A-pi0} finds $n= 3.48\pm0.11$ based upon two values of $W$.  The $W$ dependence is consistent with what was observed for $\rho$  electroproduction  \cite{morrow}, i.e. the cross section decreases with $W$ compatibly with the Regge-model predictions \cite{Laget} for the exclusive reactions. The parameters of the fit are given in Table \ref{table-wdep}. 

\subsection{The $t$-dependent differential cross section  ${d\sigma_U}/{dt}$}

Integrating only over $\phi_\pi$ yields the $t$-dependent differential cross section 

\begin{equation}
\frac{d\sigma_{U}}{dt}=\int  \frac{d^2 \sigma}{ dt d\phi_\pi}  d\phi_\pi.
\end{equation}
\noindent
The correction factor for  the region where
the CLAS detector has zero acceptance was calculated  as
\begin{equation}
\eta^\prime =\frac{\int_{\Omega^*} \ \frac{d^2 \sigma}{ dt d\phi}^{MC}  d\phi_\pi}
                   {\int _{\Omega}\frac{d^2 \sigma}{ dt d\phi_\pi}^{MC}  d\phi_\pi},
\end{equation}                   
\noindent
in which  $\Omega$ is the full phase space and $\Omega^*$ is the phase space where CLAS has non-zero acceptance.

\begin{figure*}
\begin{center}
\includegraphics[width=\textwidth]{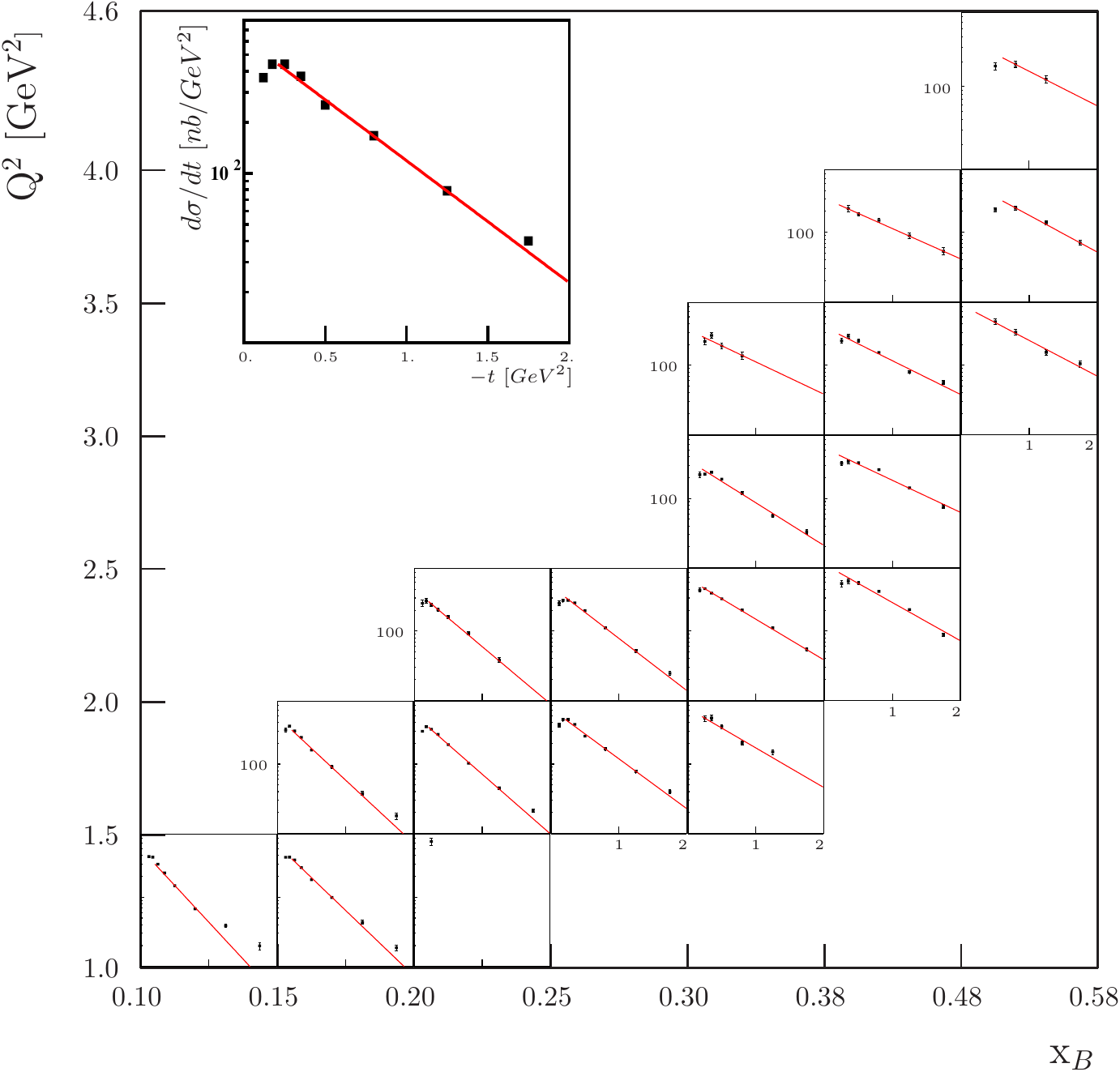}
\end{center}
\caption{(Color online)
The differential cross section $d\sigma_U/dt$=$d\sigma_T/dt+\epsilon d\sigma_L/dt$  for the reaction
$\gamma^*p\to p^\prime\pi^0$. The curves are fits to the exponential function $e^{bt}$. The insert is  an enlarged copy of the panel centered at $Q^2$=1.75 GeV$^2$ and 
$x_B$=0.275. Systematic uncertainties, including the estimated systematic uncertainty in the integration correction factor $\eta$ of 15\%, as discussed in the text, are not shown.} 

\label{fig:dsdt_log} 
\end{figure*}

Fig. \ref{fig:dsdt_log} shows the cross section  $d\sigma_T/dt+\epsilon d \sigma_L/dt$ for intervals of $Q^2$ for the different values of $x_B$. The presented cross sections were calculated only for the kinematics where the factor $\eta^\prime$ was greater than 0.45. The general feature of these distributions is that in a small interval near $|t|=|t|_{min}$ they are not diffractive. There, the cross sections cannot be described by simple exponential functions.  
However, for somewhat larger values of $\vert t \vert$, the cross sections appear to fall off exponentially with $-t$, and thus were fit by the function $e^{bt}$, where the exponential functions  appears to fit  the data with a good  $\chi^2$.  
This provides a qualitative  description of  the $|t|$-dependence by a slope parameter $b$.
The curves in Fig.~\ref{fig:dsdt_log} are the results of these fits.

\begin{figure}
\begin{center}
\includegraphics[width=\columnwidth]{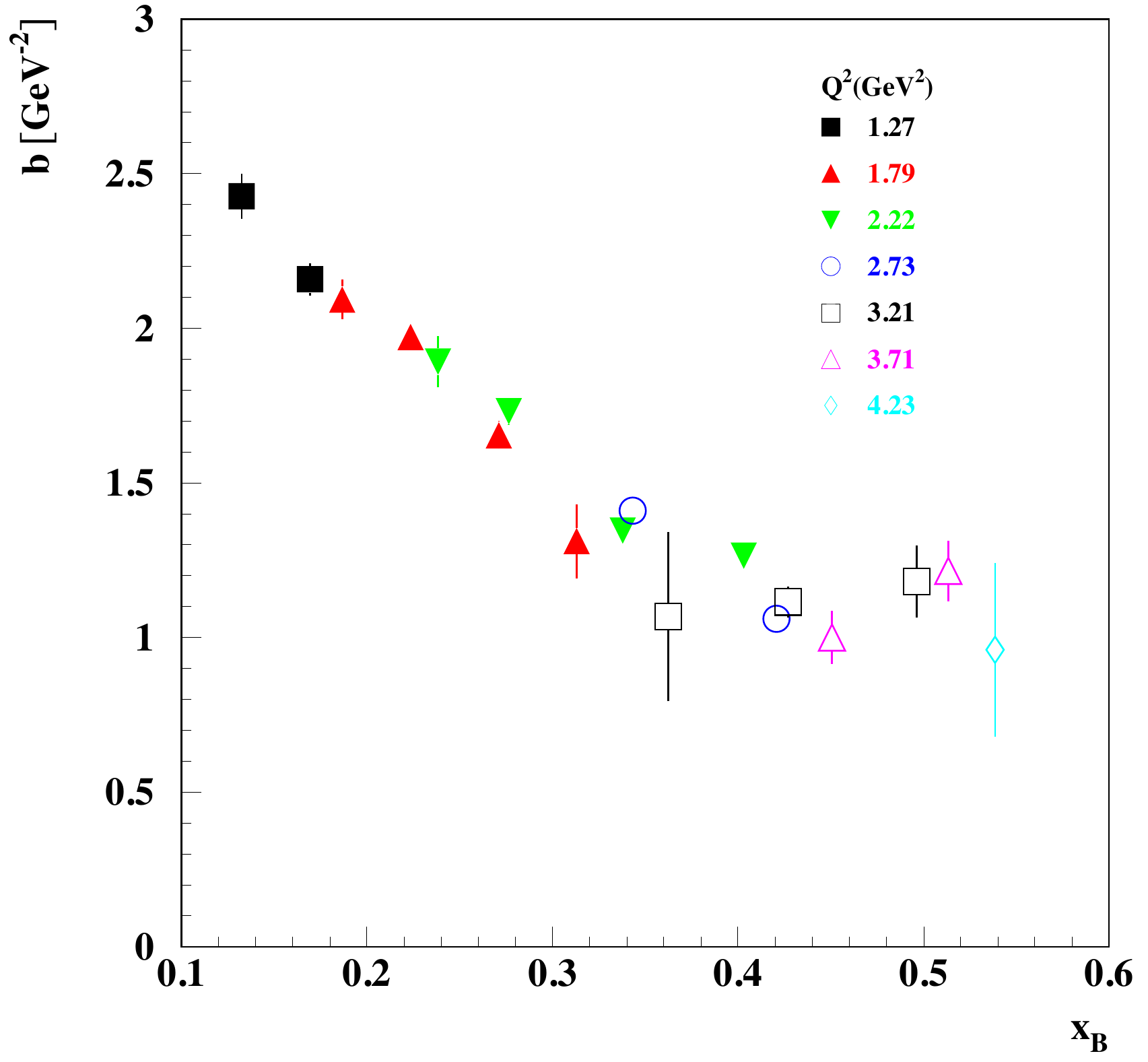}

\end{center}
\caption{
(Color online)
$t$-slope parameter $b$ for the reaction $\gamma^* p \to p^\prime \pi^0$ as a function of $x_B$ for different values of $Q^2$. 
}
\label{fig:b_slope_xb_Q2} 
\end{figure}

Fig.~\ref{fig:b_slope_xb_Q2} shows the slope parameter $b$ as a function of $x_B$ for different values of $Q^2$. 
The values of   $b$ are between 1 and 2.5 GeV$^{-2}$. The data appear to exhibit a slope parameter decrease with increasing $x_B$ for each $Q^2$ over much of the measured range, except at the highest measured regions of  $x_B$ and $Q^2$. However, the $Q^2-x_B$ correlation in the CLAS acceptance  does not permit one to make a definite conclusion about the  $Q^2$ dependences of the slope parameter for fixed $x_B$. What one can say is that at high $Q^2$ and high $x_B$  ($Q^2=4.3$~GeV$^2$, $x_B$=0.53), the slope parameter is smaller than for the lowest values of these variables ($Q^2=1.2$~GeV$^2$, $x_B$=0.12).  
The $b$ parameter in the exponential determines the  width of the transverse momentum distribution of the emerging protons, which, by  a Fourier  transform, is inversely related to the transverse size of the interaction region from which the proton emerges.  From the point of view of the handbag picture, it is inversely related to the  separation, $r_\perp$,   between the active quark and the center of momentum of the spectators (see Ref. \cite{Burkardt}). Thus the data implies that the separation is  larger at the lowest $x_B$ and $Q^2$ and becomes smaller for increasing $x_B$ and $Q^2$, as it must.

\subsection{Structure functions  }
The reduced cross sections can be expanded in terms of  structure functions 
$d\sigma_T/dt$,\  $d\sigma_L/dt$,\ $d\sigma_{LT}/dt$, and $d\sigma_{TT}/dt$ as follows:

\begin{widetext}
\begin{equation}
\frac{d^2\sigma}{dtd\phi_\pi} = \frac{1}{2\pi} 
\left [\left( \frac{d\sigma_T}{dt}+\epsilon\frac{d\sigma_L}{dt}\right) 
+ \epsilon \cos 2 \phi_\pi \frac{d\sigma_{TT}}{dt} + 
\sqrt{2\epsilon(1+\epsilon)} \cos \phi_\pi \frac{d\sigma_{LT}}{dt} \right],
\label{eq:sigmaphidep}
\end{equation}
\end{widetext}

\noindent
from which the three combinations of structure functions, ($\frac{d\sigma_T}{dt}+\epsilon\frac{d\sigma_L}{dt})$,
$\frac{d\sigma_{TT}}{dt}$ and 
$\frac{d\sigma_{LT}}{dt}$ can be extracted by fitting the cross sections to the $\phi_\pi$ distribution in each bin of $(Q^2,x_B,t)$.
The decomposition of the structure functions in terms of helicity amplitudes is given in Appendix~\ref{section:helicity_amp}, Eqs. ~\ref{B10} to~\ref{B13}.

The physical significance of the structure functions is as follows:
\begin{itemize}
\item[-] $d\sigma_L/dt$  is the sum of structure functions  initiated by a longitudinal virtual photon, both with and without nucleon helicity-flip, i.e. respectively $\Delta \nu = \pm 1$ and $\Delta \nu = 0$.

\item[-] $d\sigma_T/dt$ is the sum of structure functions which are initiated by a transverse virtual photon of positive and negative helicity ($\mu = \pm 1$), with and without nucleon helicity flip, respectively $\Delta \nu = \pm 1$ and  $0$.

\item[-] $d\sigma_{LT}/dt$ corresponds to interferences involving products  of amplitudes for longitudinal and transverse photons.

\item[-] $d\sigma_{TT}/dt$ corresponds to interferences involving products  of transverse positive and negative photon helicity amplitudes.

\end{itemize}

\begin{figure}
\begin{center}
\includegraphics[scale=0.5]{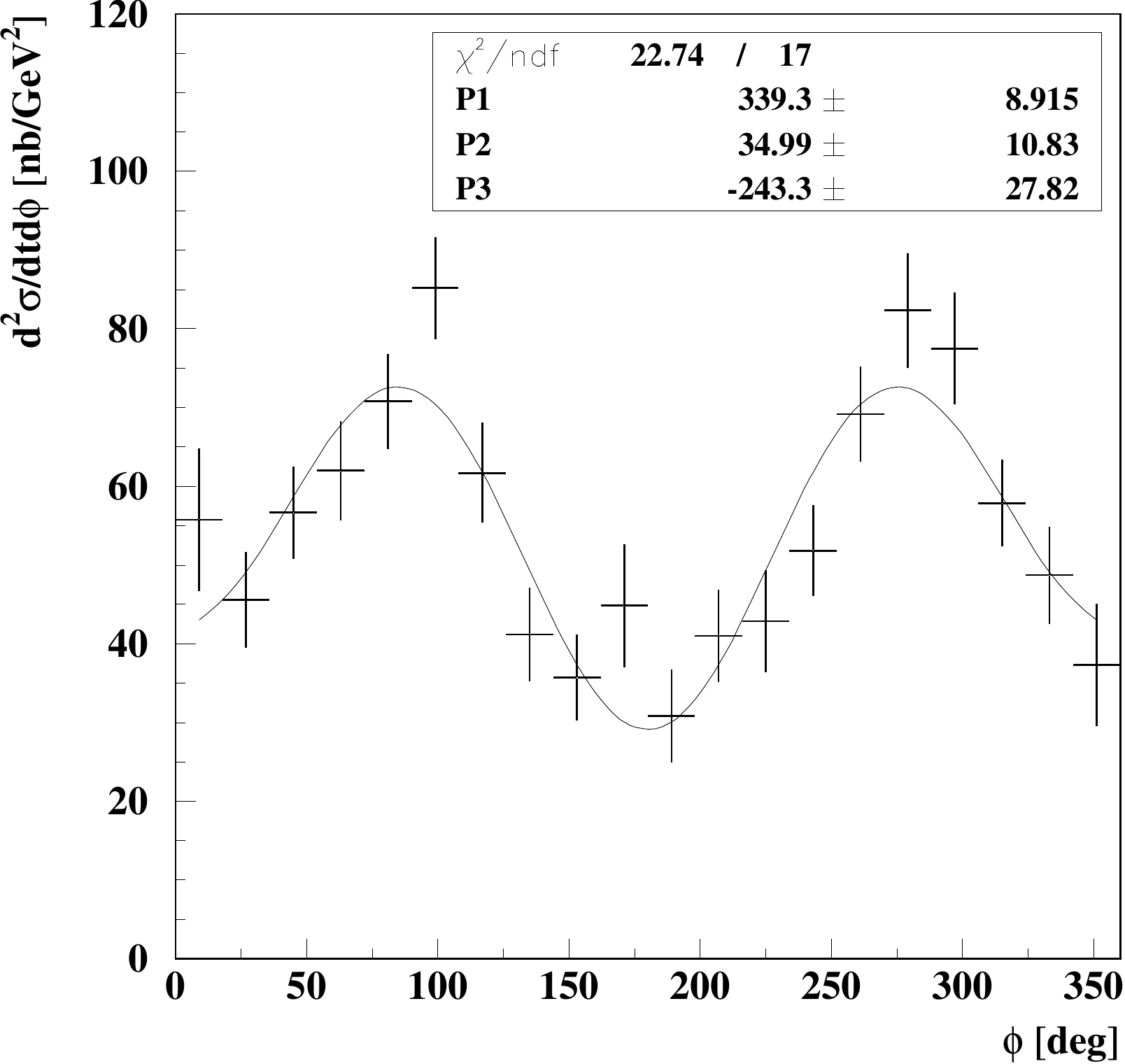}
\end{center}
\caption{
Example  of  the $\phi_\pi$ distribution of $d^2\sigma/dtd\phi_\pi$. The solid curve is a fit of the function in Eq.~\ref{eq:sigmaphidep}.
The kinematic bin corresponding to  this figure is at $t=-0.18$ GeV$^2$, $x_B=0.22$ and $Q^2=$1.75 GeV$^2$ and the data is listed in  Table~\ref{numerical-cross-sections}. 
Error bars are  statistical. 
The complete listing of all differential cross sections for all kinematic settings are found in Ref.~\cite{full_table}.}
\label{fig:phi_dist}
\end{figure}

Figure \ref{fig:phi_dist} shows a typical $\phi_\pi$-distribution of the virtual photon cross sections with a fit using the form of   Eq.~\ref{eq:sigmaphidep}.  These data are listed in Table \ref{numerical-cross-sections} as well. The complete listing of all differential cross sections for all kinematic settings are found in Ref.~\cite{full_table}.

\begin{figure*}
\begin{center}
\includegraphics[width=\textwidth]{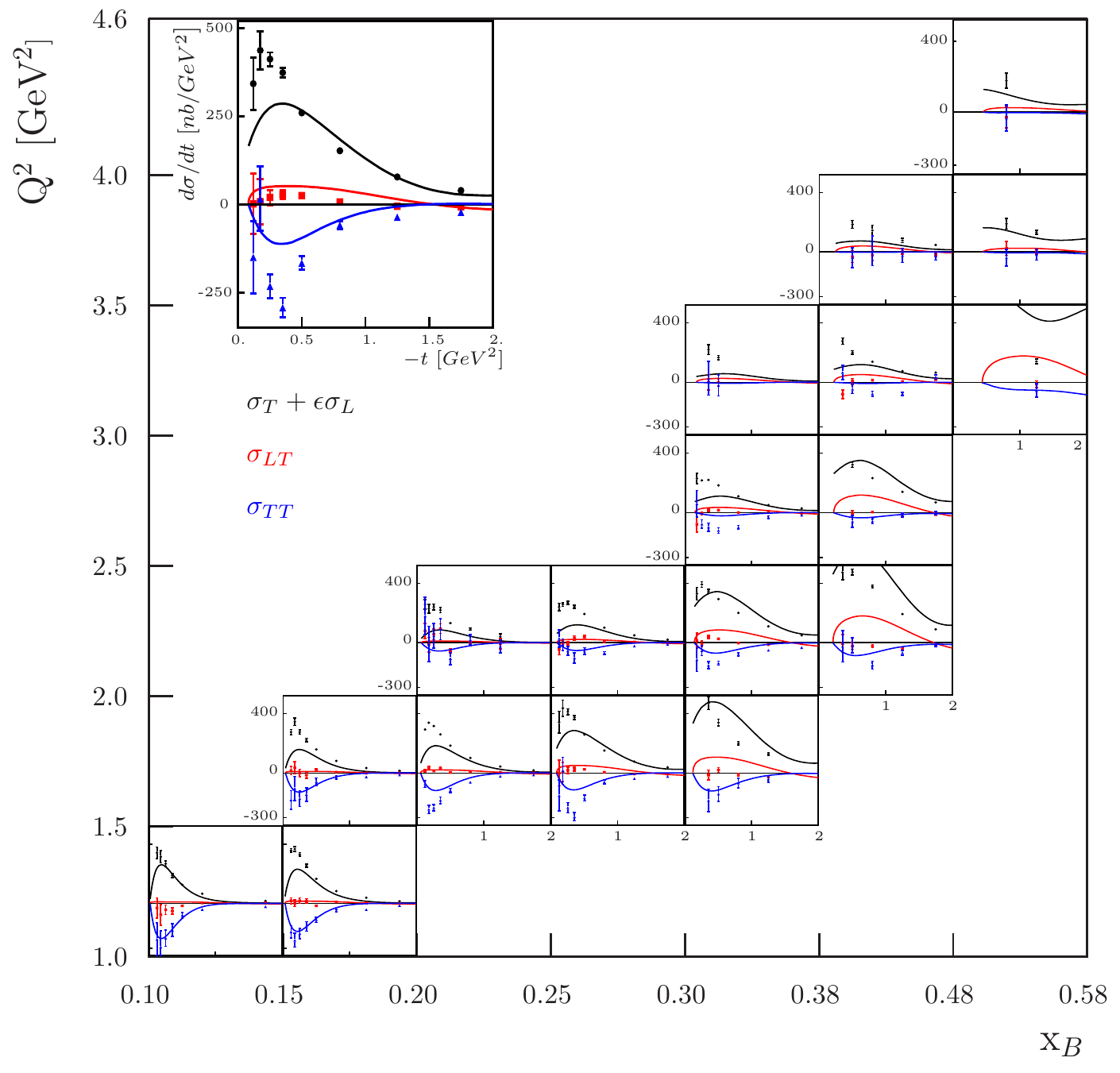}
\caption{
(Color online) Structure functions $d\sigma_U = d\sigma_T/dt+\epsilon d\sigma_L/dt $ (black circles), $ d\sigma_{TT}/dt$ (blue triangles)
and $d\sigma_{LT}/dt$ (red squares) as a function of $-t$ for different  $Q^2$ and $x_B$ for the reaction $\gamma^* p \to p^\prime \pi^0$.  All the structure functions are numerically given in Appendix~\ref{strfun_table}. 
The error bars are statistical only. The point-by-point propagated systematic uncertainties for all the structure functions are given  in Appendix~\ref{strfun_table}. The curves are the results of a Regge-based calculation \cite{Laget}: black (positive)-$d\sigma_U/dt $,\  blue (negative)-$d\sigma_{TT}/dt$,  and  red (small)-$d\sigma_{LT}/dt$. Note that  in the higher-$x_B$/lower-$Q^2$ bins that the black curves ($d\sigma_U$) from the model are much higher than the data and become off-scale.}
\label{fig:structure_functions}
\end{center}
\end{figure*}

Fig.~\ref{fig:structure_functions}  shows the extracted structure functions for  all    kinematical bins in $(Q^2,x_B,t)$.  The values of the structure functions are given numerically in Table~\ref{strfun_table}. The results of a Regge-based calculation \cite{Laget} are also shown in Fig.~\ref{fig:structure_functions}.

 A number of observations can be made independently of the model predictions. The 
$d\sigma_{TT}/dt$ structure function is negative  and  $|d\sigma_{TT}/dt|$
 is comparable  in magnitude with the unpolarized structure function 
 ($d\sigma_T/dt+\epsilon d\sigma_L/dt$).
  However, $d\sigma_{LT}/dt$ is small in comparison with $d\sigma_U/dt$ and $d\sigma_{TT}/dt$.
  This reinforces the conclusion that the asymptotic leading-order handbag approach  for which $d\sigma_L/dt$ is dominant is not applicable at the present values of $Q^2$.

\section{Comparisons with Theoretical Models}
\subsection{Regge model}

\begin{figure}
\includegraphics[width=\columnwidth]{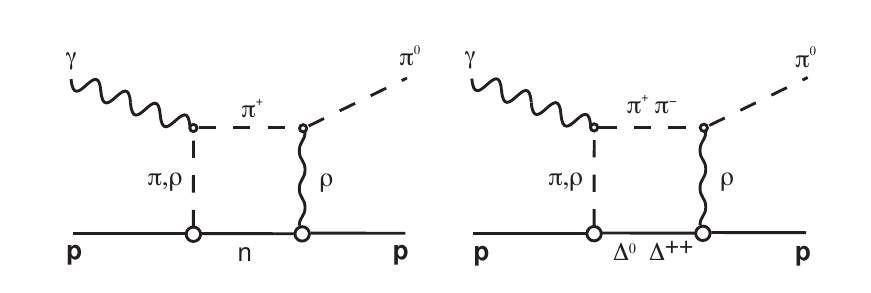}
\caption{ \label{rescatter-pi-1}Rescattering diagrams with the pion charge-exchange processes included in Ref.~\cite{Laget}. The vertical dashed and wavy lines represent the exchange of Regge trajectories. The horizontal lines correspond to on-shell  meson nucleon rescattering processes. }
\end{figure}

\begin{figure}
\includegraphics[width=\columnwidth]{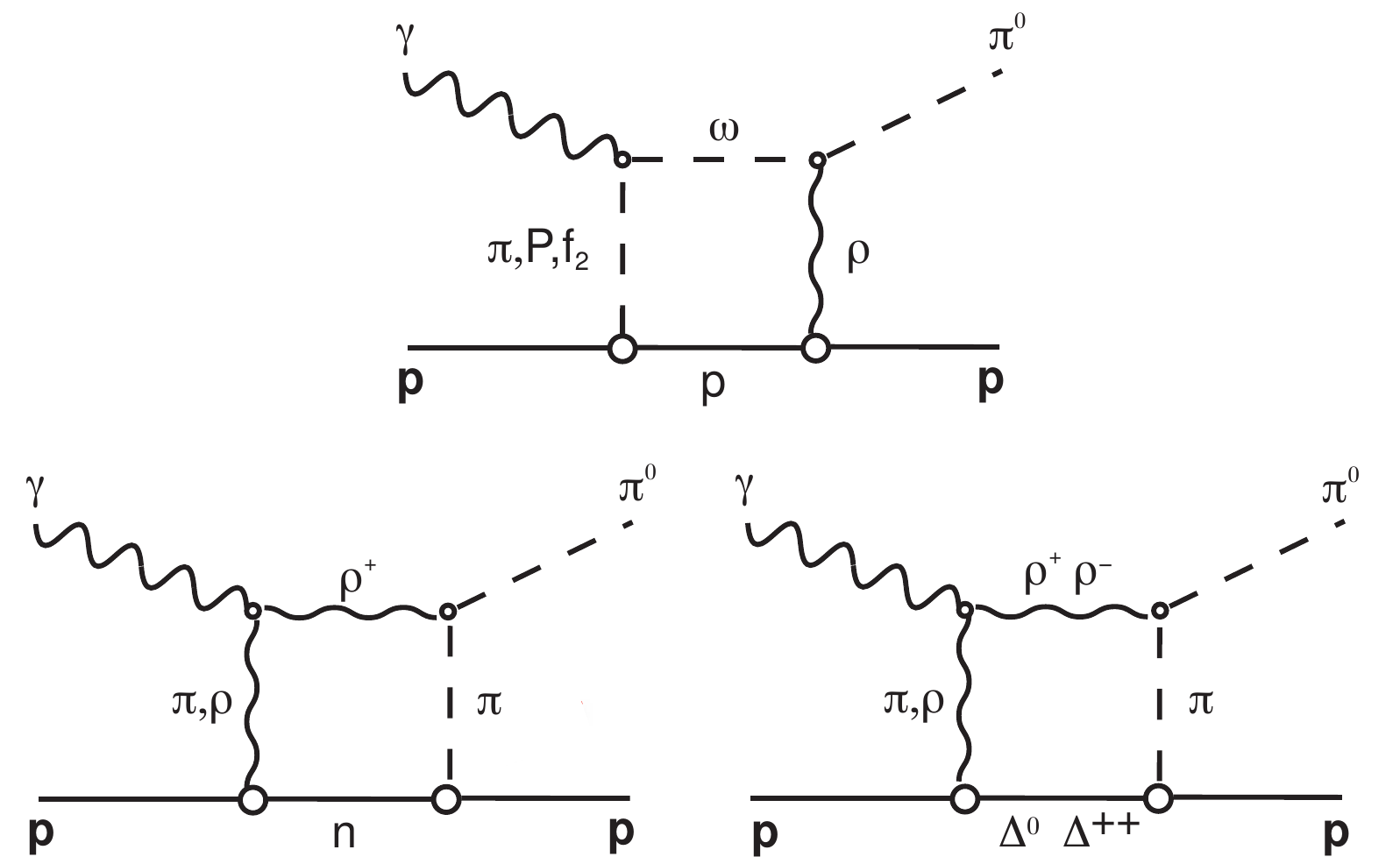}
\caption{ \label{rescatter}Rescattering diagrams with vector meson processes included in  Ref.~\cite{Laget} .}
\end{figure}

The Regge model with  charge exchange and  $\pi^\pm$ final state interactions, in addition to pole terms and elastic $\pi^0$  rescattering,  had been successfully applied in Refs.~\cite{Laget-2006,Laget-2010}  to  $\pi^0$ electroproduction at DESY  at  $Q^2$ = 0.25, 0.50 and 0.85  GeV$^2$. This mechanism, which is illustrated schematically in Fig.~\ref{rescatter-pi-1}, includes a charged-pion rescattering amplitude (see Fig.~\ref{rescatter}). Schematically, the amplitude can be written as a product of two terms: 
$$T_{\pi N} \propto \int d\Omega T_{\gamma p \to \pi^+N}(t_\gamma ) T_{\pi N \to \pi^0p}(t_\pi),$$
\noindent in which $t_\gamma = \left( k_\gamma - P_\pi \right)^2$.  The first term in the integral is the amplitude for production of a charged off-shell meson by a virtual photon and the second characterizes its rescattering. The amplitudes are largest where the intermediate mesons become on-shell. 

However, when  this scheme was applied to the Jefferson Lab Hall A kinematics~\cite{Hall-A-pi0}  at 
$Q^2$ = 2.35~GeV$^2$, the calculated cross sections were found to be an order of magnitude too low (see Ref.~\cite{Laget}). In fact, it was very difficult to understand why the experimental cross section at $Q^2$=2.35 GeV$^2$ is comparable in magnitude to the cross section  at much lower $Q^2$ values.

Then, Ref.~\cite{Laget}  included a vector-meson rescattering amplitude (see Fig.~\ref{rescatter}) taking  the form
$$T_{VN}\propto \int d\Omega T_{\gamma p\to VN}(t_\gamma ) T_{VN \to \pi^0p}(t_\pi).$$
It was found that the contributions of the $\rho^+ \Delta^0$ and $\rho^- \Delta^{++}$  rescattering (Fig.~\ref{rescatter} lower-right) are the most important, far more important than the $\omega p$ or $\rho^0p$  terms because the cross section of the $N(\rho^+,\pi)N$ reaction is larger than the $N(\omega,\pi)N$ cross section, and $N(\rho^0,\pi^0)N$  cannot occur.  These comparisons were only carried out in a narrow range of kinematics  corresponding to the available Hall A data. 

The comparison of the present data with the predictions of the Regge model \cite{Laget} is shown in Fig.~\ref{fig:structure_functions}.
Although the Regge model managed to describe the Hall A cross-section data in a narrow region of $Q^2$ and $t$, the situation here, with the large kinematic acceptance, is much more complex. In some regions of $Q^2$ and $t$ the predictions appear better than in others. This model does predict the correct signs and values of $\sigma_{TT}$ and the small value of $\sigma_{LT}$  in almost all the data intervals. 

\vspace{0.1in}

\subsection{Handbag model}
\begin{figure*}
\begin{center}
\includegraphics[scale=0.8]{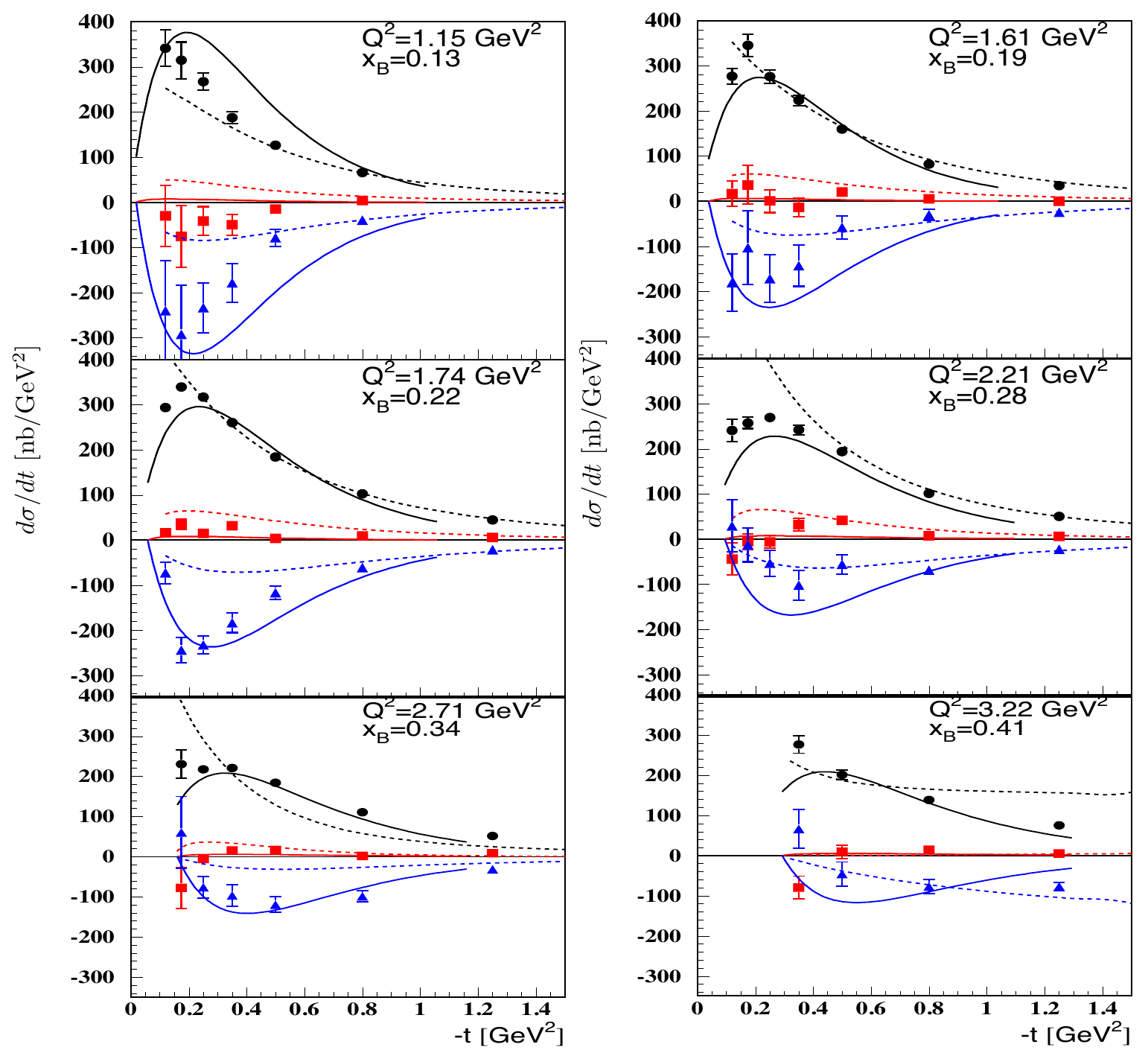}
\end{center}
\caption{ \label{fig:GK-GL} (Color online) 
The extracted structure functions vs. $t$ for the  bins  with the best kinematic coverage and  for which there are theoretical calculations. The data and curves are as follows:  black (filled circles)  - $d\sigma_U/dt =d\sigma_T/dt +\epsilon d\sigma_L/dt$,\  blue (triangles) - $d\sigma_{TT}/dt$ ,  and  red (squares) - $d\sigma_{LT}/dt$. All the structure functions are numerically given in Appendix~\ref{strfun_table}. 
The error bars are statistical only. The point-by-point propagated systematic uncertainties for all the structure functions are given in  Appendix~\ref{strfun_table}. The curves are theoretical predictions produced  with the models of Refs.~\cite{GK-ps-11} (solid) and ~\cite{Goldstein:2010gu} (dashed). In particular: black (positive) - $d\sigma_U/dt (=d\sigma_T/dt +\epsilon d\sigma_L/dt)$,\  blue (negative) - $d\sigma_{TT}/dt$,  and  red (small) - $d\sigma_{LT}/dt$
}
\end{figure*} 

Fig.~\ref{fig:GK-GL} shows the experimental structure functions at selected values of $Q^2$ and $x_B$.
The results of two GPD-based models which include transversity GPDs \cite{GK-ps-11,Goldstein:2010gu}  are  superimposed in Fig.~\ref{fig:GK-GL}. 
The primary contributing GPDs in meson production for transverse photons are  $H_T$, which characterizes the quark distributions involved in nucleon helicity-flip, and   $\bar E_T (= 2\widetilde H_T + E_T) $ which characterizes the quark distributions  involved in  nucleon  helicity-non-flip processes~\cite{diehl_haegler, Goekeler}. As a reminder, in both cases the active quark undergoes a helicity-flip.

Reference~\cite{GK-ps-11} obtains the following relations (see the Appendix for more details):
\noindent

\begin{align}\label{sigmat}
\frac{d\sigma_{T}}{dt} &= \frac{4\pi\alpha}{2k^\prime}\frac{\mu_\pi^2}{Q^8} \left[ \left(1-\xi^2\right) \left|\GPDHT\right|^2 - \frac{t'}{8m^2} \left|\GPDETbar\right|^2\right]
\end{align}

\begin{align}\label{sigmatt}
\frac{d\sigma_{TT}}{dt} = \frac{4\pi\alpha}{k^\prime}\frac{\mu_\pi^2}{Q^8}\frac{t'}{16m^2}\left|\GPDETbar\right|^2.
\end{align}

\noindent Here $\kappa^\prime(Q^2,x_B)$ is a phase space factor, $t^\prime =t-t_{min}$, where $|t_{min}|$ is the minimum value of $|t|$ corresponding to $\theta_\pi =0$, and the brackets $\langle  H_T \rangle$ and $\langle \bar E_T \rangle$ denote
the convolution of the elementary process with the GPDs $H_T$ and $\bar E_T$. The GPD $\bar E_T$  describes the spatial density  of transversely  polarized quarks in an unpolarized nucleon~\cite{diehl_haegler,Goekeler}.

Note that for the case of nucleon helicity-non-flip, characterized by  the GPD $\bar E_T$, overall helicity from the initial to the final state is not conserved. However,  angular momentum  is conserved, the difference being absorbed by the orbital motion of the scattered 
$\pi^0-N$ pair.  This accounts for the additional $t^\prime (= t-t_{min})$ factor multiplying the $\bar E_T$ terms in Eqs. \ref{sigmat} and \ref{sigmatt}.

In both calculations the contribution of $\sigma_L$  accounts for only a small fraction (typically less than  a few  percent) of the unseparated  structure functions $d\sigma_T/dt+ \epsilon d\sigma_L/dt$ in the kinematic regime under investigation. This is because  the contributions from $\tilde H$  and $\tilde E$,  the GPDs which are responsible for the leading-twist structure function $\sigma_L$,  are very small compared with the contributions from $\bar E_T$  and $H_T$,  which contribute to $d\sigma_T/dt$ and $d\sigma_{TT}/dt$.
In addition,  the transverse cross sections are strongly enhanced by the chiral condensate through the parameter $\mu_\pi=m^2_\pi/(m_u+m_d)$, where $m_u$ and $m_d$ are current quark masses \cite{G-K-09}.

 With the inclusion of the quark-helicity non-conserving  chiral-odd GPDs, which  contribute primarily to $d\sigma_T/dt$ and  $d\sigma_{TT}/dt$  and, to a lesser extent, to $d\sigma_{LT}/dt$, the model of Ref.~\cite{GK-ps-11} agrees rather well with the data. Deviations in shape become greater at smaller $-t$ for the unseparated cross section $d\sigma_U/dt$. The behavior of the cross section as $|t| \to |t|_{min}$ is determined by the interplay between $H_T$ and $\bar E_T$. 
For the GPDs of Ref.~\cite{GK-ps-11}  the parameterization was guided by the lattice calculation results of Ref.~\cite{Goekeler},  while Ref.~\cite{Goldstein:2010gu} used a GPD Reggeized diquark-quark model to obtain the GPDs.
The results  in Fig.~\ref{fig:GK-GL} 
for  the model of Ref.~\cite{GK-ps-11} (solid curves), in which  $\bar E_T$  is dominant,  agree rather well with the data. In particular, the  structure function $\sigma_U$ begins to decrease as $|t |\to |t|_{min}$, showing the effect of $\bar E_T$.    In the model of 
Ref.~\cite{Goldstein:2010gu}  (dashed curves) $H_T$ is dominant, which leads to a large rise in cross section as $-t$ becomes small so that the contribution of $\bar E_T$  relative to $H_T$ appears to be underestimated.
One can make a similar  conclusion from  the comparison between data and model predictions for  
$\sigma_{TT}$. This  shows the sensitivity of the measured  $\pi^0$ structure functions  for constraining  the transversity GPDs.
From Eq.~\ref{sigmat} for $d\sigma_T/dt$ and Eq.~\ref{sigmatt} for $d\sigma_{TT}/dt$ one can conclude that $|d\sigma_{TT}/dt|<d\sigma_T/dt<d\sigma_U/dt$.
One sees from Fig.~\ref{fig:GK-GL} that $-d\sigma_{TT}/dt$ is a sizable fraction of the unseparated cross section while $d\sigma_{LT}/dt$ is very small, which implies that contributions from transversity GPDs play a dominant  role in the $\pi^0$ electroproduction process.

\begin{figure*}
\begin{center}
\includegraphics[width=\textwidth]{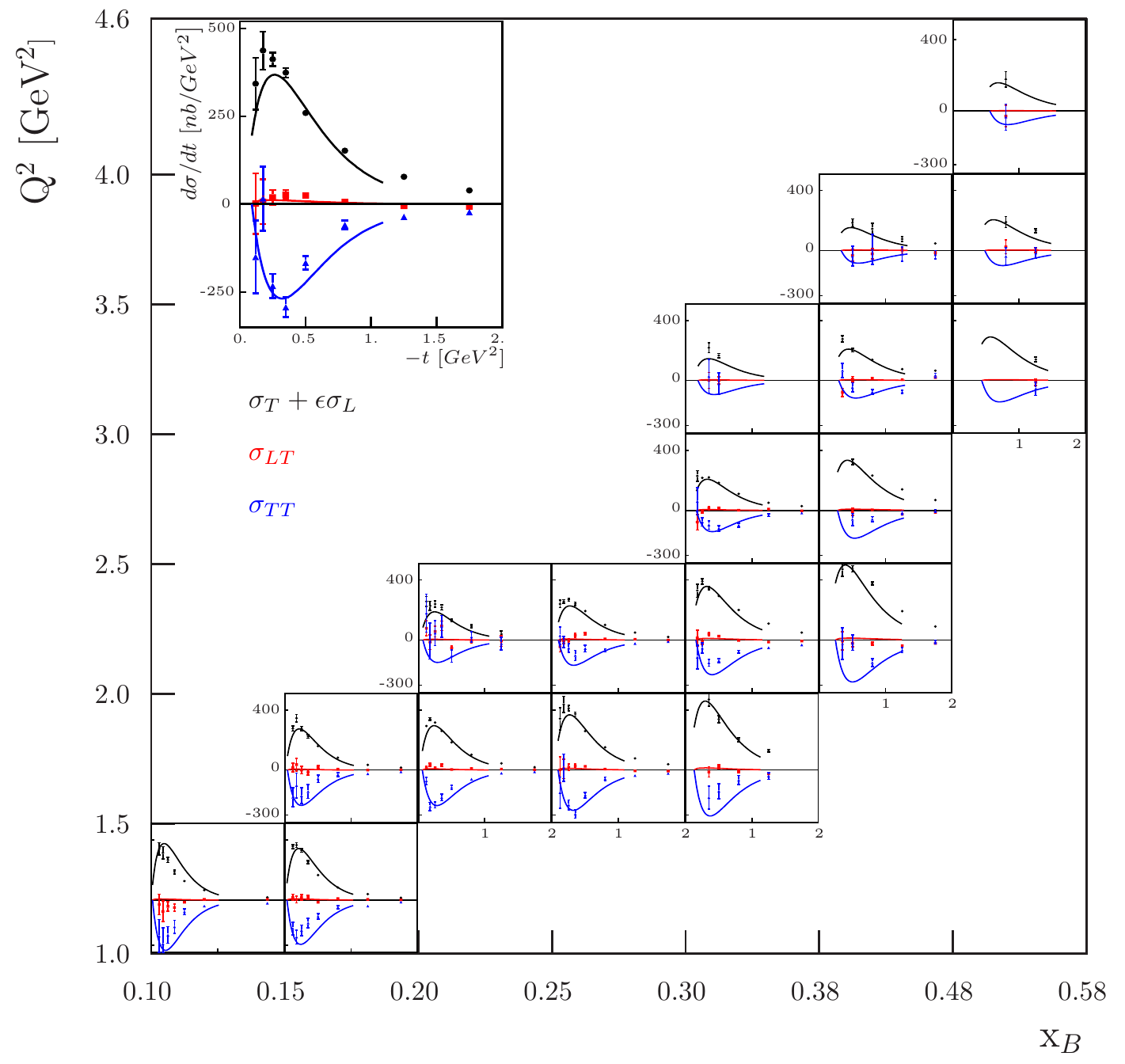}

\end{center}
\caption{\label{GK_full_kin} 
(Color online) The extracted structure functions vs. $t$ as in  Fig.~\ref{fig:structure_functions} for all kinematic bins. The data and curves are as follows:  black (positive)-$d\sigma_U/dt =d\sigma_T/dt +\epsilon d\sigma_L/dt$,\  blue (negative)-$d\sigma_{TT}/dt$,  and  red (small)-$d\sigma_{LT}/dt$.  All the structure functions are numerically given in  Appendix~\ref{strfun_table}. 
The error bars are statistical only. The point-by-point propagated systematic uncertainties are given in the table~in Appendix~\ref{strfun_table}. The curves are theoretical predictions for these structure functions obtained in the framework of the handbag model  by Ref.~\cite{GK-ps-11}. As before, black (positive)-$d\sigma_U/dt =d\sigma_T/dt +\epsilon d\sigma_L/dt$,\  blue (negative)-$d\sigma_{TT}/dt$,  and  red (small)-$d\sigma_{LT}/dt$.}
\end{figure*}

Fig.~\ref{GK_full_kin} shows the extracted structure functions vs. $t$ for all kinematic bins, 
but this time compared to the GPD calculations
of  Ref.~\cite{GK-ps-11}. While $\sigma_{LT}$ is very small in all kinematic bins, $\sigma_{TT}$ remains substantial, which is what one would expect for a transverse photon  dominated process.

\section{Conclusion}
Differential cross sections of exclusive neutral-pion electroproduction have been obtained in the few-GeV region at more than 1800 kinematic points in  bins of $Q^2, x_B$, $t$ and $\phi_\pi$. 
Virtual photon structure functions  $d\sigma_U/dt$, $d\sigma_{TT}/dt$ and $d\sigma_{LT}/dt$ have been obtained. It is found that $d\sigma_U/dt$ and $d\sigma_{TT}/dt$ are comparable in magnitude with each other, while $d\sigma_{LT}/dt$ is very much smaller than either. The $t$-dependent distributions of the structure functions have been compared with calculations based upon the Regge trajectory and handbag approaches. 
In each case, it is found that the cross sections are dominated by transverse photons.

In the Regge model~\cite{Laget}, in order to account for the magnitude of the cross section, it has been necessary to add vector meson rescattering amplitudes (Fig.~\ref{rescatter}) to the original pole terms and pseudoscalar  rescattering amplitudes (Fig.~\ref{rescatter-pi-1}).  

Within the handbag interpretation, there are two independent theoretical calculations~\cite{GK-ps-11, Goldstein:2010gu}. They confirm that the  measured unseparated cross sections are much larger than expected from leading-twist handbag calculations which are dominated by longitudinal photons.
The same conclusion can be made in an almost model independent way by noting that the structure functions $d\sigma_U/dt$ and $d\sigma_{TT}/dt$ are  comparable to each other while $d\sigma_{LT}$ is quite small in comparison.  
In the calculation of Ref.~\cite{Goldstein:2010gu} 
the dominant  GPD is $H_T$, which involves a nucleon helicity-flip, while that of   Ref.~\cite{GK-ps-11} has a larger contribution of  $\bar E_T$, which involves a nucleon non-helicity-flip. The data at $t$ near $t_{min}$ appear to favor the calculation of Ref.~\cite{GK-ps-11}.  In Eqs.~\ref{ST},~\ref{SLT} and ~\ref{STT} one  can make two observations. First, note that cross section contributions due to $\bar E_T$ vanish as $|t|\to |t|_{min}$.  There is no such constraint on terms involving $H_T$.  The observed  $d\sigma_U/dt$ does appear to turn over as  $|t| \to |t|_{min}$, which is expected when the contribution of $\bar E_T$ is relatively large, as in Ref.~\cite{GK-ps-11}. Second, the structure function $d\sigma_{TT}/dt$, which  depends on $\bar E_T$, is relatively large in the data. 

However, one must be very cautious not to over-interpret the results at this time. Detailed interpretations are model dependent and quite dynamic in that they are strongly influenced by new data as they become available. In particular, calculations are in progress to compare the theoretical models  with the 
beam-spin asymmetries obtained earlier with CLAS~\cite{demasi} and longitudinal target spin asymmetries, also obtained with CLAS, which are currently under analysis~\cite{kim}. 

In the near future new data on  $\eta$ production and ratios of $\eta$ to $\pi^0$ cross sections are expected to further constrain GPD models.
Extracting $d\sigma_L/dt$ and  $d\sigma_T/dt$  and performing new measurements with transversely  and longitudinally polarized targets would also be very useful, and are planned for the future Jefferson Lab at 12 GeV.

\begin{acknowledgements}

We thank  the staff of the Accelerator and Physics Divisions at Jefferson Lab for making the experiment possible. We also thank G. Goldstein, S. Goloskokov, P. Kroll, J. M. Laget and S. Liuti  for many informative discussions and clarifications of their work, and making available the results of their calculations. 
This work was supported in part by 
the U.S. Department of Energy and National Science Foundation, 
the French Centre National de la Recherche Scientifique and Commissariat  \`a l'Energie Atomique, the French-American Cultural Exchange (FACE),
the Italian Istituto Nazionale di Fisica Nucleare, 
the Chilean Comisi\'on Nacional de Investigaci\'on Cient\'ifica y Tecnol\'ogica (CONICYT),
the National Research Foundation of Korea, 
and the UK Science and Technology Facilities
Council (STFC).
The Jefferson Science Associates (JSA) operates the Thomas Jefferson National Accelerator Facility for 
the United States Department of Energy under contract DE-AC05-06OR23177. 
\end{acknowledgements}

\appendix
\section{Kinematics}
\label{section:kinematics}
The kinematic variables of the process
$$
e(k) + p(p) \rightarrow e^\prime(k^\prime) +  p^\prime(p^\prime) + \pi^0(v)
$$
are defined as follows. The four--momenta of the incident and outgoing electrons are denoted by $k$ and $k^\prime$ and
the four-momentum of the virtual photon $q$ is defined as $q=k-k^\prime$. In the laboratory system $\theta$ is the scattering angle between the incident and outgoing electrons, with energies $E$ and $E^\prime$, respectively.
 The photon virtuality, given by
\begin{equation}
Q^2=-q^2=-(k-k^\prime)^2 \approx 4 \ E \  E^\prime \sin^2\frac{\theta}{2}
\end{equation}
is positive. 
The four--momenta of the incident and outgoing protons are denoted by $p$ and $p^\prime$.
The energy of the virtual photon is  
\begin{equation}
\nu=\frac{p \cdot q}{m_p}=E\ - \ E ^\prime, 
\end{equation}
where $m_p$ is the proton mass.
The Bjorken scaling variable $x_B$ is defined as
\begin{equation}
x_B=\frac{Q^2}{2p\cdot q}=\frac{Q^2}{2m_p\nu}.
\end{equation}
The squared invariant mass of the photon--proton system is given by
\begin{equation}
W^2=(p+q)^2=m_p^2+2m_p\nu-Q^2.
\end{equation}
The momentum transfer $t$ to the proton is defined by the relation
\begin{equation}
t=(p-p^\prime)^2=(q-p_\pi)^2,
\end{equation}
where $p_\pi$ is the four--momentum of the $\pi^0$ meson.  The minimum momentum transfer for a given $Q^2$ and $W$ (or $x_B$)  is denoted by $t_{min}$. 

\begin{figure}
\vspace{0.20 in}
\includegraphics[width=\columnwidth]{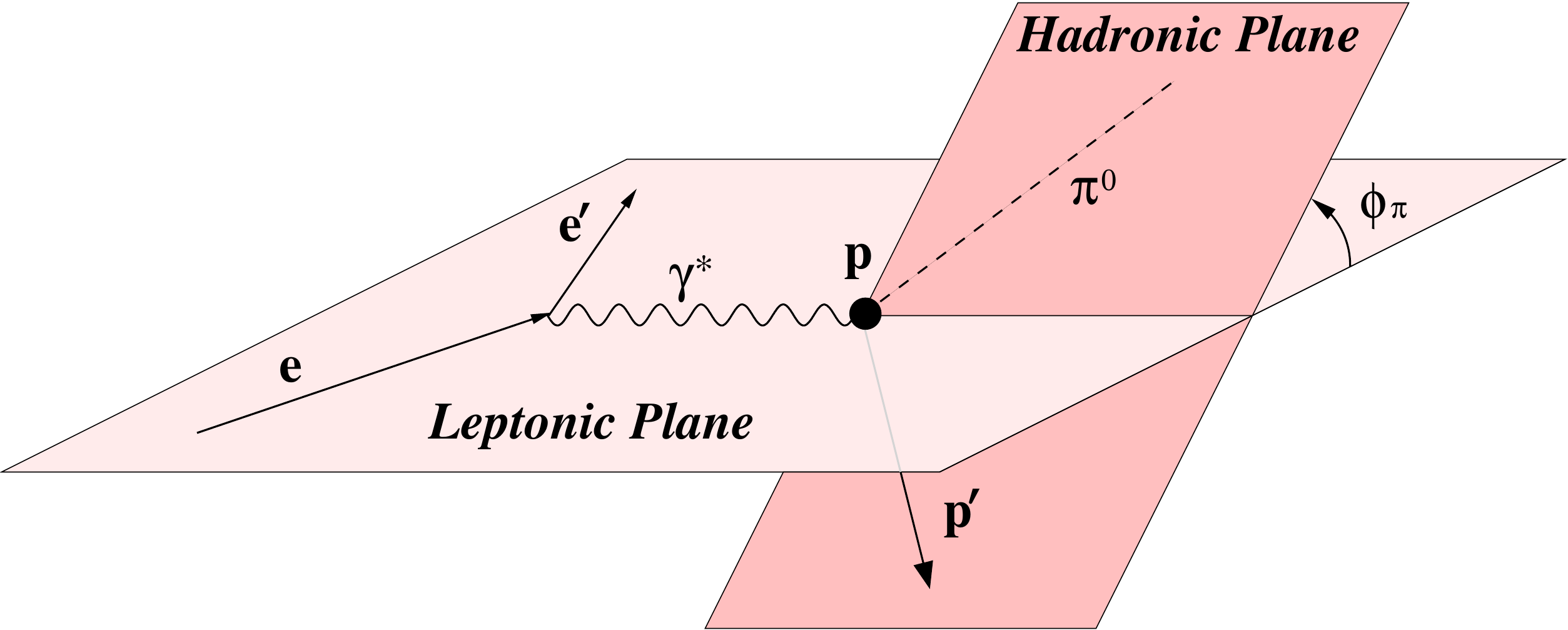}
\caption{\label{fig:phi_def} The kinematics  of $\pi^0$ electroproduction. $\phi_{\pi}$ is the angle between the lepton and hadron planes. The lepton plane is defined by the incident and the scattered electron. The hadron plane  is defined by the $\pi^0$ and the scattered proton.}
\end{figure}

The angle  $\phi_\pi$ between the leptonic and hadronic planes is defined according to the Trento convention \cite{trento} (see Fig.~\ref{fig:phi_def}).

\section{Helicity amplitudes and Generalized Parton Distributions}
\label{section:helicity_amp}

Under the assumption of single-photon exchange, the differential cross section of the reaction $ep\to e'p'\pi^0$ for an unpolarized electron beam and proton target
can be written as    \cite{G-K-09}
\begin{widetext}
\begin{equation}
 \frac {d^4\sigma}    {dQ^2 dx_B dt d\phi_\pi}  =  \Gamma(Q^2,x_B,E)  \frac{1}{2\pi}
\left[\left( \frac{d\sigma_T}{dt}+\epsilon \frac{d\sigma_L}{dt}\right)
+ \epsilon \cos 2 \phi_\pi  \frac{d\sigma_{TT}}{dt} + \sqrt{2\epsilon(1+\epsilon)} \cos \phi_\pi  \frac{d\sigma_{LT}}{dt} 
\right],
\label{eq:d4sigma}
\end{equation}
\end{widetext}
\noindent 
where $\Gamma(Q^2,x_B,E)$ is the flux of transverse virtual photons
 and $\sigma_T$, $\sigma_L$, $\sigma_{TT}$ and $\sigma_{LT}$ are the structure functions. 
They depend in general on the variables $Q^2$, $x_B$ and $t$.
The Hand convention~\cite{Hand} was adopted for the definition of the virtual photon flux factor 
$\Gamma(Q^2,x_B,E)$:

\begin{equation}
\Gamma (Q^2,x_B,E)= \frac{\alpha}{8\pi} \frac{Q^2}{m_p^2 E^2} 
\frac{1-x_B}{x_B^3} \frac{1}{1-\epsilon},
\label{eq:GammaV}
\end{equation}
\noindent and $\alpha$ is the standard electromagnetic coupling constant.
The variable $\epsilon$ represents the ratio of fluxes of longitudinally  and transversely polarized virtual photons and is given by
\begin{equation}
\epsilon=\frac {1-y-\frac{Q^2}{4E^2}}   {1-y+\frac{y^2}{2}+\frac{Q^2}{4E^2} }
\label{epsilon},
\end{equation}
with $y=p \cdot q/q \cdot k=\nu/E$.

The reduced cross section is defined as
\begin{widetext}
\begin{equation}
 \frac {d^2\sigma}    {dt d\phi_\pi}  =  \frac{1}{2\pi}
 \left[\left( \frac{d\sigma_T}{dt}+\epsilon \frac{d\sigma_L}{dt}\right)
+ \epsilon \cos 2 \phi_\pi  \frac{d\sigma_{TT}}{dt} + \sqrt{2\epsilon(1+\epsilon)} \cos \phi_\pi  \frac{d\sigma_{LT}}{dt} 
\right].
\label{eq:d2sigma}
\end{equation}
\end{widetext}

Six independent helicity amplitudes $M_{\mu^\prime\nu^\prime\mu\nu}$ describe the   $\pi^0$ electroproduction  process
$\gamma^*p\to\pi^0p^\prime$. With reference to Fig.~\ref{fig:handbag-pi0},  
$\mu$ and $\mu^\prime$ label the helicities of the virtual photon ($\mu$=0,+1,-1)  and $\pi^0$ ($\mu^\prime=0$).
The helicities of  protons before and after the interaction are labeled $\nu$ and $\nu^\prime$, respectively. We will denote  $``+"$ for the $\nu=1/2$ and $``-"$ for $\nu=-1/2$. The unmeasured helicities of the emitted and absorbed quarks are denoted $\lambda$ and $\lambda^\prime$ as in Fig.~\ref{fig:handbag-pi0}.
Four of these amplitudes describe the reaction initiated by transversely  polarized photons:  
$M_{0-++}$, $M_{0--+}$,  $M_{0+++}$, $M_{0+-+}$. The first two correspond to nucleon helicity flip and the latter two to nucleon helicity non-flip.  There are two amplitudes which describe 
the reaction due to  longitudinally polarized photons ($M_{0+0+}$, $M_{0-0+}$), with nucleon helicity non-flip and helicity flip, respectively.
It is convenient to introduce two new amplitudes with so-called natural $M^N_{0\nu^\prime\mu\nu}$ and unnatural 
$M^U_{0\nu^\prime\mu\nu}$ exchanges

\begin{align}
M^N_{0\nu^\prime\mu\nu}=\frac{1}{2}[M_{0\nu^\prime\mu\nu}+M_{0\nu^\prime-\mu\nu}],
\end{align}
\begin{align}
M^U_{0\nu^\prime\mu\nu}=\frac{1}{2}[M_{0\nu^\prime\mu\nu}-M_{0\nu^\prime-\mu\nu}].
\end{align}
The former does not change sign upon photon helicity  reversal, and the latter changes sign upon photon helicity reversal. 

The inverse equations are

\begin{align}
	M_{0\nu^\prime\mu\nu}= M^N_{0\nu^\prime\mu\nu}  + M^U_{0\nu^\prime\mu\nu},
\\	M_{0\nu^\prime -\mu\nu}= M^N_{0\nu^\prime\mu\nu} - M^U_{0\nu^\prime\mu\nu}.
\end{align}

For $t'\rightarrow 0$ a helicity amplitude vanishes (at least) as 
$M_{\mu '\nu '\mu\nu} \propto \sqrt{-t'}^{|\mu-\nu-\mu '+\nu '|}$ 
as a consequence of angular momentum conservation, where $t^\prime=t-t_{min}$. 
Thus, for transverse photons, for nucleon helicity flip ($\nu^\prime = -\nu$) the cross sections may remain finite at $t'\to 0$, while for nucleon helicity non-flip ($\nu^\prime = \nu$), the cross section should approach 0 as $t'\to 0$.
According to the findings in Refs.~\cite{G-K-09},\cite{G-K-11} and the HERMES measurement of the transverse-spin asymmetry $A_{UT}$, as well as the CLAS measurement of the $\pi^0$ cross section \cite{bedlinskiy}, it seems that  the following hierarchy of the amplitudes for transversely polarized photons holds
\begin{equation}
 \label{eq:hierarchy}
|M_{0--+}|, |M^U_{0+++}| \ll |M_{0-++}|, |M^N_{0+++}|.
\end{equation}
\noindent
The structure functions can be written in terms of the helicity amplitudes,  neglecting the  smallest amplitudes: in Eq.~\ref{eq:hierarchy} above.

The  longitudinal structure function $\sigma_L$ is connected to longitudinally polarized photons:

\begin{align}
\frac{d\sigma_L}{dt} = \frac{1}{k} \left[ \left|M_{0+0+}\right|^2 + \left|M_{0-0+}\right|^2 \right].
\label{B10}
\end{align}

The structure function $\sigma_T$ involves  transversely polarized photons:
\begin{widetext}
\begin{align}
\frac{d\sigma_T}{dt} &= \frac{1}{2k} \left[ \left|M_{0-++}\right|^2 + \left|M_{0--+}\right|^2 + \left|M_{0+++}\right|^2 + \left|M_{0+-+}\right|^2 \right] \notag
\\ & \simeq \frac{1}{2k} \left[ \left|M_{0-++}\right|^2 + 2\left|M^N_{0+++}\right|^2 \right].
\label{B11}
\end{align}

The structure function  $\sigma_{LT}$ involves the interference between the longitudinal  and transverse amplitudes
\begin{align}
\frac{d\sigma_{LT}}{dt} &= -\frac{1}{\sqrt{2}k} \text{Re} \left[ M^*_{0-0+} \left( M_{0-++} - M_{0--+} \right) + 2 M^*_{0+0+} M_{0+-+} \right] \notag\\ & \simeq -\frac{1}{\sqrt{2}k} \text{Re} \left( M^*_{0-++} M_{0-0+} \right).
\label{B12}
\end{align}
\end{widetext}

Likewise, the transverse-transverse interference cross section $\sigma_{TT}$ is
\begin{align}
\frac{d\sigma_{TT}}{dt} &= -\frac{1}{k} \text{Re} \left[ M^*_{0-++} M_{0--+} + M^*_{0+++} M_{0+-+} \right] \notag
\\ & \simeq -\frac{1}{k} \left| M^N_{0+++} \right|^2.
\label{B13}
\end{align}

The quantity $k$ is the phase space factor, which depends on $W^2, Q^2, m_p^2$ and $x_B$, and varies approximately as $Q^4$.
\begin{widetext}
\begin{align}
k		&= 16\pi \left( W^2 - m^2 \right) \sqrt{\Lambda\left(W^2, -Q^2, m^2\right)}
\\
	&= 16\pi Q^2 \left( \frac{1}{x_B} - 1 \right) \sqrt{\left( W^2-m^2 \right)^2 + Q^4 + 2W^2Q^2 +2Q^2m^2} \notag
	\\
	&= Q^4k' \notag
\end{align}
\end{widetext}

In the GPD-handbag approximation,  exclusive $\pi^0$ electroproduction can be 
decomposed into a hard part, describing the partonic subprocess and a soft part that contains the GPDs.
This factorization occurs at large photon virtualities $Q^2$ and small momentum transfer to the nucleon, $-t$.
Following the notation of Ref.~\cite{G-K-11}, the connection between the helicity amplitudes and GPDs is 

\begin{align}
M_{0+0+} = \sqrt{1-\xi^2} \frac{e_0}{Q} \left[ \langle \tilde{H} \rangle - \frac{\xi^2}{1-\xi^2} \langle \tilde{E} \rangle \right]
\end{align}

\begin{align}
M_{0-0+} = - \frac{e_0}{Q} \frac{\sqrt{-t'}}{2m} \xi \langle \tilde{E} \rangle
\end{align}

\begin{align}
M_{0-++} = e_0\frac{\mu_\pi}{Q^2}\sqrt{1-\xi^2}\langle H_T\rangle
\end{align}

\begin{align}
M^N_{0+++} = -e_0\frac{\mu_\pi}{Q^2}\frac{\sqrt{-t'}}{4m}\langle\bar{E}_T\rangle.
\end{align}

The variable $\xi\simeq x_B/(2-x_B)$, 
$\mu_\pi=m^2_\pi/(m_u+m_d)$, where $m_u$ and $m_d$ are current quark masses \cite{G-K-09} and $\bar E_T \equiv 2 \widetilde{H}_T+E_T$.
$\langle F \rangle$ denotes a convolution of GPD $F$ with the hard-scattering kernel, $\mathcal{H}_{\mu^\prime\lambda^\prime\mu\lambda}$, 
where $\lambda$ and $\lambda^\prime$ are the (unmeasured) helicities of the incoming and outgoing quarks, $\mu$ is the virtual-photon helicity and $\mu=0$ is the neutral-pion helicity, and is given by

\begin{equation}
\langle{F}\rangle\equiv\sum_\lambda \int^1_{-1} dx \mathcal{H}_{\mu^\prime\lambda^\prime\mu\lambda}F.
\end{equation}

\noindent
 $\langle H_T\rangle$ arises  primarily from nucleon helicity flip processes, while $\langle\bar{E}_T\rangle$ describes nucleon helicity non-flip processes.

Note that a factor $1/Q$ in the longitudinal amplitudes  and a factor $\mu_\pi/Q^2$ in the transverse amplitudes has been factored  in order to explicitly show the leading $Q^2$ dependence. The convolutions $ \langle F \rangle$  are still $Q^2$ dependent due to evolution, the  running of $\alpha_s$ and other effects. In the transverse convolutions there is also a summation over the parton helicities. 

Combining the above finally yields  the GPD dependence of the structure functions:
\onecolumngrid
\begin{align}
\label{SL}
\frac{d\sigma_{L} }{dt}= \frac{4\pi\alpha}{k^\prime}\frac{1}{Q^6}\left\{ \left( 1-\xi^2 \right) \left|\GPDtH\right|^2 -2\xi^2\text{Re}\left[ \GPDtH^* \GPDtE \right] - \frac{t^\prime}{4m^2} \xi^2 \left| \GPDtE \right|^2 \right\},
\end{align}

\begin{align}
\label{ST}
\frac{d\sigma_{T}}{dt} &= \frac{4\pi\alpha}{2k^\prime}\frac{\mu_\pi^2}{Q^8} \left[ \left(1-\xi^2\right) \left|\GPDHT\right|^2 - \frac{t'}{8m^2} \left|\GPDETbar\right|^2\right],
\end{align}

\begin{align}
\label{SLT}
\frac{\sigma_{LT}}{dt} &= \frac{4\pi\alpha}{\sqrt{2}k^\prime} \frac{\mu_{\pi}}{Q^7} \xi \sqrt{1-\xi^2} \frac{\sqrt{-t'}}{2m} \text{ Re} \left[ \langle H_T\rangle^* \langle\tilde{E}\rangle \right],
\end{align}

\begin{align}
\label{STT}
\frac{\sigma_{TT}}{dt} = \frac{4\pi\alpha}{k^\prime}\frac{\mu_\pi^2}{Q^8}\frac{t'}{16m^2}\left|\GPDETbar\right|^2.
\end{align}


\section{Structure functions}
The structure functions are presented in this table. The first error is statistical and the second  is the systematic uncertainty.
\label{strfun_table}
\squeezetable
\setlength\LTleft{0pt}
\setlength\LTright{0pt}
\begin{longtable}{ccc@{\extracolsep{1cm}}d@{\extracolsep{0pt}}cdcd@{\extracolsep{1cm}}d@{\extracolsep{0pt}}cdcd@{\extracolsep{1cm}}d@{\extracolsep{0pt}}cdcd}
\hline\hline
\multicolumn{1}{c}{\textbf{$Q^2$,}} &
\multicolumn{1}{c}{\textbf{$x_B$}} &
\multicolumn{1}{c}{\textbf{$-t$,}} &
\multicolumn{5}{c}{\textbf{$\frac{d\sigma_T}{dt}+\epsilon\frac{d\sigma_L}{dt}$,}} &
\multicolumn{5}{c}{\textbf{$\frac{d\sigma_{LT}}{dt}$,}} &
\multicolumn{5}{c}{\textbf{$\frac{d\sigma_{TT}}{dt}$,}} \\
\multicolumn{1}{c}{\textbf{$GeV^2$}} &
\multicolumn{1}{c}{\textbf{}} &
\multicolumn{1}{c}{\textbf{$GeV^2$}} &
\multicolumn{5}{c}{\textbf{$nb/GeV^2$}} &
\multicolumn{5}{c}{\textbf{$nb/GeV^2$}} &
\multicolumn{5}{c}{\textbf{$nb/GeV^2$}}
\\\hline
\endhead 

\hline
\endfoot

\hline \hline
\endlastfoot
1.14 & 0.131 & 0.12 	&	 341	 &  $\pm$	& 40 &	 $\pm$	& 59 &	 -30 &	$\pm$& 	 68 &	$\pm$& 	 114 & -240 & $\pm$ & 111 & 	$\pm$ & 156\\
1.15 & 0.132 & 0.17 	&	 314	 &  $\pm$	& 40 &	 $\pm$	& 75 &	 -76 &	$\pm$& 	 69 &	$\pm$& 	 126 & -292 & $\pm$ & 108 & 	$\pm$ & 215\\
1.15 & 0.132 & 0.25 	&	 267	 &  $\pm$	& 19 &	 $\pm$	& 15 &	 -42 &	$\pm$& 	 32 &	$\pm$& 	 37 & -233 & $\pm$ & 55 & 	$\pm$ & 21\\
1.15 & 0.132 & 0.35 	&	 188	 &  $\pm$	& 13 &	 $\pm$	& 33 &	 -50 &	$\pm$& 	 23 &	$\pm$& 	 43 & -179 & $\pm$ & 43 & 	$\pm$ & 66\\
1.15 & 0.132 & 0.49 	&	 126.3	 &  $\pm$	& 4.7 &	 $\pm$	& 10 &	 -15.0 &	$\pm$& 	 8.0 &	$\pm$& 	 5.5 & -78 & $\pm$ & 19 & 	$\pm$ & 8.1\\
1.15 & 0.132 & 0.77 	&	 66.0	 &  $\pm$	& 2.0 &	 $\pm$	& 7.9 &	 3.8 &	$\pm$& 	 3.1 &	$\pm$& 	 6.4 & -39.8 & $\pm$ & 7.8 & 	$\pm$ & 16\\
1.16 & 0.133 & 1.71 	&	 17.8	 &  $\pm$	& 2.0 &	 $\pm$	& 1.6 &	 4.3 &	$\pm$& 	 1.2 &	$\pm$& 	 2.0 & -21.2 & $\pm$ & 6.6 & 	$\pm$ & 7.7\\
1.38 & 0.169 & 0.12 	&	 357	 &  $\pm$	& 13 &	 $\pm$	& 35 &	 19 &	$\pm$& 	 19 &	$\pm$& 	 30 & -191 & $\pm$ & 42 & 	$\pm$ & 47\\
1.38 & 0.169 & 0.17 	&	 366	 &  $\pm$	& 15 &	 $\pm$	& 24 &	 2 &	$\pm$& 	 22 &	$\pm$& 	 21 & -247 & $\pm$ & 46 & 	$\pm$ & 53\\
1.38 & 0.169 & 0.25 	&	 331	 &  $\pm$	& 12 &	 $\pm$	& 16 &	 19 &	$\pm$& 	 18 &	$\pm$& 	 17 & -202 & $\pm$ & 36 & 	$\pm$ & 49\\
1.38 & 0.169 & 0.35 	&	 254	 &  $\pm$	& 10 &	 $\pm$	& 13 &	 17 &	$\pm$& 	 15 &	$\pm$& 	 24 & -153 & $\pm$ & 32 & 	$\pm$ & 25\\
1.38 & 0.169 & 0.49 	&	 166.2	 &  $\pm$	& 5.1 &	 $\pm$	& 12 &	 -15.4 &	$\pm$& 	 7.1 &	$\pm$& 	 12 & -109 & $\pm$ & 18 & 	$\pm$ & 18\\
1.38 & 0.169 & 0.77 	&	 83.4	 &  $\pm$	& 3.3 &	 $\pm$	& 4.1 &	 9.7 &	$\pm$& 	 4.4 &	$\pm$& 	 10 & -48.5 & $\pm$ & 9.6 & 	$\pm$ & 5.4\\
1.38 & 0.169 & 1.21 	&	 39.6	 &  $\pm$	& 1.7 &	 $\pm$	& 3.8 &	 4.0 &	$\pm$& 	 1.7 &	$\pm$& 	 1.9 & -40.8 & $\pm$ & 4.5 & 	$\pm$ & 3.0\\
1.38 & 0.170 & 1.71 	&	 15.3	 &  $\pm$	& 1.4 &	 $\pm$	& 1.5 &	 0.81 &	$\pm$& 	 0.80 &	$\pm$& 	 1.6 & -13.6 & $\pm$ & 4.0 & 	$\pm$ & 5.1\\
1.61 & 0.186 & 0.12 	&	 276	 &  $\pm$	& 17 &	 $\pm$	& 46 &	 17 &	$\pm$& 	 29 &	$\pm$& 	 58 & -180 & $\pm$ & 64 & 	$\pm$ & 71\\
1.61 & 0.186 & 0.18 	&	 345	 &  $\pm$	& 25 &	 $\pm$	& 57 &	 36 &	$\pm$& 	 42 &	$\pm$& 	 102 & -103 & $\pm$ & 82 & 	$\pm$ & 87\\
1.61 & 0.187 & 0.25 	&	 276	 &  $\pm$	& 15 &	 $\pm$	& 7.0 &	 0 &	$\pm$& 	 26 &	$\pm$& 	 21 & -171 & $\pm$ & 52 & 	$\pm$ & 41\\
1.61 & 0.187 & 0.35 	&	 223	 &  $\pm$	& 12 &	 $\pm$	& 11 &	 -14 &	$\pm$& 	 20 &	$\pm$& 	 11 & -143 & $\pm$ & 46 & 	$\pm$ & 46\\
1.61 & 0.187 & 0.49 	&	 159.8	 &  $\pm$	& 6.3 &	 $\pm$	& 11 &	 20 &	$\pm$& 	 10 &	$\pm$& 	 11 & -58 & $\pm$ & 25 & 	$\pm$ & 19\\
1.61 & 0.187 & 0.78 	&	 82.4	 &  $\pm$	& 3.2 &	 $\pm$	& 7.1 &	 5.6 &	$\pm$& 	 4.8 &	$\pm$& 	 19 & -30 & $\pm$ & 12 & 	$\pm$ & 27\\
1.61 & 0.187 & 1.21 	&	 34.5	 &  $\pm$	& 2.3 &	 $\pm$	& 3.0 &	 0.1 &	$\pm$& 	 3.3 &	$\pm$& 	 1.7 & -24.9 & $\pm$ & 6.4 & 	$\pm$ & 6.6\\
1.61 & 0.187 & 1.71 	&	 16.0	 &  $\pm$	& 1.9 &	 $\pm$	& 1.6 &	 2.3 &	$\pm$& 	 1.8 &	$\pm$& 	 2.2 & -12.2 & $\pm$ & 6.2 & 	$\pm$ & 4.6\\
1.74 & 0.223 & 0.25 	&	 316.7	 &  $\pm$	& 6.7 &	 $\pm$	& 9.2 &	 14.9 &	$\pm$& 	 8.5 &	$\pm$& 	 19 & -232 & $\pm$ & 20 & 	$\pm$ & 44\\
1.75 & 0.223 & 0.12 	&	 293.3	 &  $\pm$	& 7.8 &	 $\pm$	& 24 &	 16.2 &	$\pm$& 	 9.8 &	$\pm$& 	 12 & -72 & $\pm$ & 23 & 	$\pm$ & 13\\
1.75 & 0.223 & 0.17 	&	 339.3	 &  $\pm$	& 8.9 &	 $\pm$	& 26 &	 35 &	$\pm$& 	 11 &	$\pm$& 	 8.3 & -243 & $\pm$ & 28 & 	$\pm$ & 26\\
1.75 & 0.224 & 0.35 	&	 260.5	 &  $\pm$	& 7.0 &	 $\pm$	& 13 &	 32.1 &	$\pm$& 	 9.2 &	$\pm$& 	 5.0 & -183 & $\pm$ & 22 & 	$\pm$ & 20\\
1.75 & 0.224 & 0.49 	&	 184.4	 &  $\pm$	& 5.0 &	 $\pm$	& 8.6 &	 3.6 &	$\pm$& 	 6.3 &	$\pm$& 	 3.7 & -116 & $\pm$ & 15 & 	$\pm$ & 20\\
1.75 & 0.224 & 0.78 	&	 102.2	 &  $\pm$	& 2.4 &	 $\pm$	& 5.4 &	 9.2 &	$\pm$& 	 3.1 &	$\pm$& 	 5.0 & -61.0 & $\pm$ & 7.3 & 	$\pm$ & 12\\
1.75 & 0.224 & 1.22 	&	 44.5	 &  $\pm$	& 1.4 &	 $\pm$	& 3.0 &	 6.3 &	$\pm$& 	 1.3 &	$\pm$& 	 2.2 & -21.2 & $\pm$ & 4.1 & 	$\pm$ & 6.0\\
1.75 & 0.224 & 1.72 	&	 19.00	 &  $\pm$	& 1.00 &	 $\pm$	& 4.4 &	 2.24 &	$\pm$& 	 0.85 &	$\pm$& 	 3.2 & -12.3 & $\pm$ & 3.0 & 	$\pm$ & 5.4\\
1.87 & 0.270 & 0.12 	&	 342	 &  $\pm$	& 74 &	 $\pm$	& 108 &	 1 &	$\pm$& 	 86 &	$\pm$& 	 72 & -150 & $\pm$ & 103 & 	$\pm$ & 101\\
1.87 & 0.271 & 0.18 	&	 437	 &  $\pm$	& 54 &	 $\pm$	& 90 &	 7 &	$\pm$& 	 64 &	$\pm$& 	 74 & 16 & $\pm$ & 91 & 	$\pm$ & 167\\
1.87 & 0.271 & 0.25 	&	 412	 &  $\pm$	& 19 &	 $\pm$	& 32 &	 20 &	$\pm$& 	 21 &	$\pm$& 	 20 & -233 & $\pm$ & 34 & 	$\pm$ & 39\\
1.87 & 0.271 & 0.35 	&	 374	 &  $\pm$	& 14 &	 $\pm$	& 26 &	 27 &	$\pm$& 	 13 &	$\pm$& 	 20 & -293 & $\pm$ & 28 & 	$\pm$ & 41\\
1.87 & 0.271 & 0.49 	&	 259.5	 &  $\pm$	& 7.3 &	 $\pm$	& 13 &	 25.1 &	$\pm$& 	 7.2 &	$\pm$& 	 6.1 & -167 & $\pm$ & 19 & 	$\pm$ & 14\\
1.87 & 0.271 & 0.78 	&	 151.8	 &  $\pm$	& 4.1 &	 $\pm$	& 7.8 &	 6.4 &	$\pm$& 	 4.2 &	$\pm$& 	 5.7 & -59 & $\pm$ & 12 & 	$\pm$ & 4.6\\
1.87 & 0.271 & 1.22 	&	 77.7	 &  $\pm$	& 3.0 &	 $\pm$	& 5.5 &	 -5.7 &	$\pm$& 	 2.3 &	$\pm$& 	 2.8 & -36.4 & $\pm$ & 7.4 & 	$\pm$ & 5.6\\
1.87 & 0.272 & 1.72 	&	 39.2	 &  $\pm$	& 2.1 &	 $\pm$	& 3.5 &	 -7.0 &	$\pm$& 	 1.9 &	$\pm$& 	 1.9 & -22.9 & $\pm$ & 4.6 & 	$\pm$ & 3.8\\
1.95 & 0.313 & 0.35 	&	 470	 &  $\pm$	& 44 &	 $\pm$	& 82 &	 -13 &	$\pm$& 	 34 &	$\pm$& 	 18 & -183 & $\pm$ & 77 & 	$\pm$ & 58\\
1.95 & 0.313 & 0.49 	&	 339	 &  $\pm$	& 23 &	 $\pm$	& 21 &	 21 &	$\pm$& 	 15 &	$\pm$& 	 34 & -140 & $\pm$ & 50 & 	$\pm$ & 43\\
1.95 & 0.313 & 0.78 	&	 202	 &  $\pm$	& 12 &	 $\pm$	& 13 &	 -11.1 &	$\pm$& 	 9.4 &	$\pm$& 	 5.8 & -67 & $\pm$ & 31 & 	$\pm$ & 23\\
1.96 & 0.313 & 1.22 	&	 129.4	 &  $\pm$	& 9.6 &	 $\pm$	& 17 &	 -24.8 &	$\pm$& 	 8.3 &	$\pm$& 	 6.7 & -39 & $\pm$ & 22 & 	$\pm$ & 21\\
2.10 & 0.238 & 0.12 	&	 258	 &  $\pm$	& 33 &	 $\pm$	& 81 &	 79 &	$\pm$& 	 51 &	$\pm$& 	 109 & 179 & $\pm$ & 126 & 	$\pm$ & 218\\
2.10 & 0.238 & 0.35 	&	 219	 &  $\pm$	& 18 &	 $\pm$	& 8.1 &	 95 &	$\pm$& 	 31 &	$\pm$& 	 10 & 91 & $\pm$ & 72 & 	$\pm$ & 46\\
2.10 & 0.238 & 0.49 	&	 132.5	 &  $\pm$	& 8.9 &	 $\pm$	& 13 &	 -53 &	$\pm$& 	 15 &	$\pm$& 	 9.0 & -105 & $\pm$ & 41 & 	$\pm$ & 28\\
2.10 & 0.238 & 0.78 	&	 92.6	 &  $\pm$	& 8.9 &	 $\pm$	& 9.2 &	 -8 &	$\pm$& 	 13 &	$\pm$& 	 12 & 21 & $\pm$ & 35 & 	$\pm$ & 32\\
2.10 & 0.238 & 1.21 	&	 40	 &  $\pm$	& 21 &	 $\pm$	& 16 &	 -6 &	$\pm$& 	 35 &	$\pm$& 	 31 & -23 & $\pm$ & 43 & 	$\pm$ & 27\\
2.10 & 0.239 & 0.17 	&	 228	 &  $\pm$	& 29 &	 $\pm$	& 148 &	 -13 &	$\pm$& 	 49 &	$\pm$& 	 265 & -7 & $\pm$ & 119 & 	$\pm$ & 268\\
2.10 & 0.239 & 0.25 	&	 240	 &  $\pm$	& 20 &	 $\pm$	& 24 &	 57 &	$\pm$& 	 36 &	$\pm$& 	 30 & 47 & $\pm$ & 83 & 	$\pm$ & 106\\
2.21 & 0.275 & 0.12 	&	 241	 &  $\pm$	& 25 &	 $\pm$	& 11 &	 -44 &	$\pm$& 	 36 &	$\pm$& 	 9.0 & 29 & $\pm$ & 58 & 	$\pm$ & 17\\
2.21 & 0.276 & 0.17 	&	 257	 &  $\pm$	& 12 &	 $\pm$	& 18 &	 -6 &	$\pm$& 	 17 &	$\pm$& 	 13 & -13 & $\pm$ & 38 & 	$\pm$ & 41\\
2.21 & 0.276 & 0.25 	&	 268.8	 &  $\pm$	& 9.8 &	 $\pm$	& 19 &	 -6 &	$\pm$& 	 13 &	$\pm$& 	 20 & -54 & $\pm$ & 29 & 	$\pm$ & 30\\
2.21 & 0.276 & 0.35 	&	 242	 &  $\pm$	& 11 &	 $\pm$	& 11 &	 32 &	$\pm$& 	 14 &	$\pm$& 	 12 & -102 & $\pm$ & 34 & 	$\pm$ & 22\\
2.21 & 0.276 & 0.49 	&	 193.5	 &  $\pm$	& 7.1 &	 $\pm$	& 17 &	 41.1 &	$\pm$& 	 9.4 &	$\pm$& 	 20 & -56 & $\pm$ & 22 & 	$\pm$ & 47\\
2.21 & 0.276 & 0.78 	&	 101.4	 &  $\pm$	& 3.0 &	 $\pm$	& 6.6 &	 7.3 &	$\pm$& 	 4.3 &	$\pm$& 	 7.0 & -69 & $\pm$ & 10 & 	$\pm$ & 10\\
2.21 & 0.277 & 1.22 	&	 50.0	 &  $\pm$	& 2.0 &	 $\pm$	& 3.3 &	 5.8 &	$\pm$& 	 2.3 &	$\pm$& 	 3.9 & -22.5 & $\pm$ & 6.9 & 	$\pm$ & 2.4\\
2.21 & 0.277 & 1.72 	&	 20.8	 &  $\pm$	& 1.5 &	 $\pm$	& 3.1 &	 -0.1 &	$\pm$& 	 1.8 &	$\pm$& 	 2.3 & -10.1 & $\pm$ & 4.8 & 	$\pm$ & 5.3\\
2.24 & 0.332 & 0.18 	&	 330	 &  $\pm$	& 44 &	 $\pm$	& 31 &	 14 &	$\pm$& 	 53 &	$\pm$& 	 37 & -114 & $\pm$ & 80 & 	$\pm$ & 118\\
2.24 & 0.337 & 0.25 	&	 392	 &  $\pm$	& 19 &	 $\pm$	& 44 &	 -8 &	$\pm$& 	 20 &	$\pm$& 	 34 & -53 & $\pm$ & 34 & 	$\pm$ & 27\\
2.24 & 0.338 & 0.49 	&	 293.7	 &  $\pm$	& 6.5 &	 $\pm$	& 15 &	 26.4 &	$\pm$& 	 5.5 &	$\pm$& 	 13 & -137 & $\pm$ & 14 & 	$\pm$ & 12\\
2.25 & 0.337 & 0.35 	&	 346	 &  $\pm$	& 12 &	 $\pm$	& 14 &	 40 &	$\pm$& 	 11 &	$\pm$& 	 12 & -152 & $\pm$ & 24 & 	$\pm$ & 15\\
2.25 & 0.338 & 0.78 	&	 200.8	 &  $\pm$	& 3.8 &	 $\pm$	& 13 &	 -2.1 &	$\pm$& 	 3.3 &	$\pm$& 	 5.0 & -78.6 & $\pm$ & 9.7 & 	$\pm$ & 10\\
2.25 & 0.339 & 1.22 	&	 110.2	 &  $\pm$	& 2.6 &	 $\pm$	& 5.4 &	 -13.3 &	$\pm$& 	 2.3 &	$\pm$& 	 4.2 & -50.4 & $\pm$ & 6.5 & 	$\pm$ & 6.1\\
2.25 & 0.339 & 1.73 	&	 49.9	 &  $\pm$	& 1.7 &	 $\pm$	& 4.6 &	 -6.5 &	$\pm$& 	 1.8 &	$\pm$& 	 5.7 & -32.3 & $\pm$ & 3.7 & 	$\pm$ & 5.8\\
2.34 & 0.403 & 0.35 	&	 472	 &  $\pm$	& 48 &	 $\pm$	& 53 &	 -6 &	$\pm$& 	 60 &	$\pm$& 	 79 & -24 & $\pm$ & 105 & 	$\pm$ & 210\\
2.34 & 0.403 & 0.49 	&	 475	 &  $\pm$	& 20 &	 $\pm$	& 39 &	 -22 &	$\pm$& 	 23 &	$\pm$& 	 27 & -17 & $\pm$ & 51 & 	$\pm$ & 53\\
2.34 & 0.404 & 0.78 	&	 377	 &  $\pm$	& 11 &	 $\pm$	& 17 &	 -22 &	$\pm$& 	 10 &	$\pm$& 	 5.8 & -150 & $\pm$ & 26 & 	$\pm$ & 19\\
2.34 & 0.404 & 1.22 	&	 192.8	 &  $\pm$	& 7.4 &	 $\pm$	& 13 &	 -37.3 &	$\pm$& 	 7.9 &	$\pm$& 	 4.4 & -67 & $\pm$ & 16 & 	$\pm$ & 43\\
2.35 & 0.404 & 1.73 	&	 90.5	 &  $\pm$	& 6.6 &	 $\pm$	& 3.1 &	 -22.4 &	$\pm$& 	 7.4 &	$\pm$& 	 5.7 & -13 & $\pm$ & 12 & 	$\pm$ & 8.4\\
2.71 & 0.336 & 0.18 	&	 230	 &  $\pm$	& 35 &	 $\pm$	& 29 &	 -78 &	$\pm$& 	 52 &	$\pm$& 	 84 & 60 & $\pm$ & 90 & 	$\pm$ & 188\\
2.71 & 0.343 & 0.25 	&	 217.3	 &  $\pm$	& 8.1 &	 $\pm$	& 10 &	 -6 &	$\pm$& 	 10 &	$\pm$& 	 4.3 & -76 & $\pm$ & 27 & 	$\pm$ & 22\\
2.71 & 0.343 & 0.35 	&	 220.5	 &  $\pm$	& 8.1 &	 $\pm$	& 8.0 &	 15.5 &	$\pm$& 	 9.8 &	$\pm$& 	 7.6 & -97 & $\pm$ & 27 & 	$\pm$ & 28\\
2.71 & 0.343 & 0.49 	&	 183.8	 &  $\pm$	& 6.0 &	 $\pm$	& 9.4 &	 17.0 &	$\pm$& 	 7.4 &	$\pm$& 	 12 & -120 & $\pm$ & 19 & 	$\pm$ & 31\\
2.71 & 0.343 & 1.22 	&	 51.3	 &  $\pm$	& 2.4 &	 $\pm$	& 4.5 &	 9.0 &	$\pm$& 	 2.7 &	$\pm$& 	 5.0 & -31.5 & $\pm$ & 9.7 & 	$\pm$ & 16\\
2.72 & 0.344 & 0.78 	&	 110.4	 &  $\pm$	& 3.6 &	 $\pm$	& 5.8 &	 1.8 &	$\pm$& 	 4.7 &	$\pm$& 	 5.8 & -99 & $\pm$ & 14 & 	$\pm$ & 20\\
2.72 & 0.344 & 1.73 	&	 28.7	 &  $\pm$	& 1.9 &	 $\pm$	& 3.5 &	 -2.9 &	$\pm$& 	 2.2 &	$\pm$& 	 2.0 & -17.2 & $\pm$ & 5.6 & 	$\pm$ & 9.2\\
2.75 & 0.423 & 0.50 	&	 323	 &  $\pm$	& 19 &	 $\pm$	& 21 &	 -8 &	$\pm$& 	 23 &	$\pm$& 	 16 & -60 & $\pm$ & 40 & 	$\pm$ & 16\\
2.75 & 0.423 & 0.78 	&	 232.4	 &  $\pm$	& 6.9 &	 $\pm$	& 17 &	 4.3 &	$\pm$& 	 6.4 &	$\pm$& 	 16 & -58 & $\pm$ & 17 & 	$\pm$ & 24\\
2.75 & 0.424 & 1.23 	&	 140.7	 &  $\pm$	& 4.9 &	 $\pm$	& 9.0 &	 -25.8 &	$\pm$& 	 5.6 &	$\pm$& 	 5.8 & -16 & $\pm$ & 13 & 	$\pm$ & 12\\
2.75 & 0.424 & 1.73 	&	 69.3	 &  $\pm$	& 4.6 &	 $\pm$	& 2.9 &	 -12.8 &	$\pm$& 	 5.3 &	$\pm$& 	 3.7 & -2.7 & $\pm$ & 9.6 & 	$\pm$ & 12\\
3.12 & 0.362 & 0.35 	&	 219	 &  $\pm$	& 33 &	 $\pm$	& 139 &	 1 &	$\pm$& 	 53 &	$\pm$& 	 213 & 27 & $\pm$ & 114 & 	$\pm$ & 398\\
3.12 & 0.362 & 0.50 	&	 167	 &  $\pm$	& 14 &	 $\pm$	& 20 &	 1 &	$\pm$& 	 23 &	$\pm$& 	 59 & -21 & $\pm$ & 71 & 	$\pm$ & 56\\
3.22 & 0.431 & 0.78 	&	 138.4	 &  $\pm$	& 6.2 &	 $\pm$	& 6.5 &	 15.0 &	$\pm$& 	 7.9 &	$\pm$& 	 5.5 & -77 & $\pm$ & 17 & 	$\pm$ & 16\\
3.23 & 0.428 & 0.35 	&	 277	 &  $\pm$	& 22 &	 $\pm$	& 15 &	 -80 &	$\pm$& 	 29 &	$\pm$& 	 16 & 67 & $\pm$ & 48 & 	$\pm$ & 20\\
3.23 & 0.430 & 0.50 	&	 201	 &  $\pm$	& 12 &	 $\pm$	& 17 &	 10 &	$\pm$& 	 16 &	$\pm$& 	 17 & -46 & $\pm$ & 30 & 	$\pm$ & 31\\
3.23 & 0.432 & 1.23 	&	 75.5	 &  $\pm$	& 3.8 &	 $\pm$	& 9.2 &	 5.6 &	$\pm$& 	 4.3 &	$\pm$& 	 12 & -77 & $\pm$ & 11 & 	$\pm$ & 32\\
3.23 & 0.432 & 1.73 	&	 65.4	 &  $\pm$	& 5.0 &	 $\pm$	& 6.7 &	 18.8 &	$\pm$& 	 5.7 &	$\pm$& 	 6.2 & 35 & $\pm$ & 14 & 	$\pm$ & 15\\
3.29 & 0.496 & 1.23 	&	 140	 &  $\pm$	& 17 &	 $\pm$	& 18 &	 -12 &	$\pm$& 	 23 &	$\pm$& 	 9.7 & -54 & $\pm$ & 45 & 	$\pm$ & 12\\
3.67 & 0.451 & 0.78 	&	 145	 &  $\pm$	& 36 &	 $\pm$	& 23 &	 -22 &	$\pm$& 	 35 &	$\pm$& 	 28 & 8 & $\pm$ & 101 & 	$\pm$ & 56\\
3.67 & 0.451 & 1.23 	&	 77	 &  $\pm$	& 15 &	 $\pm$	& 1.8 &	 2 &	$\pm$& 	 17 &	$\pm$& 	 2.9 & -24 & $\pm$ & 48 & 	$\pm$ & 8.8\\
3.68 & 0.451 & 0.49 	&	 185	 &  $\pm$	& 26 &	 $\pm$	& 18 &	 -32 &	$\pm$& 	 39 &	$\pm$& 	 29 & -38 & $\pm$ & 66 & 	$\pm$ & 57\\
3.68 & 0.451 & 1.73 	&	 47.0	 &  $\pm$	& 6.9 &	 $\pm$	& 3.9 &	 -14.7 &	$\pm$& 	 9.4 &	$\pm$& 	 7.3 & -27 & $\pm$ & 27 & 	$\pm$ & 7.9\\
3.76 & 0.513 & 0.78 	&	 190	 &  $\pm$	& 37 &	 $\pm$	& 40 &	 24 &	$\pm$& 	 46 &	$\pm$& 	 37 & -39 & $\pm$ & 56 & 	$\pm$ & 41\\
3.76 & 0.514 & 1.23 	&	 132	 &  $\pm$	& 13 &	 $\pm$	& 11 &	 1 &	$\pm$& 	 14 &	$\pm$& 	 8.4 & -17 & $\pm$ & 37 & 	$\pm$ & 40\\
4.23 & 0.539 & 0.78 	&	 178	 &  $\pm$	& 42 &	 $\pm$	& 45 &	 -28 &	$\pm$& 	 60 &	$\pm$& 	 57 & -34 & $\pm$ & 74 & 	$\pm$ & 64\\
\end{longtable}

\twocolumngrid

\

\end{document}